\documentclass[a4paper,11pt]{article}
\usepackage{jheppub} 

\usepackage[bitstream-charter]{mathdesign}
\usepackage[amsmath]{empheq}
\numberwithin{equation}{section}
\DeclareSymbolFont{usualmathcal}{OMS}{cmsy}{m}{n}
\DeclareSymbolFontAlphabet{\mathcal}{usualmathcal}

\usepackage{comment}
\usepackage{braket}
\usepackage{bbm}
\usepackage{bm}
\usepackage{tikz}
\usepackage{color}
\usepackage{graphicx}

\usepackage{subcaption}

\newcommand{\be}{\begin{equation}}
	\newcommand{\ee}{\end{equation}}
\newcommand{\ba}{\begin{aligned}}
	\newcommand{\ea}{\end{aligned}}

\newcommand{\bs}[1]{\boldsymbol{\mathbf{#1}}}

\newcommand{\tp}{\gamma_+}
\newcommand{\tm}{\gamma_-}
\newcommand{\tpm}{\gamma_{\pm}}
\newcommand{\tonly}{\gamma}

\newcommand{\tr}{\mathrm{tr}}

\newcommand{\B}{\mathrm M}

\title{\boldmath Universality in the tripartite information after global quenches: (generalised) quantum XY models}

\author[a]{Vanja Mari\'c,}
\author[a]{Maurizio Fagotti}
\affiliation[a]{Universit\'e Paris-Saclay, CNRS, LPTMS, 91405, Orsay, France}

\emailAdd{maurizio.fagotti@universite-paris-saclay.fr}

\abstract{We consider the R\'enyi-$\alpha$ tripartite information $I_3^{(\alpha)}$ of three adjacent subsystems in the stationary state emerging after global quenches in noninteracting spin chains from both homogeneous and bipartite states. We identify settings in which $I_3^{(\alpha)}$ remains nonzero also in the limit of infinite lengths and develop an effective quantum field theory description of free fermionic fields on a ladder. We map the calculation into a Riemann-Hilbert problem with a piecewise constant matrix for a doubly connected domain. We find an explicit solution for $\alpha=2$ and an implicit one for $\alpha>2$. In the latter case, we develop a rapidly convergent perturbation theory that we use to derive analytic formulae approximating $I_3^{(\alpha)}$ with outstanding accuracy.}

\makeatletter
\gdef\@fpheader{}
\makeatother

\begin{document}
\maketitle
\flushbottom

\section{Introduction}
\label{sec:intro}
Nonequilibrium time evolution in isolated quantum many-body systems has been intensively investigated for its capability to turn a ground state into an effective finite-temperature one \cite{Polkovnikov2011Colloquium,Eisert2015Quantum,Gogolin2016Equilibration}. Such a remarkable feature, closely related to the eigenstate thermalisation hypothesis, is expected quite generally if the interactions decay sufficiently fast to zero with the distance \cite{Rigol2008,deutsch91,Srednicki1994Chaos, Deutsch_2018}. The main exception is found in integrable systems, in which the stationary properties of observables could differ even significantly from those of thermal states \cite{Kinoshita2006,Rigol2007Relaxation,Vidmar2016Generalized}. For example, in one dimensional systems with local interactions the correlation lengths are generally finite at nonzero temperature (which can be connected with the absence of phase transitions) and the entropy of subsystems is extensive (which can be traced back to the thermodynamic laws); on the other hand, at zero temperature the correlation lengths can diverge (in critical systems) and the entropy is sub-extensive \cite{Eisert2010Area_laws}. Thus, at equilibrium in 1D systems the divergence of a correlation length and the extensivity of the entropy seem to be incompatible features. 
In the stationary state emerging after a quench in an integrable system, such two properties can instead coexist \cite{Maric2022Universality}, suggesting the potential genesis of exotic states of matter. 

In the last three decades the measures of entanglement have acquired an increasingly important role in quantifying the universal properties of a quantum many-body system, especially for their mild dependence on the system's details \cite{Amico2008,Eisert2010Area_laws,Horodecki2009Quantum}. We also mention that, while probing entanglement experimentally is still a difficult problem, progress has been made also in that respect~\cite{Kaufman2016}. 
The best example of entanglement measure is arguably the von Neumann entropy of subsystems.  Restricting to 1D systems, which are the ones we consider in this paper, universality is usually associated with the zero temperature properties at quantum phase transitions; in this regard the entanglement entropy turns out to be  sensitive to criticality at the leading order of an asymptotic expansion in the subsystem's length~\cite{Calabrese2004Entanglement,Calabrese2009Entanglement,Holzhey1994Geometric,Korepin2004PRL}. It could however also happen that the universal properties of interest are subleading with respect to non-universal ones. This is the case for the topological contribution to the entanglement entropy in 2D, which is concealed behind a term proportional to the area of the subsystem \cite{Kitaev2006Topological}. It becomes then desirable to undress the quantities of the unwanted non-universal details. In the aforementioned example, this is done by considering a special combination of entanglement entropies, known as ``tripartite information''\cite{Cerf1998Information}, that in the context of topological order is usually referred to as ``topological entanglement entropy''\cite{Kitaev2006Topological}. 

The measures of entanglement have been thoroughly studied also after global quenches, establishing for example that the entropy saturates to an extensive value, that is to say, proportional to the size of the subsystem \cite{Calabrese2005Evolution,Sotiriadis2008,Bastianello2018Spreading,Bertini2022Growth,Alba2017Entanglement,Alba2018Entanglement,Casini2016spread,Liu2014,Skinner2019}. One could argue that such extensive parts are mainly due to classical correlations, therefore, it becomes of interest to study combinations of entropies that simplify such contributions, in complete analogy with the topological entanglement entropy. 
For this reason, quantities such as the mutual information have attracted special attention~\cite{Alba2019QuantumInformationScrambling,Bertini2022EntanglementNegativity,Eisler2014,Parez2022}.  

At late times after a global quench, correlations often exhibit an exponential decay in the distance \cite{Calabrese2011Quantum,Calabrese2012,Sotiriadis2008}, but this is not a strict rule (see, e.g., \cite{Maric2022Universality}) and there are many examples in which they decay algebraically, i.e., there are divergent correlation lengths. In this paper we are interested exactly in such exotic states that emerge after global quenches and that display both extensive entropy and infinite correlation lengths. Similarly to the topological case mentioned above, in those states the leading asymptotic behaviour of the entanglement entropy is not universal and is not qualitatively different from what one would find in any other stationary state with finite correlation lengths: the universal properties that we envisage to exist are subleading. Inspired by the definition of the topological entanglement entropy, 
we turn our attention to the tripartite information. This is a special combination of entanglement entropies that simplifies both extensive and boundary contributions (of the subsystem), leaving a potentially universal quantity. We show that this is indeed the case.
Specifically, we derive and generalise the results announced in Ref.~\cite{Maric2022Universality}.
We study the R\'enyi-$\alpha$ tripartite information of three adjacent intervals in the stationary states emerging after global quenches in noninteracting spin chains. We identify and discuss a wide class of global quantum quench protocols which at infinite time reach stationary states with a non-zero tripartite information. We derive the asymptotic behaviour of the R\'enyi-$\alpha$ tripartite information by taking a continuum limit and mapping the calculation to a Riemann-Hilbert problem with a piecewise constant matrix for a doubly-connected domain. We show that the R\'enyi-$\alpha$ tripartite information depends on few system details and discuss a limit, which Ref.~\cite{Maric2022Universality} dubbed ``residual tripatrite information'', that seems to be even independent of them. Technically speaking, the methods that we employ build on Refs~\cite{Casini2009reduced_density,Casini2009Entanglement,Fries2019,Blanco2022}, which we enrich to take into account the important differences~\cite{Igloi2010,Fagotti2010disjoint} between the entanglement of spins and that of fermions.

The paper is organized as follows. In section \ref{sec:Tripartite information} we introduce the tripartite information and discuss its behavior in different equilibrium settings. Section \ref{sec:model} introduces the non-interacting models and quantum quench protocols we study, discusses the differences between the entanglement of spins and fermions and introduces the essentials of the 
free fermionic techniques that we will use. Section \ref{sec:Results} is a summary of the results obtained in this work, and it has been postponed up to that point only because some quantities are defined in the preceding sections; a reader familiar with the concepts introduced in Sections~\ref{sec:Tripartite information} and \ref{sec:model} is encouraged to read Section~\ref{sec:Results} before the previous ones. Section \ref{sec:Results} shows also examples where the asymptotic formulae, which will be derived later on, are applied to concrete quench protocols. 
The derivations of the results are presented afterwards.
Section \ref{sec:Field theory description} introduces a continuum description, shows that stationary states of the studied quench protocols can be described by an effective quantum field theory of free fermionic fields on a ladder, presents a brief overview of the formalism for computing the entanglement entropies in quantum field theory and applies it 
to derive expressions that enable mapping the calculation to a Riemann-Hilbert problem. The mapping is the subject of section \ref{sec:Riemann-Hilbert problem}. Section \ref{sec:Solution to the Riemann-Hilbert problem} deals with the solution of the Riemann-Hilbert problem, which will be exact for $\alpha=2$ and perturbative for $\alpha>2$. 
Conclusions are drawn in section \ref{sec:Conclusions}.

\section{Tripartite information and its residual value}
\label{sec:Tripartite information}

The von Neumann entropy $S_1(A)$ of a subsystem $A$ described by a reduced density matrix $\rho_A$ is  the Shannon entropy of $\rho_A$'s eigenvalues:
\be
S_1(A)=-\mathrm{tr}[\rho_A\log\rho_A]\, .
\ee
If the state is pure, $S_1(A)$ measures the entanglement between $A$ and the rest. Additional information about the entanglement is provided by the R\'enyi entropies
\be
S_\alpha(A)=\frac{1}{1-\alpha}\log\mathrm{tr}[\rho_A^\alpha]\, ,
\ee
with $\alpha=2,3,4,\ldots$, which completely characterise the distribution of eigenvalues of $\rho_A$ (also known as entanglement spectrum). If the state is mixed, the von Neumann entropy and the R\'enyi entropies, which we will also generically refer to as ``entanglement entropies'', are also affected by classical correlations, that is to say, they are not good measures of entanglement.  
\begin{figure}
\centering
\begin{tikzpicture}[scale=0.4]
     \draw[black,line width=0.6pt] (-1,0) to (27,0);
    \foreach \x in {0,...,26}
    \filldraw[ball color=blue!20!white,opacity=0.75,shading=ball] (\x,0) circle (6pt);

    \node[anchor=south] at (6.5,0.6) {\large $A$};
     \node[anchor=south] at (13,0.6) {\large $B$};
      \node[anchor=south] at (19.5,0.6) {\large $C$};
    \draw[ rounded corners = 1,thick] (3.6,-0.5) rectangle ++(5.8,1);
     \draw[ rounded corners = 1,thick] (9.6,-0.5) rectangle ++(6.8,1);
 \draw[rounded corners = 1,thick] (16.6,-0.5) rectangle ++(5.8,1);
    \end{tikzpicture}
    \caption{By $A$, $B$ or $C$ we denote a connected block of spins, i.e. an interval, in an infinite spin chain. In this work we study the tripartite information of three adjacent blocks. }
    \label{fig blocks}
\end{figure}
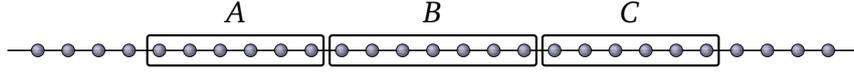

In this work we will indicate the subsystems with words of letters, $A$, $B$, and $C$, that refer to connected blocks of spins, and we will always assume that $A$, $B$, and $C$ are adjacent, as shown in Figure \ref{fig blocks}. In such a geometry, at finite temperature the entropies  become extensive with the size of the subsystem and carry a leading classical contribution. 
Such extensive terms can be cancelled in combinations of entropies, such as in the mutual information.  The mutual information of two subsystems $A$ and $C$ is defined as~\cite{Cerf1998Information}
\be
I_2(A,C)=S_1(A)+S_1(C)-S_1(A\cup C)
\ee
and its R\'enyi analogue reads
\be
I_2^{(\alpha)}(A,C)=S_\alpha(A)+S_\alpha(C)-S_\alpha(A\cup C)\, .
\ee
The simplification of the extensive contributions is manifest in thermal states when $C$ is the complement of $A$ (i.e., $B=\emptyset$ and $A$ and $C$ extend over the entire chain in Fig.~\ref{fig blocks}), where it was shown that  the mutual information satisfies an area law ~\cite{Wolf2008,Kuwahara2021,Eisert2010Area_laws,Bernigau2015,Lemm2022,Alhambra2022}.

The mutual information is a natural quantity to study also in quantum field theory, and it has received substantial attention both in 1+1~\cite{Casini2005,Calabrese2009Entanglement1,Casini2009reduced_density,Casini2007mutual,Furukawa2009Mutual,Headrick2010,Chen2013JHEP,Chen2014Holographic,Headrick2015,Casini2015Mutual,Blanco2011,Fries2019,Arias2018,Blanco2019,Blanco2022,Ares2022,Asplund2014,Molina-Vilaplana2011,Lepori2022} and higher dimensions \cite{Casini2009Remarks,Shiba2012,Cardy2013,Allais2012,Agon2016,Agon2016Large,Chen2017PRD,Casini2015area,Tonni2011}: while the entanglement entropies are ultraviolet divergent, such divergences cancel out in the mutual information, provided that the subsystems are not adjacent.

A related quantity is the tripartite information. For three subsystems $A,B,C$ the tripartite information~\cite{Cerf1998Information} is defined as
\begin{multline}\label{eq:tripartite information definition}
	I_3(A,B,C)=I_2(A,B)+I_2(A,C)-I_2(A,B\cup C)=\\
	S_1(A)+S_1(B)+S_1(C)-S_1(A\cup B)-S_1(B\cup C)-S_1(A\cup C)+S_1(A\cup B\cup C)
\end{multline}
and its R\'enyi analogue reads
\be\label{eq:tripartite information renyi definition}
I_3^{(\alpha)}(A,B,C)=I_2^{(\alpha)}(A,B)+I_2^{(\alpha)}(A,C)-I_2^{(\alpha)}(A,B\cup C)\, ,
\ee
thus it can be viewed as a measure of the extensivity of the mutual information.
While the mutual information can be interpreted~\cite{Groisman2005} as a measure of the total amount of correlations, classical and quantum, between the two subsystems, to the best of our knowledge there is no such a simple interpretation for the tripartite information.
There have been numeorus investigations into the tripartite information in disparate settings.
Besides the remarkable role in quantifying topological properties~\cite{Kitaev2006Topological}, which we have already mentioned, it has received a substantial attention in quantum field theories in 1+1 dimensions \cite{Casini2009Remarks,Caraglio2008Entanglement,Furukawa2009Mutual,Rajabpour2012,Calabrese2009Entanglement1,Calabrese2011Entanglement,Coser2014OnRenyi,Ruggiero2018,Alba2010Entanglement,Blanco2011,Alba2011Entanglement,Fagotti2010disjoint,Fagotti2012New,Fries2019,Balasubramanian2011,Grava2021,Ares2021Crossing}, mainly conformal, as well as in higher dimensions \cite{Casini2009Remarks,Agon2016,Agon2022,AliAkbari2021}. There is also a general result \cite{Hayden2013Holographic} that in holographic theories the tripartite information is never positive, implying that the mutual information is always extensive or superextensive, and suggesting that quantum entanglement dominates over classical correlations. En passant, we note that the tripartite information was recently studied in continuously monitored chains~\cite{Carollo2022}, on Hamming graphs~\cite{Parez2022Multipartite}, and also as a diagnostic for quantum scrambling \cite{Hosur2016Chaos,Schnaack2019Tripartite,Sunderhauf2019Quantum,Kuno2022}.

The key reason why we focus on the tripartite information is that 
it remains bounded in the limit $a\ll |B|\ll |A|,|C|$, allowing us to unveil a contribution that is subleading  in the mutual information and that exhibits universal properties. Here and in the following $|A|$ stands for the size of $A$.
To be more quantitative, let us consider a connected block $A$.
In many states of interest in spin chains (e.g., at equilibrium), the entanglement entropies of such a block  are captured by the Ansatz
\be\label{eq:Ansatz}
S_\alpha(A)\sim \kappa_\alpha \frac{|A|}{a}+\omega_\alpha \log\frac{|A|}{a}+d_\alpha\, ,
\ee
where $a$ is the lattice spacing and $\sim$ denotes the limit in which the subsystem size is much larger than $a$. In the rest of the paper 
we will use  $\sim$ to denote the limit in which all subsystem's sizes 
are much larger than the lattice spacing. In the ground state of a local Hamiltonian $\kappa_\alpha$ vanishes, while $\omega_\alpha$ is nonzero only if there are divergent correlation lengths \cite{Hastings2007}.  In equilibrium at finite temperature $\kappa_\alpha$ becomes nonzero and $\omega_\alpha$ is expected to vanish  (this was explicitly shown in conformal field theories~\cite{Calabrese2004Entanglement}, and, using the approach of Ref.~\cite{Jin2004}, could be rather easily proved in the absence of interactions; it can also be argued by the absence of phase transitions at finite temperature, but we are not aware of a general proof).
A non-zero value of both $\kappa_\alpha$ and $\omega_\alpha$ was instead observed in non-equilibrium steady states~\cite{Eisler2014,Fraenkel2021,FagottiMaricZadnik2022} and in excited states~\cite{Alba2009Entanglement,Ares2014}, as well as after global quenches from ground states of critical systems~\cite{Maric2022Universality}.

Assuming translational invariance, from~\eqref{eq:Ansatz} it follows that, if $A$ and $B$ are adjacent intervals, the mutual information behaves as
\be\label{eq:I2AB}
I_2^{(\alpha)}(A,B)\sim \omega_\alpha \log\frac{|A||B|}{a(|A|+|B|)}+d_\alpha\, ,
\ee
which explicitly shows the dependency on the UV cut-off $a$ as well as on the boundary contributions included in $d_\alpha$.  Note that the mutual information diverges in the limit of infinite size of the blocks $A$ and $B$. For three adjacent intervals $A,B,C$ (see Figure \ref{fig blocks}),
the behaviour of the tripartite information follows from that of the mutual information (cf. \eqref{eq:I2AB} and \eqref{eq:tripartite information renyi definition})
\begin{equation}
I_3^{(\alpha)}(A,B,C)\sim \omega_\alpha\log(1-x)+I_2^{(\alpha)}(A,C),
\end{equation}
where $x=|A||C|/((|A|+|B|)(|B|+|C|))$ is the cross ratio.
In equilibrium at nonzero temperature, $I_2(A,C)$ vanishes exponentially fast with $|B|$ irrespective of $A,C$ \cite{Bluhm2022exponentialdecayof,Alhambra2022} hence, since $\omega_\alpha$ is zero, $I_3^{(\alpha)}(A,B,C)$ approaches zero in the limit of large lengths. This applies also to the ground state of noncritical systems, which satisfy the area law~\cite{Hastings2007} and exhibit exponential decay of correlations \cite{Hastings2006}. On the other hand, in the ground state of a conformal critical system, $\omega_\alpha$ is proportional to the central charge and we need more refined arguments to infer the values of the tripartite information. In a 1+1 dimensional conformal field theory for the configuration of Figure \ref{fig blocks} one has~\cite{Calabrese2009Entanglement1}
\begin{equation}
	\tr\bs \rho_{A\cup C}^\alpha \sim c_\alpha^2\left(\frac{a^2(|A|+|B|)(|B|+|C|)}{|A||B||C|(|A|+|B|+|C|)}\right)^{\frac{c}{6}(\alpha-\frac{1}{\alpha})}\mathcal{F}_\alpha(x),
\end{equation}
where $c$ is the central charge, $c_\alpha$ is a non-universal normalization factor and $\mathcal{F}_\alpha(x)$ is a universal function of the cross ratio $x=|A||C|/((|A|+|B|)(|A|+|C|))$ which depends on the full operator content of the theory. Consequently, the tripartite information is a universal function of $x$, directly related to $\mathcal{F}$ through
\be
I_{3,\, CFT}^{(\alpha)}(A,B,C)\sim \mathcal G_\alpha(x)\, , \quad \mathcal{G}_\alpha(x)=\frac{1}{\alpha-1}\log[\mathcal{F}_\alpha(x)].
\ee
We mention that $\mathcal G_\alpha(x)$ can be positive, zero, or negative \cite{Casini2009Remarks}, the latter case occurring,  in particular, when the central charge is large enough~\cite{Fagotti2012New}. 

Generally, however, $\mathcal G_\alpha(x)$ exhibits two main properties: $\mathcal G_\alpha(0^+)=0$, which manifests cluster decomposition, and $\mathcal G_\alpha(x)=\mathcal G_\alpha(1-x)$, which is a consequence of crossing symmetry (it follows from the fact that, in a pure state, the entropy of a subsystem matches the entropy of its complement). These two properties together imply $\mathcal G_\alpha(1^-)= 0$.  
That is to say, in all the situations covered in this brief summary the tripartite information is zero in the limit $a\ll |B|\ll |A|,|C|$, which corresponds to the limit $x\to 1^-$. Note also that the value of the tripartite information for $|B|=0$ is zero by definition\footnote{This follows from the convention that the entanglement entropy of an empty set is zero. Defining it differently would lead to a violation of some desirable properties, such as the strong subadditivity of the Von Neumann entanglement entropy. Moreover, we want to keep the property that in pure states the entropy of a subsystem is equal to the entropy of its complement. Since the entropy of the whole system is zero, we should also set it to zero for its complement, the empty set.}.
A nonzero value of $I_3^{(\alpha)}(A,B,C)$ in the limit $a\ll |B|\ll |A|,|C|$, which was called ``residual tripartite information''~\cite{Maric2022Universality},  is therefore an indicator of  properties that are not generally found in systems at equilibrium.
Ref.~\cite{Maric2022Universality} has identified some nonequilibrium settings which are expected to exhibit a residual tripartite information.
Focusing on those quench protocols, which we introduce in the next section, we will analytically compute the R\'enyi-$\alpha$ tripartite information of adjacent subsystems in the limit in which the subsystem's lengths approach infinity.

\begin{figure}
	\centering
	\includegraphics[width=0.7\textwidth]{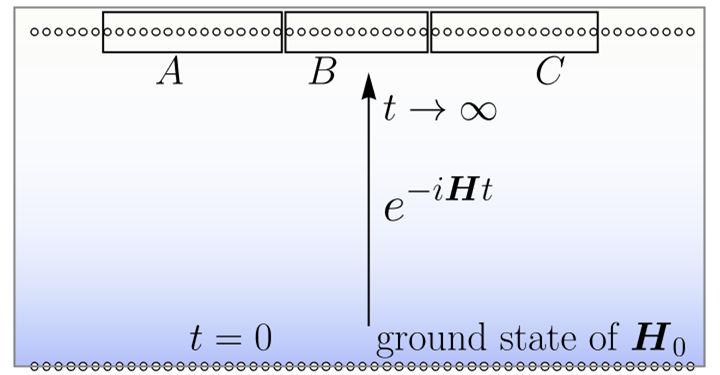}
	\caption{Schematic representation of a global quench from the ground state of a local Hamiltonian $\bs H_0$. We study the tripartite information of $A$, $B$, and $C$ in the limit of infinite time. }
	\label{fig:schematic_critical}
\end{figure}

\section{The model}
\label{sec:model}
We consider a class of Hamiltonians, known as generalised XY model \cite{Suzuki1971The,Lieb1961}, which are mapped into quadratic forms of fermions by the Jordan-Wigner transformation 
\be\label{Jordan Wigner transformation}
\bs a_{2\ell-1}=\prod_{j<\ell}\bs \sigma_j^z \bs \sigma_\ell^x\qquad
\bs a_{2\ell}=\prod_{j<\ell}\bs \sigma_j^z \bs \sigma_\ell^y\, .
\ee 
The fermions defined by Eq.~\eqref{Jordan Wigner transformation} are self-adjoint (Majorana) and satisfy the algebra
\begin{equation}
    \{\bs a_{j}, \bs a_\ell\}= 2\delta_{j\ell} \; .
\end{equation}
The most studied example belonging to this class is the quantum XY model
\be\label{eq:XY}
\bs H=\sum\nolimits_\ell\left( J_x\bs\sigma_\ell^x\bs\sigma_{\ell+1}^x+J_y\bs\sigma_\ell^y\bs\sigma_{\ell+1}^y+h\bs\sigma_\ell^z\, \right),
\ee
of which the XX model ($J_x=J_y$) and the transverse-field Ising model ($J_y=0$) are special cases. 
The generalisation, which is described in Appendix \ref{appendix generalized XY models}, can be dealt with analogously to the XY model. It allows for interactions of any range, though with a particular structure. For the sake of simplicity we shall assume that the Hamiltonian is local, i.e., that the interactions have a finite range\footnote{All the results we are going to present are expected to hold true even with  quasilocal Hamiltonians in which the coupling constants decay exponentially fast with the range.}.

By translational invariance, $\bs H$ can be expressed as follows\footnote{More precisely, with periodic boundary conditions the mapping would result in two distinct noninteracting sectors, which however turn out to be equivalent for our purposes.}
\begin{equation}\label{eq:symbol of the Hamiltonian definition}
	\bs H=\frac{1}{4}\sum_{j,\ell} \vec{\bs a}_j^\dagger \int_{-\pi}^{\pi}\frac{dk}{2\pi}\mathcal{H}(k)e^{ik(j-l)} \ \vec{\bs a}_\ell,  \quad \vec{\bs a}_\ell\equiv \begin{pmatrix}
		\bs a_{2\ell-1} \\ \bs a_{2\ell} 
	\end{pmatrix},
\end{equation}
where the $2\times2$ matrix $\mathcal{H}(k)$ generates the coupling constants through its Fourier coefficients and, from now on, will be referred to as the symbol of the Hamiltonian \cite{Fagotti2016Charges}. To be explicit, Hamiltonian \eqref{eq:XY} has the following symbol
\begin{equation}
	\mathcal{H}(k)=2(J_x-J_y)\sin(k)\sigma^x+[2h-2(J_x+J_y)\cos(k)]\sigma^y \; .
\end{equation}
The properties of the symbol in view of Hamiltonian's locality are given in Table \ref{t:list}. In particular, local Hamiltonians are characterized by a smooth symbol.

\begin{figure}
	\centering
	\includegraphics[width=0.7\textwidth]{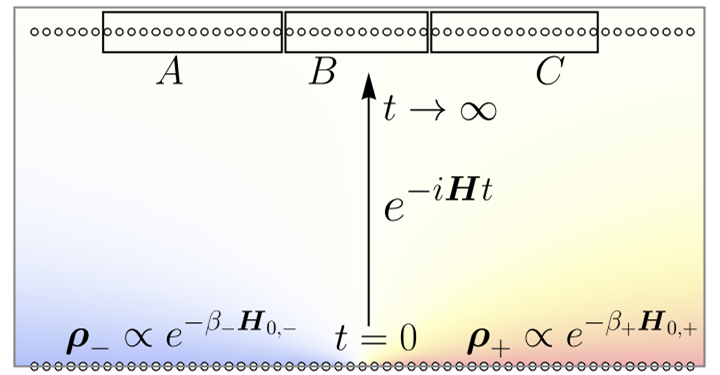}
	\caption{Schematic representation of a bipartitioning protocol in which the quench is global. We study  the limit of infinite time in the tripartite information of $A$, $B$, and $C$  in which the lengths $|A|, |B|, |C|$ will be assumed large but not enough to be affected by the inhomogeneity of the state. }
	\label{fig:schematic_bipartitioning}
\end{figure}

We consider two kinds of quenches: 
\begin{description}
	\item[Quench from a critical point:] The initial state is the ground state of a critical Hamiltonian $\bs H_0$ with local interactions, such as the XX model. 
	In the symbol of $\bs H_0$ criticality is manifested in the existence of (at least) a momentum $\bar k$ for which the kernel of $\mathcal H_0(\bar k)$ is non-empty.
	
	The state is let to evolve with a different local Hamiltonian $\bs H$ for an arbitrarily long time (see Fig.~\ref{fig:schematic_critical} for a sketch).
	
	If we relax the constraint of criticality in the initial state, this is the standard example of global quench and, as such, has been thoroughly investigated (we refer the reader to Refs~\cite{Calabrese2016Quantum,Essler2016Quench,Vidmar2016Generalized,Caux2016} for some reviews on the topic that are relevant to our work). In integrable systems, like the generalised XY model, local observables approach stationary values described by generalised Gibbs ensembles (GGEs), which retain memory of infinitely many integrals of motion. 
	
	In this paper we will assume that the time is so large that the observables in the subsystems can be described by a GGE.  
	
	\item[Bipartitionig protocol:] The initial state is separated in two parts with macroscopically different properties, for example, a domain-wall of spins or two thermal states of a local Hamiltonian $\bs H_0$ at different temperatures. 
	
	The state is let to evolve with a different local Hamiltonian $\bs H$ for an arbitrarily long time (see Fig.~\ref{fig:schematic_bipartitioning} for a sketch).
	
	Similar bipartitioning protocols in integrable systems have recently received a lot of attention, we mention Refs~\cite{Sabetta2013,Eisler2014Entanglement,Mazza2018,Bertini2018Entanglement,Alba2019Entanglement,Gruber2019Magnetization,Collura2020Domain} as an incomplete list of works focussed on integrable systems including also investigations into the entanglement properties; the interested reader is encouraged to consult the Collection~\cite{Bastianello2022Introduction} for an overview (Ref.~\cite{Alba2021Generalized} in particular reviews the case considered here in which the quench is global). 
	
	In integrable systems, the abrupt change of properties of the initial state experiences a ballistic spreading in time in which the expectation values of local observables approach values that depend on the ratio between the distance from the inhomogeneity and the time.
	This is the paradigm of quench protocol captured by the theory of generalised hydrodynamics \cite{Bertini2016Determination,Bertini2016Transport,Castro-Alvaredo2016Emergent}. The limit of infinite time, in particular, is described by a nonequilibrium stationary state, usually called NESS, which can be represented by a generalised Gibbs ensemble notwithstanding the entire state not being translationally invariant.
	
	In this paper we will assume that the time is so large that the observables in the subsystems can be described by a NESS.
	
\end{description}

\subsection{Correlation matrices and filling functions}
\begin{table}[!t]
	\begin{center}
		\begin{tabular}{l|l}
			$2$-by-$2$ symbol of $M$ ($M_{2\ell+i, 2n+j}=M_{2(\ell-n)+i, j}$, $\ell,n\in\mathbb Z$):& $M_{i j}(k)=\sum\limits_{n\in \mathbb Z} e^{i k n} M_{2(\ell-1)+i,2(\ell+n-1)+j}$\\
			\hline
			Jordan-Wigner transformation:&$\vec{\bs a}_\ell\equiv \begin{pmatrix}
			\bs a_{2\ell-1} \\ \bs a_{2\ell} 
			\end{pmatrix}=\begin{pmatrix}
			\prod_{j<\ell} \bs \sigma_j^z \bs \sigma_\ell^x \\ \prod_{j<\ell}\bs \sigma_j^z \bs \sigma_\ell^y
			\end{pmatrix} $\\
			\hline
			Hamiltonian and its symbol:&$\bs H=\frac{1}{4}\sum_{j,\ell} \vec{\bs a}_j^\dagger \int_{-\pi}^{\pi}\frac{dk}{2\pi}\mathcal{H}(k)e^{ik(j-l)} \ \vec{\bs a}_\ell$\\
			quasilocality:& $\mathcal H(k)$ is smooth\\
			locality:&$\mathcal H(k)$ has a finite number of nonzero\\
			&Fourier coefficients\\
			dispersion relation (reflection symmetric $\bs H$):&$\varepsilon(k)=\frac{1}{2}\sqrt{\tr[\mathcal H^2(k)]}$\\
			\hline
			correlation matrix:&$\Gamma_{2(\ell-1)+i, 2(n-1)+j}=\delta_{\ell n}\delta_{i j}-\braket{\vec{\bs a}_\ell\vec{\bs a}_n^\dagger}_{i j}$\\
			\hline
			ground state of $\bs H$ (w/o symmetry breaking):&$\Gamma(k)=-\mathrm{sgn}\left(\mathcal{H}(k)\right)$\\
			thermal state of $\bs H$:&$\Gamma(k)=-\tanh\left(\frac{\beta\mathcal{H}(k)}{2}\right)$\\
			GGE after a quench $\Gamma_0\rightarrow \bs H$ (refl. sym. $\bs H$):&$\Gamma(k)=\frac{\{\Gamma_0(k),\mathcal{H}(k)\}}{2\varepsilon^2(k)}\mathcal{H}(k)$\\
			NESS after a quench
			$\Gamma_-\otimes\Gamma_+\rightarrow\bs H$ (refl. sym. $\bs H$):&
			
			$\Gamma(k)=\sum_{s=\pm 1}\bigg(\frac{\Gamma_s(k)}{2}$\\
			
			& $-\mathrm{sgn}[\varepsilon'(k)]s\frac{\tr[\Gamma_s(k)\mathcal{H}(k)]+\tr[\Gamma_s(k)]\mathcal{H}(k)}{4\varepsilon(k)}\bigg)$
			
			\\
		\end{tabular}
	\end{center}\caption{Free-fermion dictionary. A list of the main ingredients of the free-fermion techniques used in this paper. The symbol of a reflection-symmetric Hamiltonian of the generalised XY model can be any traceless 2-by-2 matrix function $\mathcal H(k)$ satisfying the usual property of symbols, i.e., $\mathcal H^\dag(k)=\mathcal H(k)=-\mathcal H^t(-k)$.}\label{t:list}
\end{table}

Since we restrict ourselves to noninteracting quenches, the states are Gaussian at any time, i.e., their density matrix is the exponential of a quadratic form of fermions. In such states the expectation values of operators are completely determined by the fermionic two-point correlation matrix $\Gamma_{2\ell+i, 2n+j}=\delta_{\ell n}\delta_{i j}-\braket{\vec{\bs a}_\ell\vec{\bs a}_n^\dagger}_{i j}$. In a one-site shift invariant state the correlation matrix is block Toeplitz and can be expressed in terms of a $2$-by-$2$ matrix function $\Gamma(k)$, its symbol,  by Fourier transform
\begin{equation}
	\Gamma_{2\ell+i,2n+j}=\int_{-\pi}^\pi\frac{dk}{2\pi}e^{ik(\ell-n)}\Gamma_{i j}(k) \; .
\end{equation}
We remind the reader that the algebra of the Majorana fermions implies the property $\Gamma^\dag(k)=\Gamma(k)=-\Gamma^t(-k)$, where ${\bullet}^t$ denotes transposition. 

The correlation matrices for the quenches of interest are shown in Table~\ref{t:list}. In all quenches considered, the late time dynamics can be captured by a generalised Gibbs ensemble; in this respect, we remind the reader that a GGE is locally equivalent to a family of excited states, which, in this context, are called ``representative states'' \cite{Caux2013Time,Essler2016Quench}.
Thus, we briefly introduce also the standard language used to describe excited states in noninteracting models, as well as in a class of interacting integrable systems. 
Given an excited state, the filling function $\vartheta(k)$ is a regularised characteristic function of the set of occupied momenta, i.e., it represents the average occupation of momenta around $k$\footnote{ 
	We warn the reader that the ambiguity in the definition of what is a particle and what is a hole allows for alternative definitions of  filling functions. 
	In particular, one is in principle free to choose any excited state as a reference state characterised by $\vartheta(k)=0$. In order to attach some physical meaning to $\vartheta(k)$, it is however convenient to choose a reference state with low entanglement. 
	If the ground state is non-critical, the most natural choice is to promote the ground state to the reference state (the other possibility being the state with maximal energy). If the ground state is critical, instead, choosing the ground state of a conservation law with a non-critical ground state would do the job. For example, the ground state of the XX model in $0$ field is critical. Since, however, the total magnetization in the $z$ direction commutes with the Hamiltonian, we can choose as reference state the ground state of the model with a large enough magnetic field, which is the state in which all spins are aligned in the $z$ direction. This is a customary choice in models with a $U(1)$ symmetry. In this paper we will use such standard conventions.}. Filling functions, together with the closely related root densities\footnote{In noninteracting systems, in particular, the filling function $\vartheta(k)$ and the root density $\rho(k)$ satisfy $\vartheta(k)=2\pi \rho(k)$.}, are the main ingredients of the theory of generalised hydrodynamics. Since they satisfy simple dynamical equations (see Ref.~\cite{Fagotti2020} for a thorough discussion in noninteracting spin chains), they are convenient to work with especially in the presence of inhomogeneities, where other formalisms struggle with.
Assuming that there is a smooth unitary matrix function $ U(k)$ of $k$ diagonalising the symbol of the correlation matrix of a stationary state, the filling function can be identified with $\Gamma(k)$'s eigenvalues as follows:
\begin{equation}\label{eq:correlation matrix symbol eigenvalues and the filling function}
	U^\dagger(k) \Gamma(k)U(k)=\mathrm{diag}\left[2\vartheta(k)-1,-2\vartheta(-k)+1\right].
\end{equation}
We refer the reader to Section 2 of Ref.~\cite{Alba2021Generalized} for a pedagogical introduction to filling functions (or root densities) in noninteracting systems; for the purpose of this paper, it is sufficient to regard \eqref{eq:correlation matrix symbol eigenvalues and the filling function} as a definition of $\vartheta(k)$.

\subsection{R\'enyi entropies}\label{ss:Renyi}
The tripartite information of adjacent subsystems is a linear combination of entropies of spin blocks with the entropy of two disjoint blocks~\eqref{eq:tripartite information definition}. In our system the entropy of a spin block is equal to the entropy of the corresponding block of Jordan-Wigner fermions \cite{Jin2004,Vidal2003} and reads
\be\label{eq:Ssingleblock}
S_\alpha(A)=\tfrac{1}{1-\alpha}\tfrac{1}{2}\tr \Bigl[\log\Bigl(\bigl(\tfrac{\mathrm I+\Gamma_A}{2}\bigr)^\alpha+\bigl(\tfrac{\mathrm I-\Gamma_A}{2}\bigr)^\alpha\Bigr)\Bigr]\, ,
\ee
where $\Gamma_A$ is the correlation matrix in $A$. 
In the case of disjoint blocks, instead, the non-locality of the Jordan-Wigner transformation complicates the correspondence. A procedure for computing the R\'enyi entropies of two disjoint blocks on the lattice that takes into account these non-locality effects has been proposed in \cite{Fagotti2010disjoint} and is briefly reviewed  below.

The entropy $S_\alpha(A \cup C)$ is expressed in terms of four correlation matrices: $\Gamma_1\equiv \Gamma_{A\cup C, A\cup C}$, $\Gamma_2\equiv \mathrm P \Gamma_1 \mathrm P$, $\Gamma_3\equiv \Gamma_1-\Gamma_{A\cup C,B}\Gamma_{B,B}^{-1}\Gamma_{B,A\cup C}$ and $\Gamma_4\equiv P\Gamma_3 P$, where $\Gamma_{A',A''}$ for some subsystems $A',A''$ is the correlation matrix with the indices running in $A',A''$ respectively,
\begin{equation}
\begin{split}
    & \left(\Gamma_{A',A''}\right)_{2(\ell-1)+i,2(n-1)+j}=\delta_{A'(\ell),A''(n)}\delta_{ij}-\braket{\vec{\bs a }_{A'(\ell)} \vec{\bs a}^\dagger _{A''(n)}}_{ij} \; ,\\  \quad & \ell\in\{1,2,\ldots,|A'|\},\ n\in\{1,2,\ldots,|A''|\}, \  i,j\in\{1,2\}, \\ & A'(1)<A'(2)<\ldots <A'(|A'|) \textrm{ are the elements of $A'$ in ascending order (analogously for $A''$)},
\end{split}
\end{equation}
and
\be\label{eq:operator P definition}
\mathrm P_{\ell n}\equiv \delta_{\ell n}\begin{cases}
	-1& 1\leq \ell\leq 2|A|\\
	1&\text{otherwise}
\end{cases}
\ee
introduces a minus sign for each fermion in block $A$. Specifically, we have
\begin{equation}\label{eq:entropy disjoint blocks lattice exact}
	S_\alpha(A \cup C)=\frac{1}{1-\alpha}\log \left[\frac{1}{2^{\alpha}}\sum_{j_1,j_2,\ldots j_\alpha=1,2,3,4 \atop (-1)^{N_3+N_4} =1} (-1)^{N_4}|\det \Gamma_{B,B}|^{\frac{N_3+N_4}{2}}\{\Gamma_{j_1},\Gamma_{j_2},\ldots, \Gamma_{j_\alpha}\}\right].
\end{equation}
Here $N_3$ ($N_4$) is the total number of $j_1,j_2,\ldots j_\alpha$ equal to $3$ ($4$). Only the terms in which the total number of indices equal to $3$ or $4$ is even appear. The symbol $\{\ldots\}$ inside the logarithm is defined as the trace of a product of normalized Gaussians,
\begin{equation}\label{eq:trace of a product of gaussians}
	\{\Gamma_{j_1},\Gamma_{j_2},\ldots, \Gamma_{j_\alpha}\}\equiv\tr \left[\rho(\Gamma_{j_1})\rho(\Gamma_{j_2})\ldots\rho(\Gamma_{j_\alpha})\right],
\end{equation}
where $\rho(\Gamma)\propto \exp{\left(\sum_{j,\ell \in A,C} \vec{\bs a}_j^\dagger W_{j,l} \vec{\bs a}_\ell /4\right)}$ has correlation matrix $\Gamma$ ($\Gamma$ and $W$ are related to each other through $\Gamma=\tanh(W/2)$).

The entropy of two disjoint blocks of fermions consists solely of the first term,
\begin{equation}
	S_\alpha^{\mathrm{(fermions)}}(A\cup C)=\frac{1}{1-\alpha}\log \{\Gamma_1, \Gamma_1,\ldots, \Gamma_1\}.
\end{equation}
The other terms appearing in \eqref{eq:entropy disjoint blocks lattice exact} account for the non-locality of the Jordan-Wigner fermions with respect to the spin degrees of freedom. For instance, the operator $\sigma^x_{j\in A}\sigma^x_{\ell\in C}$, when expressed in terms of fermions, contains an undesired string in between the two blocks. In accordance, terms with  $N_3+N_4\neq 0$ in \eqref{eq:entropy disjoint blocks lattice exact} are accompanied by a factor equal to a power of the expectation value of the string of $\bs \sigma_\ell^z$ operators between the two blocks,
\begin{equation}
	\left| \det \Gamma_{B,B} \right|^{\frac{N_3+N_4}{2}}=\braket{\prod_{j\in B}\bs \sigma_j^z}^{N_3+N_4}\, .
\end{equation}
After global quenches in systems without semilocal charges, string order doesn't survive. That is to say, the expectation value of strings of Pauli matrices decays exponentially with the length of the string (see Ref.~\cite{FagottiMaricZadnik2022} for a more detailed discussion). Thus, it is reasonable to expect the string expectation value to kill all the contributions in \eqref{eq:entropy disjoint blocks lattice exact} from the terms with non-zero $N_3+N_4$. And indeed our numerical checks confirm the validity of this statement. In the limit in which all lengths approach infinity, the entropy of disjoint blocks after a global quench is therefore captured by the following simplified expression 
\begin{equation}\label{eq:entropy disjoint blocks lattice after global quenches}
	S_\alpha(A\cup C)\sim \frac{1}{1-\alpha}\log\left[\frac{1}{2}\sum_{j_1,j_2,\ldots, j_\alpha=1 }^2 \{\Gamma_{j_1},\Gamma_{j_2},\ldots, \Gamma_{j_\alpha}\}\right] +\log 2.
\end{equation}
Each term of the sum can be evaluated using the  recursive formula derived in  Ref.~\cite{Fagotti2010disjoint}, which expresses a product of two normalised  Gaussians as a Gaussian
\begin{equation}\label{eq:product of two gaussians}
	\rho(\Gamma)\rho(\Gamma')=\{\Gamma,\Gamma'\}\rho\left(\Gamma\times \Gamma'\right), \quad \Gamma\times \Gamma'\equiv \mathrm I-(\mathrm I-\Gamma')(\mathrm I+\Gamma\Gamma')^{-1}(\mathrm I-\Gamma)\, ,
\end{equation}
together with a formula for the trace of a product of normalised Gaussians
\begin{equation}\label{eq:trace of a product of two gaussians determinant formula}
	\{\Gamma,\Gamma'\}^2=\mathrm{tr}[\rho(\Gamma)\rho(\Gamma')]^2=\det \bigg(\frac{\mathrm I+\Gamma \Gamma'}{2}\bigg) \; .
\end{equation}
Alternatively,  \eqref{eq:product of two gaussians} could be replaced by the formula for the trace of a product of an arbitrary number of Gaussians reported in Ref.~\cite{Klich2014}. Note that \eqref{eq:trace of a product of two gaussians determinant formula} fixes the value of $\{\Gamma,\Gamma'\}$ only up to a sign; in general such a sign ambiguity can be lifted by computing the product of the eigenvalues with halved degeneracy \cite{Fagotti2010disjoint}.

Finally,  similarly to what happens in the ground state of free fermions~\cite{Casini2009Entanglement}, we anticipate that also in our nonequilibrium setting the fermionic tripartite information asymptotically vanishes. 
Consequently, the asymptotic behaviour of the spin tripartite information after the quench is captured by the following formula
\begin{equation}\label{eq:tripartite information universal ratios}
	I_3^{(\alpha)}(A,B,C)\sim \frac{1}{\alpha-1}\log\left[\frac{1}{2}\sum_{j_1,j_2,\ldots, j_\alpha=1 }^2 \frac{  \{\Gamma_{j_1},\Gamma_{j_2},\ldots, \Gamma_{j_\alpha}\}  }{\{\Gamma_{1},\Gamma_{1},\ldots, \Gamma_{1}\}}\right] -\log 2,
\end{equation}
which we will conveniently use in the analytical calculations but refrain from assuming in the numerical checks. 

\section{Results}
\label{sec:Results}

\subsection{Tripartite information and R\'enyi entropies of disjoint blocks}
\label{sec: results tripartite information}

\begin{description}
	\item[Universality.] We find that, when nonzero, the tripartite information $I_3^{(\alpha)}(A,B,C)$ of adjacent subsystems at late times after global quenches and in the limit of large subsystems ($|A|,|B|,|C|\to \infty$) depends on the quench through few system details and on the subsystems through the cross ratio
	\begin{equation}
		x=\frac{|A||C|}{(|A|+|B|)(|C|+|B|)}\, .
	\end{equation}
	As far as we can see, there is no underlying conformal symmetry that could explain the dependency on just $x$. We prove it analytically for $\alpha=2$ and for some special contributions to $I_3^{(\alpha)}(A,B,C)$ with $\alpha>2$. More generally, we show it to be consistent with our analytical and numerical analysis. 
	
	The most universal behaviour that we observe is in the limit $x\rightarrow 1^-$ (i.e., $|A|,|C|\gg |B|\gg a$, where $a$ is the lattice spacing) in which the tripartite information (see Fig \ref{fig:schematic_residual} for a sketch) is either equal to $0$ (in the trivial cases) or equal to $-\log 2$, independently of $\alpha$.
	Since the result is independent of the R\'enyi index, it also applies to the (von Neumann) tripartite information $I_3(A,B,C)$ ($\alpha\to 1^+$). Following the terminology of Ref.~\cite{Maric2022Universality}, in all the interesting cases we have then found a universal ``residual tripartite information'' equal to $-\log 2$. This should be contrasted with the zero value found in the ground states of conformal systems and in thermal states.
	
	\begin{figure}
		\centering
		\includegraphics[width=0.8\textwidth]{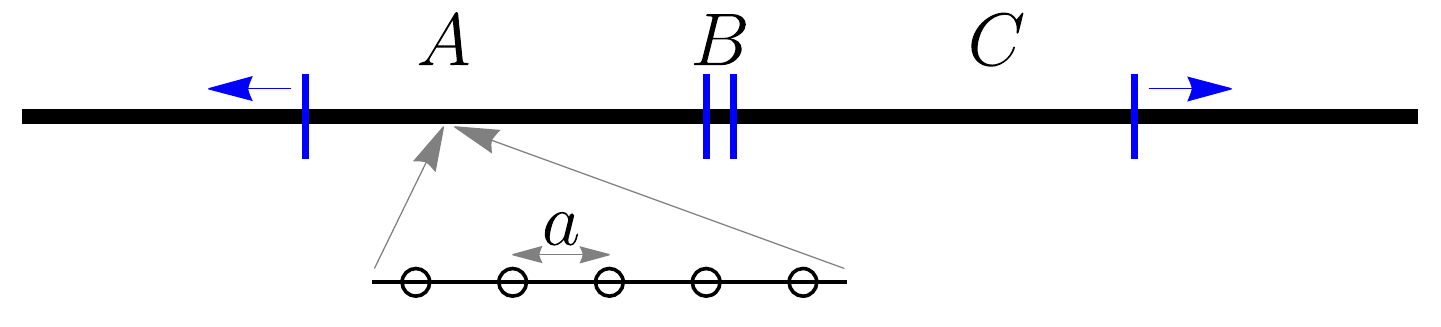}
		\caption{Residual tripartite information: $I_3(A,B,C)$ with  $a\ll |B| \ll |A|,|C|$, where $a$ is the lattice spacing.}
		\label{fig:schematic_residual}
	\end{figure}

	\item[Asymptotic predictions.] Besides asymptotically depending on the cross ratio, the tripartite information depends only on the discontinuities of the filling function characterising the stationary state after the quench. Denoting the filling function by $\vartheta(k)$, as in \eqref{eq:correlation matrix symbol eigenvalues and the filling function}, we parametrize the discontinuties about momentum $k=k_F$ as follows
	\begin{equation}
		\tpm=\lim_{k\to k_F^\pm}[2\vartheta(k)-1]\, .
	\end{equation}
Incidentally, we will also call $k_F$ Fermi momentum, for the role it plays in the ground state of critical systems.
	We obtain the analytic predictions
	\begin{empheq}{equation}\label{eq:result tripartite information alpha 2}
		I_3^{(2)}(A,B,C)=-\log 2+\log\left[1+(1-x)^{\sum_{k_F}\frac{[\arctan(\tp)-\arctan(\tm)]^2}{\pi^2}}\right]
	\end{empheq}
	for $\alpha=2$ and
	\be\label{eq:result tripartite information higher alpha abelian}
	I_3^{(\alpha)}(A,B,C)= -\log 2+\frac{1}{\alpha-1}\log\left[\frac{1}{2}\sum_{\Sigma} (1-x)^{\sum_{k_F}\tr[(R_0\Sigma)^2-R_0^2]}Q(x;\mathrm R_0,\Sigma)\right]
	\ee
	for higher integer $\alpha$. Here the sum is over all $\alpha\times \alpha$ diagonal matrices $\Sigma$ with the elements on the diagonal equal either to $1$ or $-1$ (in total $2^\alpha$ terms), $\mathrm R_0$ is the $\alpha\times\alpha$ circulant matrix defined in \eqref{eq:R0} and $Q(x;\mathrm R_0,\Sigma)$ is a ``non-abelian correction'', always very close to $1$. 
	Specifically, each ratio $\frac{  \{\Gamma_{j_1},\Gamma_{j_2},\ldots, \Gamma_{j_\alpha}\}  }{\{\Gamma_{1},\Gamma_{1},\ldots, \Gamma_{1}\}}$ appearing in \eqref{eq:tripartite information universal ratios} is described by the term in \eqref{eq:result tripartite information higher alpha abelian} with $\Sigma_{\ell n}=-\delta_{\ell n}e^{i\pi j_\ell}$. 
	Formula \eqref{eq:result tripartite information alpha 2} was first conjectured in Ref.~\cite{Maric2022Universality} and we prove here its validity. Formula \eqref{eq:result tripartite information higher alpha abelian} without $Q(x;\mathrm R_0,\Sigma)$ can be obtained as the lowest order of a rapidly convergent perturbation theory based on an implicit solution to a Riemann-Hilbert problem (Section~\ref{ss:perturbation}). We call it "Abelian approximation" since it neglects all the effects of non-commutativity of $\mathrm R_0$ and $\Sigma\mathrm R_0\Sigma$. Although a priori crude, this approximation gives surprisingly accurate results. We lack an explicit expression for the correction to the formula, parametrised here by $Q(x;\mathrm R_0,\Sigma)$, but we have computed it at the leading order in the perturbation theory. The result reads
	\be\label{eq:3rdorderapp}
	Q(x;\mathrm R_0,\Sigma)\approx k_{3,1}(x)^{-\sum_{k_F}\frac{1}{4}\tr[(i[\Sigma\mathrm R_0\Sigma,\mathrm R_0])^2]}\, ,
	\ee
	where $k_{3,1}(x)$ is worked out in Section \ref{sec:Solution to the Riemann-Hilbert problem} and is shown in Fig.~\ref{f:k31}. More generally, each term of the expansion of $\log Q(x;\mathrm R_0,\Sigma)$ is factorised in a term that depends only on $\alpha$ and $\gamma_\pm$ (e.g., the exponent of \eqref{eq:3rdorderapp}) and a term that depends only on the cross ratio $x$ (e.g., $\log k_{3,1}(x)$). In practice, going beyond the third order \eqref{eq:3rdorderapp} seems to be a mathematical problem rather than a physical one, as it is almost impossible to see a disagreement from the third-order approximation.
	
	Predictions apart, it should be noticed that our results imply a non-zero value of the tripartite information after quenches, similarly to the ground states of conformal systems and unlike any thermal (Gibbs) state. A property that makes the behavior of the tripartite information after quenches under consideration different from the former is the absence of the crossing symmetry $x\to 1-x$, which is manifest in the possibility to have a nonzero "residual tripartite information" in a state with clustering properties.
	
	\begin{figure}[!t]
		\centering
		\includegraphics[width=0.6\textwidth]{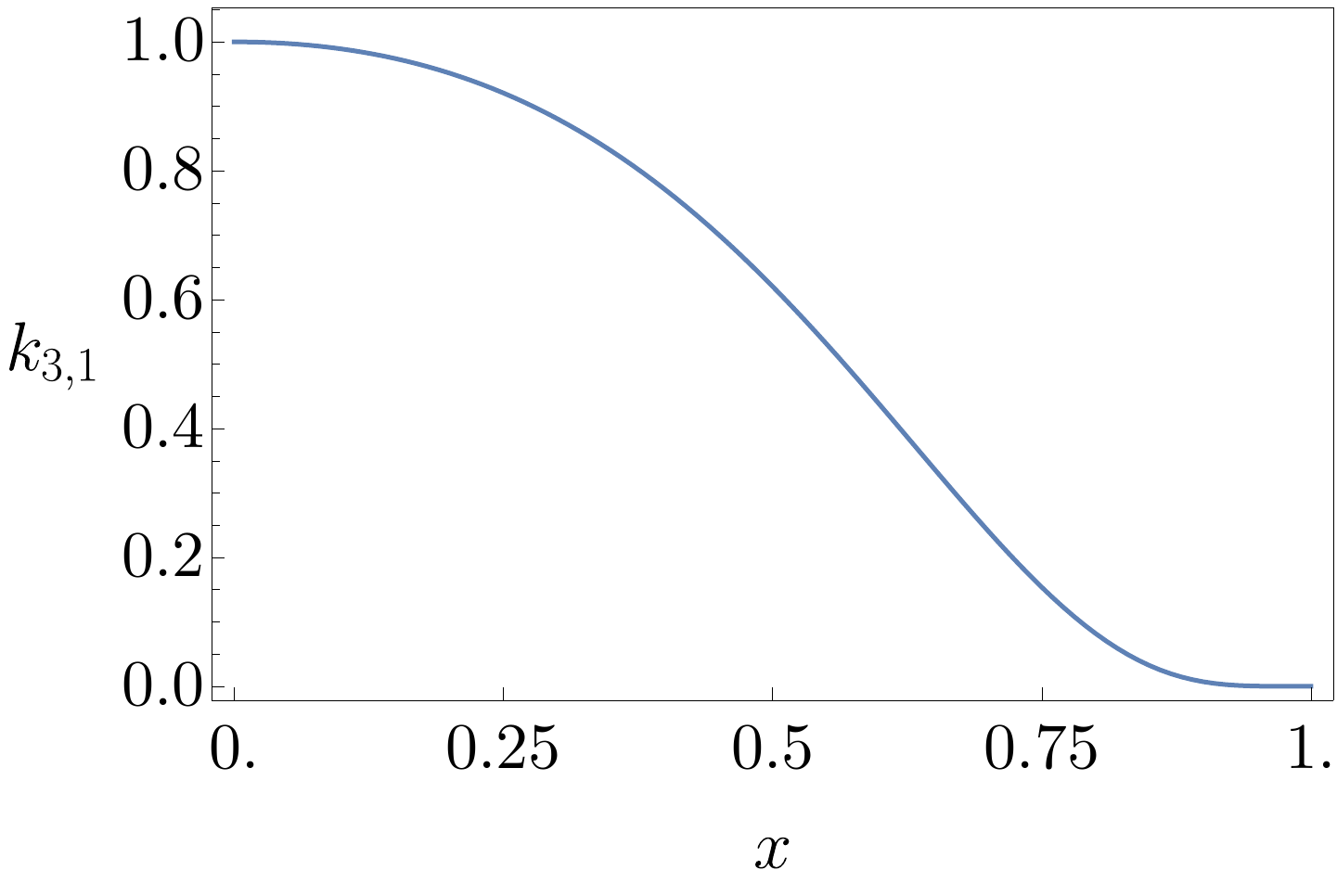}
		\caption{The function $k_{3,1}(x)$ characterizing the dependency of the leading non-Abelian correction~\eqref{eq:3rdorderapp} on the subsystem (through the cross ratio $x$).
		}
		\label{f:k31}
	\end{figure}
	
	\item[Determinant representation.]
	As a side product of this work, we produce simple formulas that can be used as an alternative to those worked out in Ref.~\cite{Fagotti2010disjoint} (discussed in section \ref{ss:Renyi}) for computing the terms with $j_1,j_2,\ldots,j_\alpha\in\{1,2\}$ in \eqref{eq:entropy disjoint blocks lattice exact}, i.e., all terms appearing in \eqref{eq:tripartite information universal ratios}.
	Specifically, we argue (see Appendix \ref{appendix generalisation and back to the chain}) that any such term can be evaluated with the formula
	\be\label{eq:trace of a product of gaussians lattice P on the left}
	\{\Gamma_{j_1},\hdots,\Gamma_{j_\alpha}\}=\sqrt{\det\left| 2^{\frac{1-\alpha}{\alpha}}\mathrm I_\alpha\otimes \mathrm I_{2(\ell_A+\ell_C)}-i 2^{\frac{1-\alpha}{\alpha}}
		\left[\frac{\B+\Sigma \B\Sigma}{2}\otimes \Gamma_{1}-\frac{\B-\Sigma \B\Sigma}{2}\otimes (P \Gamma_{1})
		\right]\right| },
	\ee
	where $\B$ is the $\alpha\times \alpha$ Hermitian circulant matrix defined in \eqref{eq:Bdef}.
	This representation is more useful than the one proposed in Ref.~\cite{Fagotti2010disjoint}: on the one hand, the matrix in the determinant is linear in $\Gamma_{1}$ and, on the other hand, it allows one to evaluate the R\'enyi entropy even for arbitrarily large values of $\alpha$ without the need of working out the recursive procedure of Ref.~\cite{Fagotti2010disjoint}.
	We expect  similar representations for any term appearing in formula \eqref{eq:entropy disjoint blocks lattice exact} for the R\'enyi entropy of disjoint
	blocks, but we have not developed further in that direction.

\end{description}

\subsection{Examples}
\begin{figure}[!p]
	\centering
	\begin{minipage}{0.48\textwidth}
		\centering
		\includegraphics[width=1\textwidth]{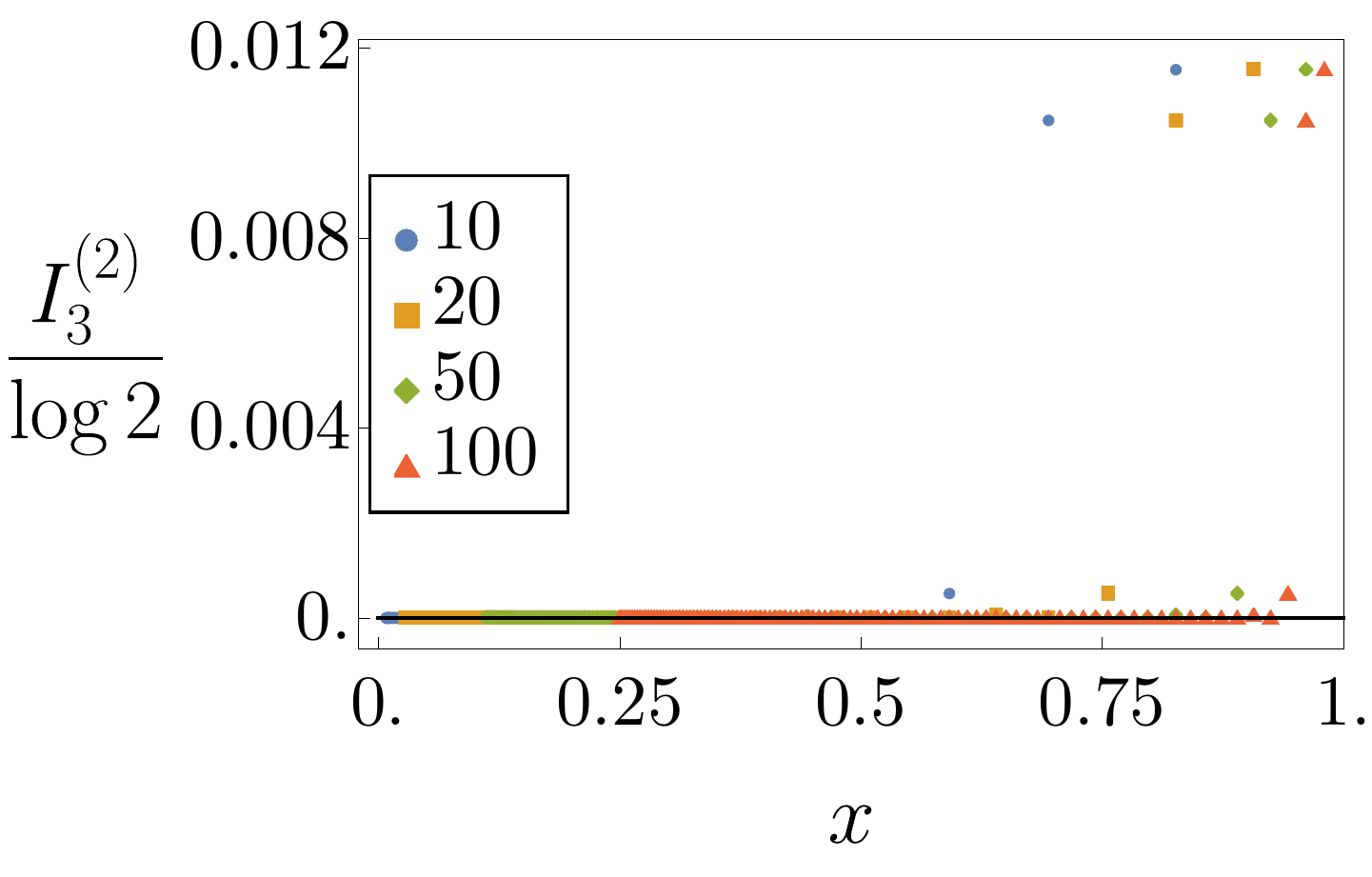}
		\subcaption{Thermal state $\bs\rho\propto \exp(-\beta \bs H) $ with $\beta=1$ in the XX model with $(J_x,J_y,h)=(1,1,0.7)$.}\label{sf:thermalXX}
	\end{minipage}\hspace{0.4 cm}
	\begin{minipage}{0.48\textwidth}
		\centering
		\includegraphics[width=1\textwidth]{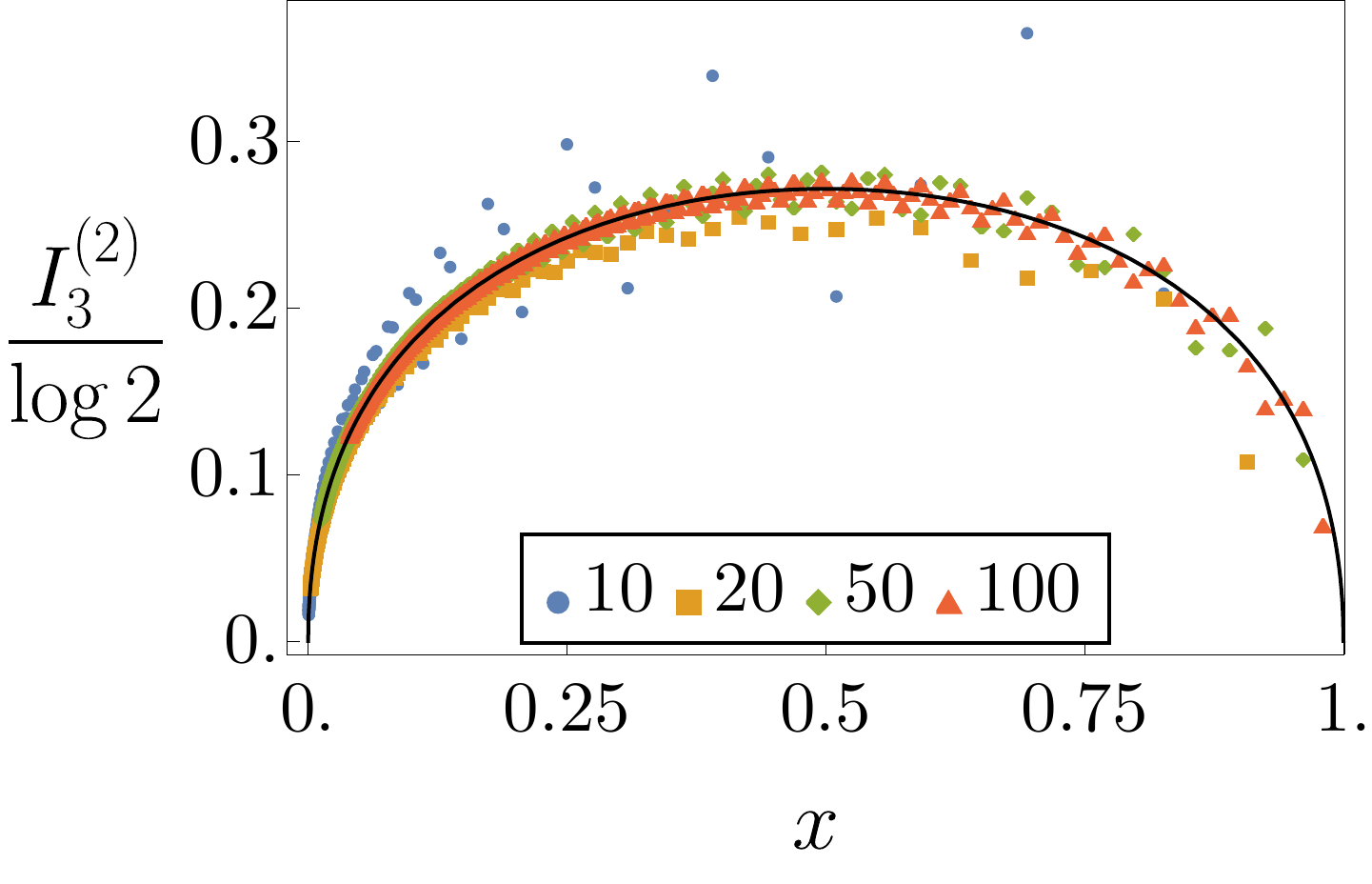}
		\subcaption{(Critical) Ground state of the XX model with $(J_x,J_y,h)=(1,1,0.7)$.}\label{sf:criticalXX}
	\end{minipage}\\
	
	\vspace{0.5 cm}
	\begin{minipage}{0.48\textwidth}
		\centering
		\includegraphics[width=1\textwidth]{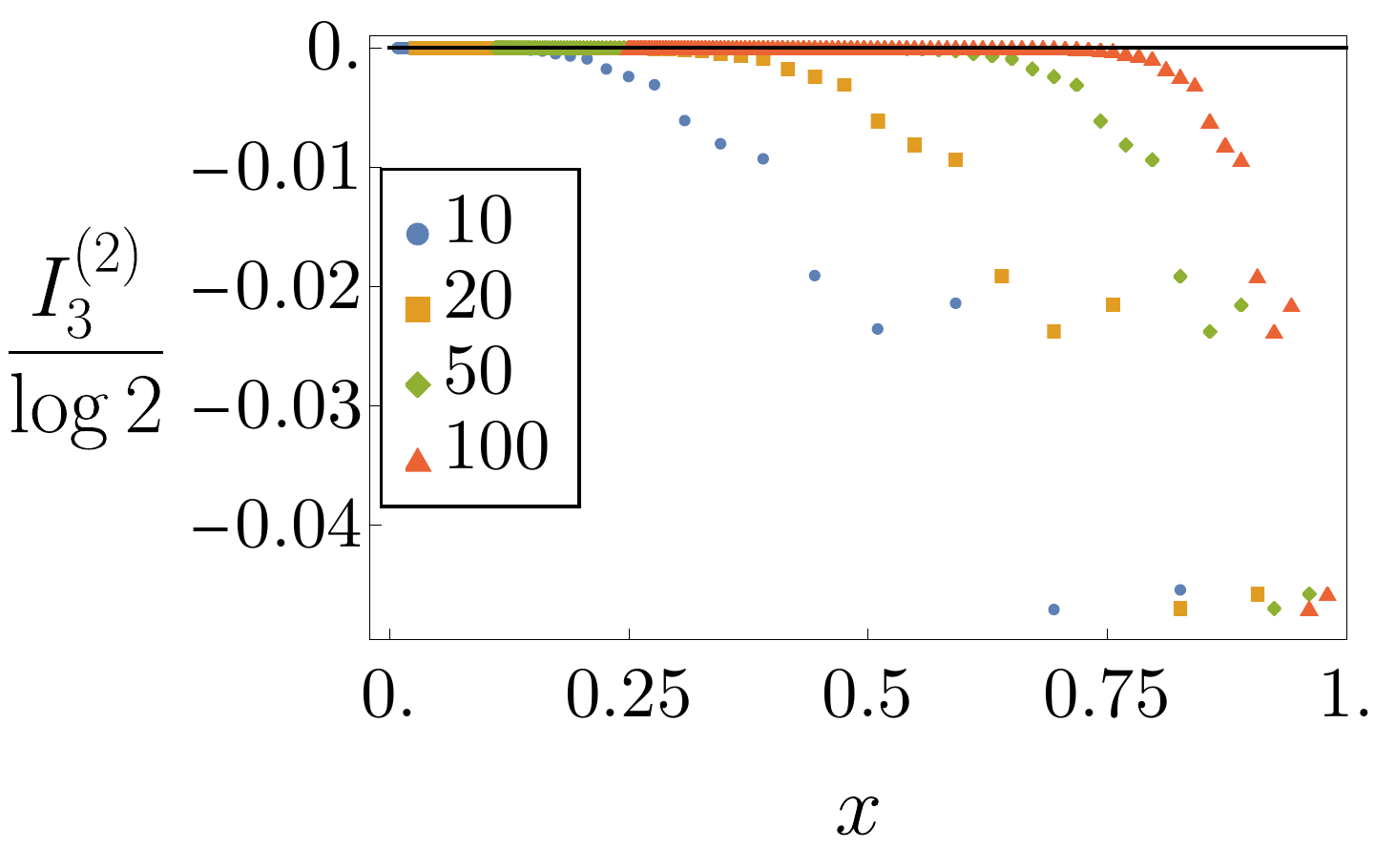}
		\subcaption{Quench from the ground state of a gapped Ising Hamiltonian with $(J_x,J_y,h)=(1,0,5)$.}\label{sf:fromGappedIsing}
	\end{minipage}\hspace{0.4 cm}
	\begin{minipage}{0.48\textwidth}
		\centering
		\includegraphics[width=1\textwidth]{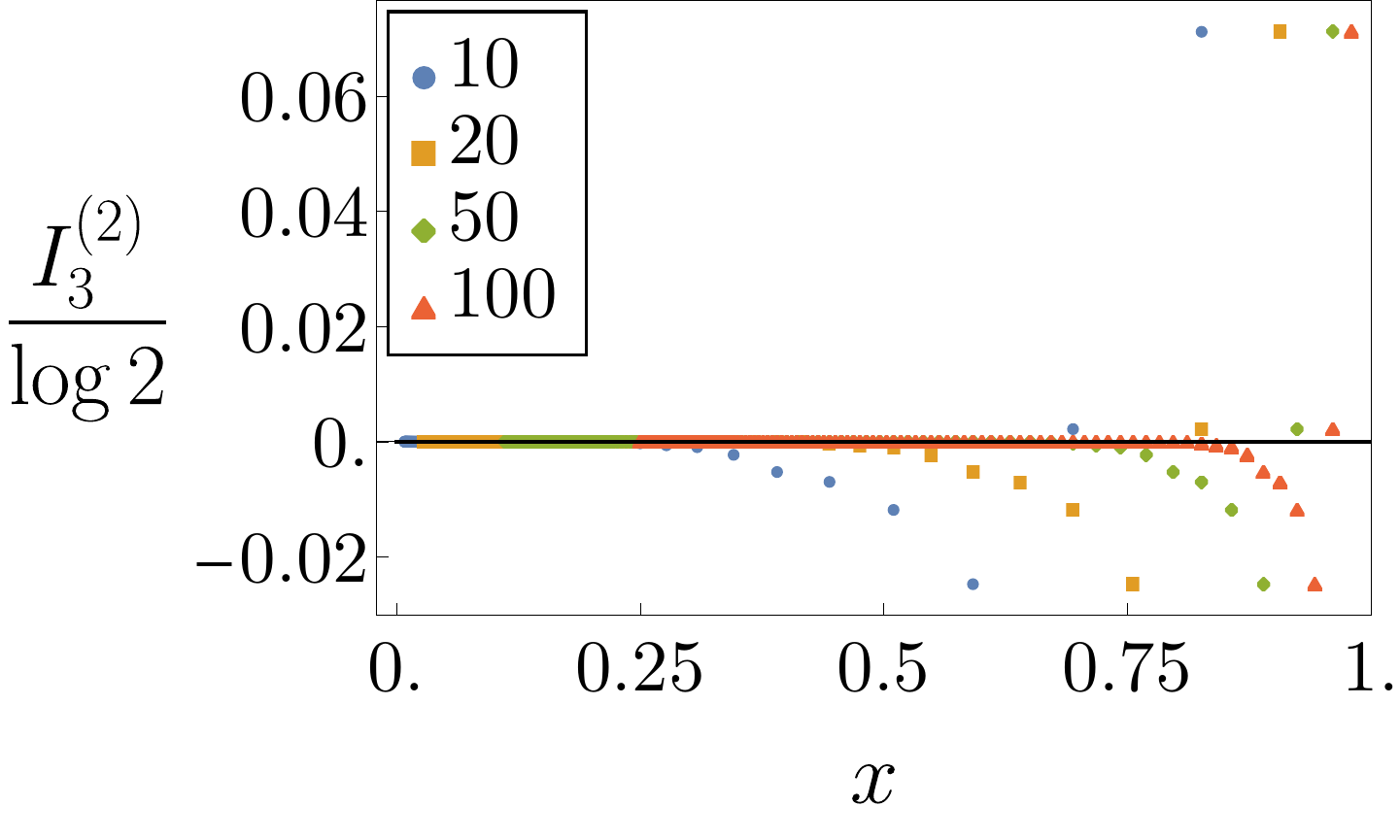}
		\subcaption{Quench from the ground state of the critical Ising model $(J_x,J_y,h)=(1,0,{1})$.}\label{sf:fromCriticalIsing}
	\end{minipage}
	
	\vspace{0.5 cm}
	\begin{minipage}{0.48\textwidth}
		\centering
		\includegraphics[width=0.9\textwidth]{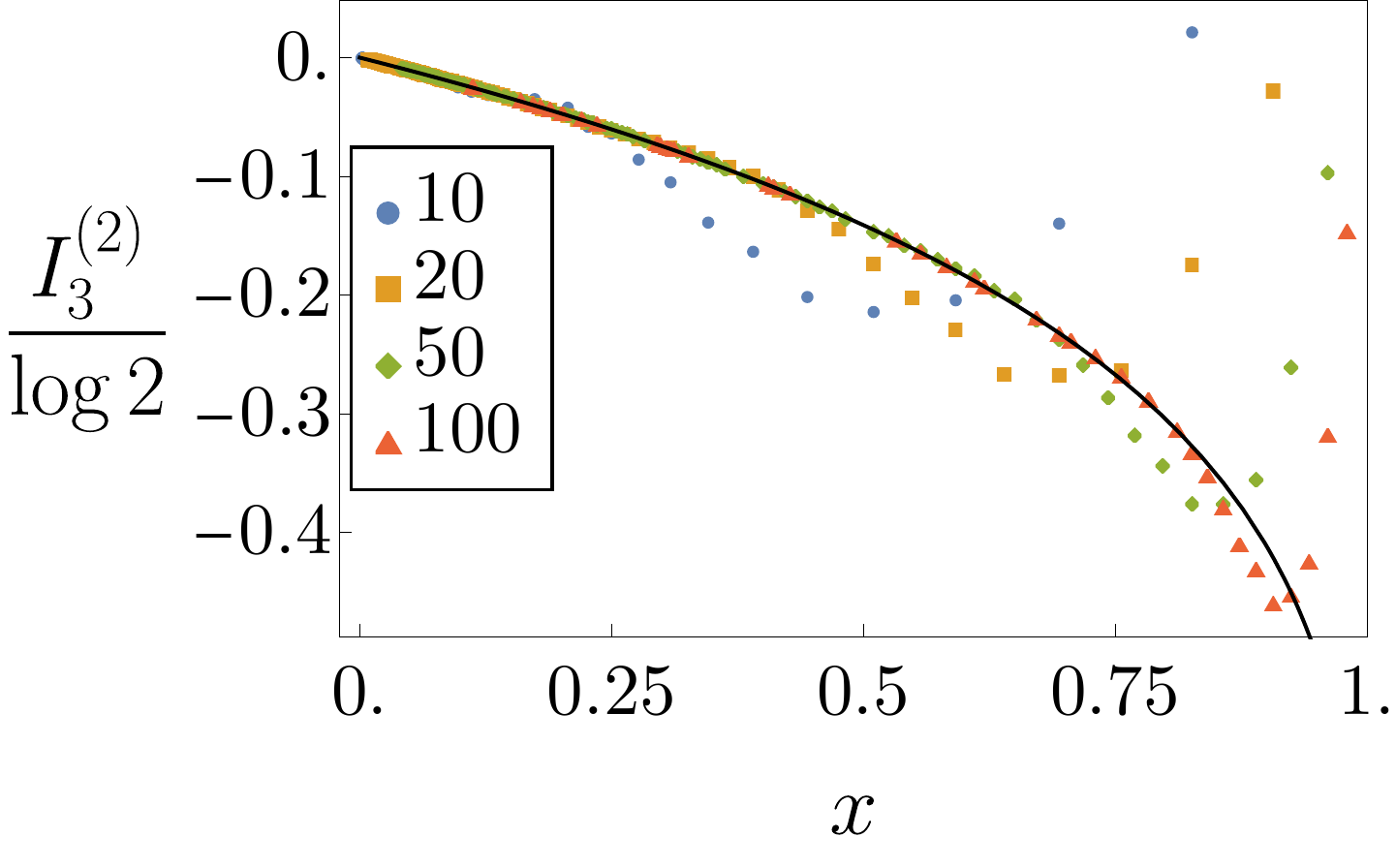}
		\subcaption{Quench from the ground state of a critical XX model with $(J_x,J_y,h)=(1,1,0.3)$.}\label{sf:fromXX}
	\end{minipage}\hspace{0.4 cm}
	\begin{minipage}{0.48\textwidth}
		\centering
		\includegraphics[width=0.9\textwidth]{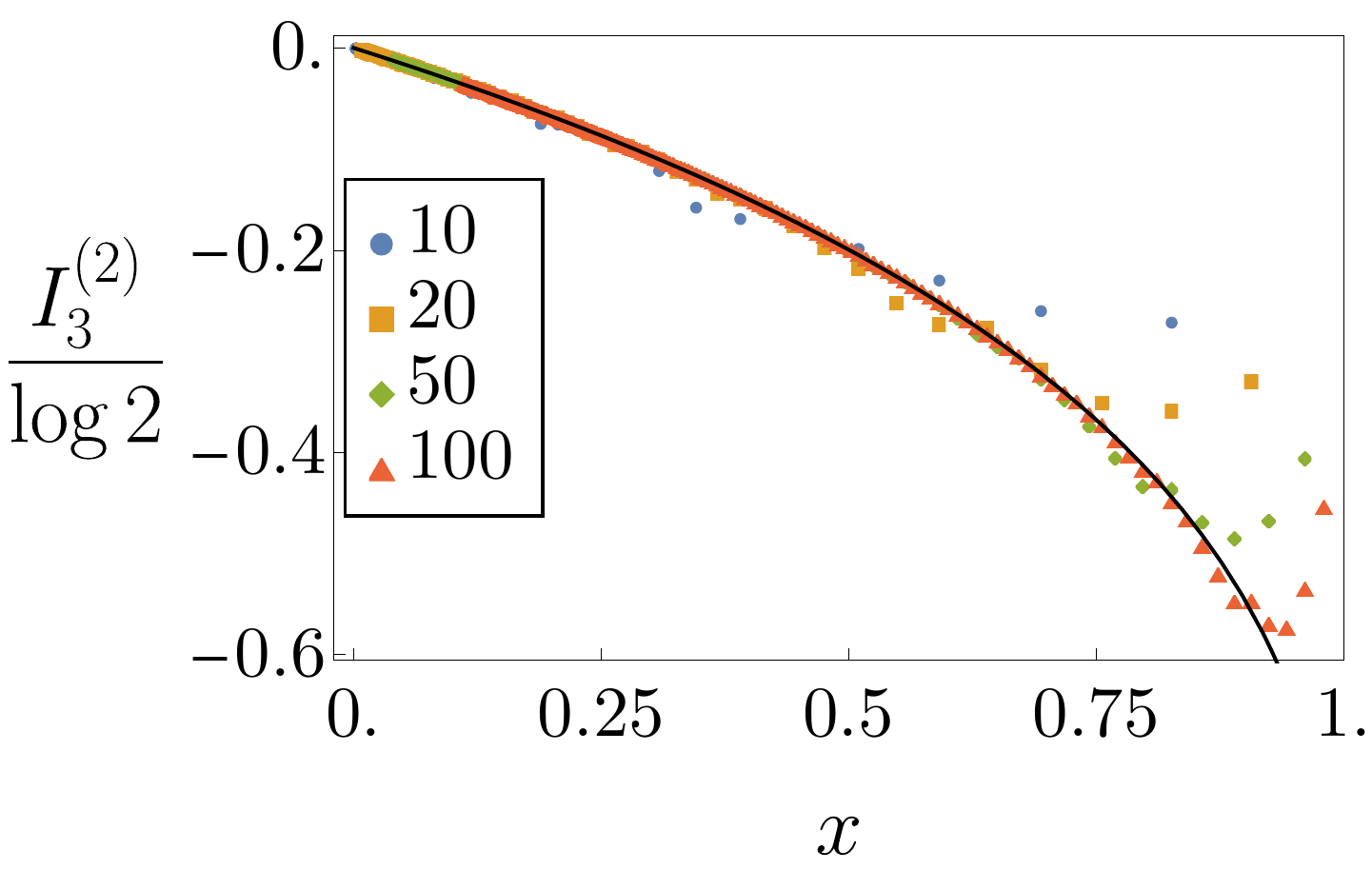}
		\subcaption{Bipartitioning protocol from $\bs \rho\propto e^{-\beta_-\sum_{\ell<0}{\bs \sigma_\ell^z}}\otimes e^{\beta_+\sum_{\ell\geq 0}{\bs \sigma_\ell^z}}$ with $\beta_-=2$ and $\beta_+=1$.}\label{sf:bipartitioning}
	\end{minipage}
	
	\caption{R\'enyi-$2$ tripartite information vs the cross ratio $x=|A||C|/((|A|+|B|)(|B|+|C|))$ for $|A|=|C|\in\{10, 20,50, 100\}$ and variable $|B|$ in different situations. Time evolution in all nonequilibrium settings is under the XY Hamiltonian \eqref{eq:XY} with $(J_x,J_y,h)=(1,0.5,0.7)$. The solid curves are exact analytical predictions for the limit of large subsystems $A,B,C$. In particular, (a) shows the standard vanishing of the tripartite information in equilibrium at nonzero temperature and (b) is the CFT prediction~\cite{Calabrese2009Entanglement1,Fagotti2012New,Furukawa2009Mutual} $I_3(A,B,C)=-\log 2 +\log(1+\sqrt{1-x}+\sqrt{x})$. The remaining predictions have been conjectured in Ref.~\cite{Maric2022Universality} and are proved in this paper.
	}
	\label{fig:different}
\end{figure}

We report  explicit examples in which we have numerically evaluated the tripartite information. The numerical data are obtained using~\eqref{eq:Ssingleblock} and \eqref{eq:entropy disjoint blocks lattice exact} with the correlation matrices  computed using the formulas collected in Table~\ref{t:list}. 

\subsubsection{Equilibrium}
According to our previous discussion, the tripartite information is expected to vanish in equilibrium at nonzero temperature as the lengths approach infinity. In the noninteracting systems we are considering, this can be argued by the fact that the symbol of the correlation matrix is smooth. We show it explicitly  in Fig.~\ref{sf:thermalXX}, where we consider a thermal state of the XX model. In contrast, Fig.~\ref{sf:criticalXX} displays the R\'enyi-$2$ tripartite information at zero temperature, which is nonzero only because we chose a model with a conformal critical ground state. 

\subsubsection{Global quenches with translational invariance}

\begin{figure}[!t]
	\centering
	\includegraphics[width=0.7\textwidth]{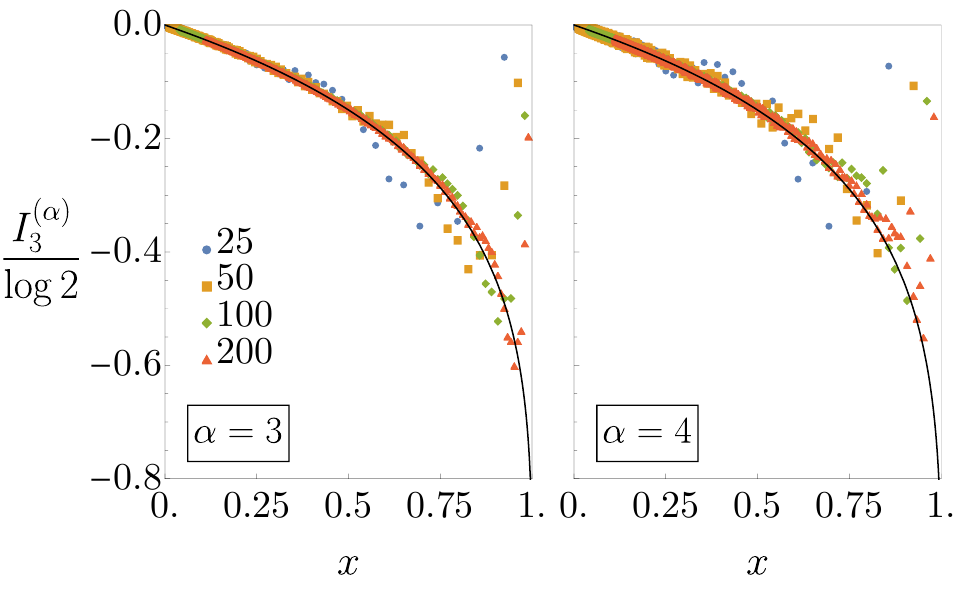}
	\caption{R\'enyi-$\alpha$ tripartite information at infinite time after the quench from the ground state of the XX model with $(J_x,J_y,h)=(1,1,0.3)$ under the XY Hamiltonian~\eqref{eq:XY} with $(J_x,J_y,h)=(1,0.5,0.7)$ for $|A|=|C|\in\{25, 50,100, 200\}$ and variable $|B|$. The solid curves are the Abelian approximation, i.e., \eqref{eq:result tripartite information higher alpha abelian} with $Q(x;\mathrm R_0,\Sigma)=1$. See Fig.~\ref{fig:fixed_x_critical} for a comparison with more accurate approximations. }
	\label{fig:fixedACcritical}
\end{figure}

\begin{figure}[!t]
	\centering
	\includegraphics[width=0.7\textwidth]{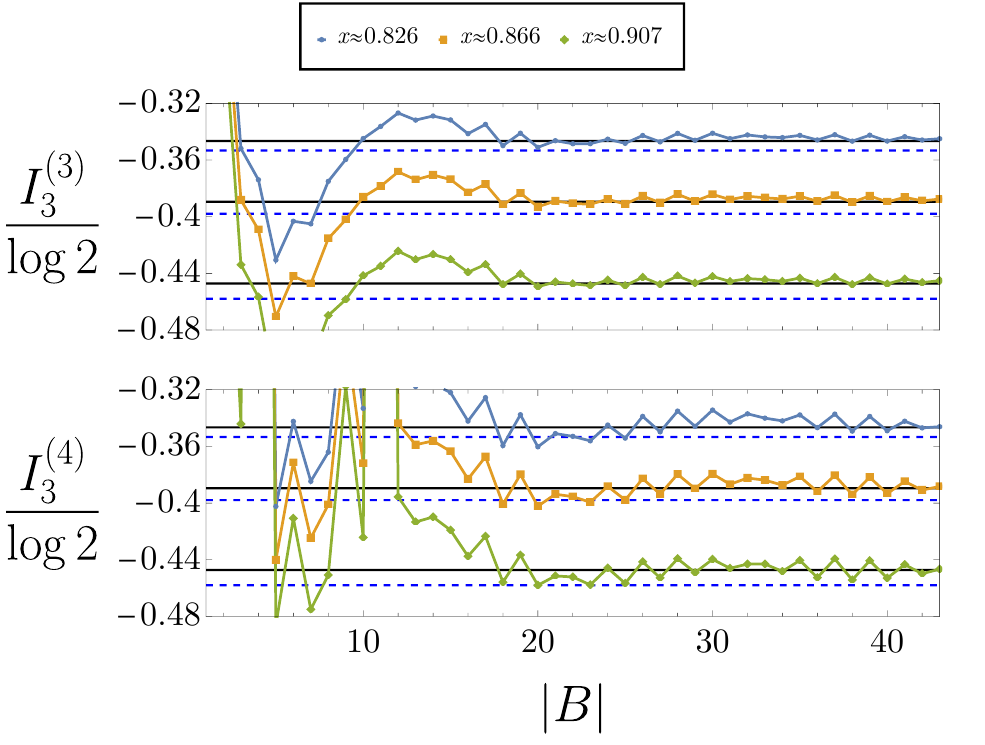}
	\caption{R\'enyi-$\alpha$ tripartite information at infinite time after the same quench as in Figure \ref{fig:fixedACcritical} vs |B|. The value of the cross ratio $x$ is fixed, while the subsystem sizes are varied as follows:  $(|A|,|B|,|C|)=n(10,1,10),n(10,1,20),n(20,1,20)$ with integer $n$. The blue dashed lines are the Abelian approximation, while the black solid lines include the leading non-Abelian correction.}
	\label{fig:fixed_x_critical}
\end{figure}

We consider here three global quenches from the ground state of local Hamiltonians. In the first case, shown in Fig.~\ref{sf:fromGappedIsing}, the pre-quench Hamiltonian is gapped, so the correlations lengths are finite. Since the post-quench Hamiltonian is local, this property is preserved even at infinite time after the quench.  We do not expect a nonzero tripartite information in this case and indeed the numerical data show quite clearly that $I_3^{(2)}(A,B,C)$ approaches zero. In the second case, shown in Fig.~\ref{sf:fromCriticalIsing}, 
the pre-quench Hamiltonian is gapless and there are correlations that decay algebraically even at infinite time after the quench. Nevertheless, the tripartite information approaches zero as the lengths are increased. As pointed out in Ref.~\cite{Maric2022Universality}, this happens because correlations do not decay to zero slowly enough. 

In the third example we replaced the initial state with the ground state of a critical Hamiltonian with central charge equal to $1$. According to our work, the R\'enyi-$\alpha$ tripartite information should be nonzero.
In Fig.~\ref{sf:fromXX} we show the R\'enyi-$2$ tripartite information, while in Figs. \ref{fig:fixedACcritical} and \ref{fig:fixed_x_critical} we discuss higher values of $\alpha$. Predictions \eqref{eq:result tripartite information alpha 2} and \eqref{eq:result tripartite information higher alpha abelian} are in excellent agreement with the numerical data.
One can clearly see that the finite size effects related to the lengths of the subsystems disappear as their size is increased.

\subsubsection{Bipartitioning protocols}

\begin{figure}[!t]
	\centering
	\includegraphics[width=0.7\textwidth]{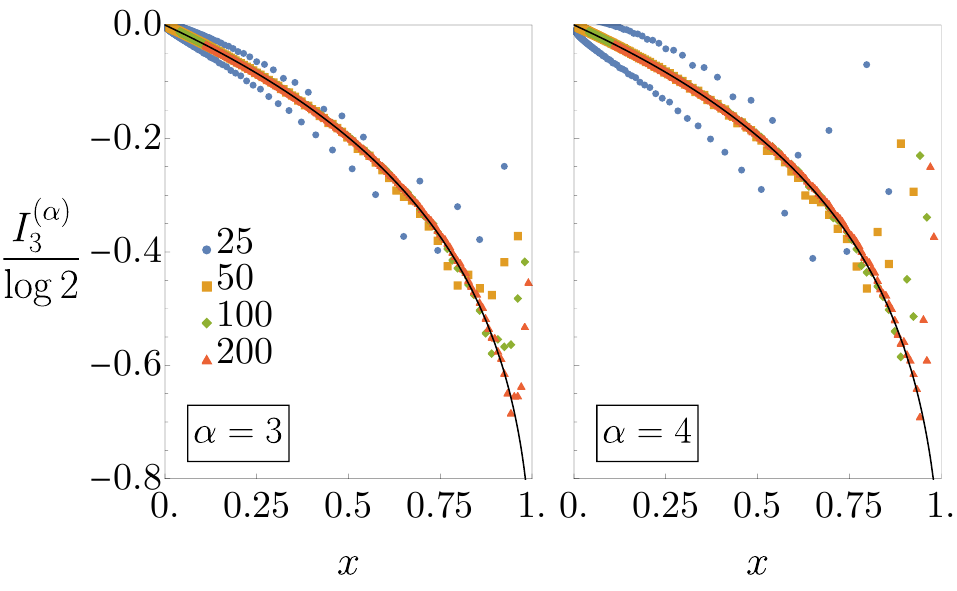}
	\caption{R\'enyi-$\alpha$ tripartite information at infinite time after the  quench from $\bs \rho\propto e^{-\beta_-\sum_{\ell<0}{\bs \sigma_\ell^z}}\otimes e^{\beta_+\sum_{\ell\geq 0}{\bs \sigma_\ell^z}}$, with $\beta_-=2$ and $\beta_+=1$, under the XY Hamiltonian~\eqref{eq:XY} with $(J_x,J_y,h)=(1,0.5,0.7)$ for $|A|=|C|\in\{25, 50,100, 200\}$ and variable $|B|$. The solid curves are the Abelian approximation, i.e., \eqref{eq:result tripartite information higher alpha abelian} with $Q(x;\mathrm R_0,\Sigma)=1$. See Fig.~\ref{fig:fixed_x_bipartition} for a comparison with more accurate approximations.}
	\label{fig:fixedACbipartition}
\end{figure}

\begin{figure}[!t]
	\centering
	\includegraphics[width=0.7\textwidth]{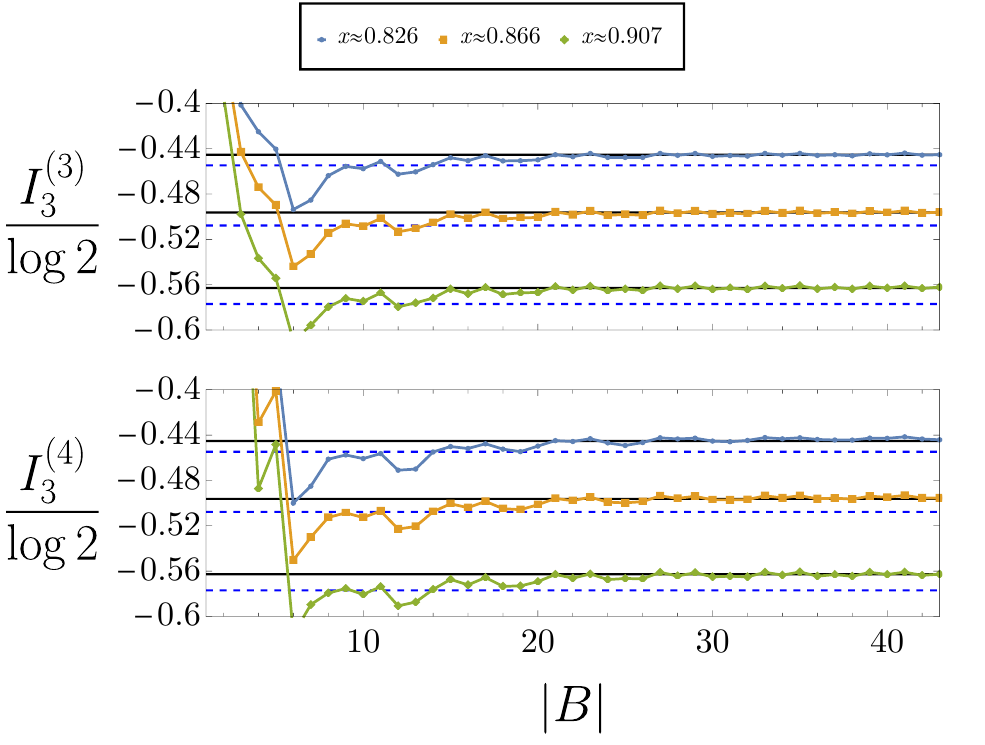}
	\caption{
		The same as in Figure~\ref{fig:fixed_x_critical} for the quench of Figure~\ref{fig:fixedACbipartition}.
	}
	\label{fig:fixed_x_bipartition}
\end{figure}
As an example of bipartitioning protocol we consider time evolution under the XY Hamiltonian \eqref{eq:XY} of the junction of two thermal states, $\bs \rho_-\propto \exp{(-\beta_-\bs H_{0,-}})$ and  $\bs \rho_+\propto \exp{(-\beta_+\bs H_{0,+}})$ described by the simplified Hamiltonian $\bs H_{0,-}=\sum_{\ell<0}\bs\sigma_{\ell}^z$ and $\bs H_{0,+}=-\sum_{\ell\geq 0}\bs\sigma_{\ell}^z$.
In similar settings correlations are expected to decay algebraically in the NESS as a manifestation of the abrupt change of properties in the initial state (independently of how correlations decay there). 
Figures \ref{sf:bipartitioning}, \ref{fig:fixedACbipartition} and \ref{fig:fixed_x_bipartition} show an excellent agreement the between predictions \eqref{eq:result tripartite information alpha 2} and \eqref{eq:result tripartite information higher alpha abelian} and the numerical data.

\section{Field theory description}
\label{sec:Field theory description}
\subsection{Fermionic correlations at late times}
\label{sec:continuum limit}
The low-energy properties of a critical spin-chain Hamiltonian can be described by an effective quantum field theory. That can be obtained by reintroducing the lattice spacing $a$ in such a way that only the low-energy part of the spectrum survives the continuum limit $a\rightarrow 0$. In particular, in noninteracting spin chains it is sufficient to expand around the momenta with the lowest excitation energies; the fermionic operators are then written in terms of fast oscillating factors and slow fields, which have a well defined continuum limit (see, e.g., \cite{Affleck1988LesHouches}).
The fast oscillating factors cancel out in the effective description of the Hamiltonian, which can indeed be expressed in terms of slow fields only.

After a global quench, an extensive number of particles is excited, and  the excitations with the lowest energy are not sufficient to describe the state.
In the next section we sketch a trick to describe anyway the problem within a low-energy field theory. Here however we adopt another approach, which allows us to identify the relevant properties of the chain system more quickly in view of our immediate access to the late-time correlation matrix. 
Specifically, we group spin sites into blocks and  define lattice fermions whose correlations are slow and for which a continuum limit exists.
In this respect, the procedure we present has some similarities to the renormalization group \cite{Kadanoff1966,Wilson1974}.

For the sake of simplicity, we start assuming that the limit of infinite time can be described by a stationary state of the XX model, whose correlation matrix has the symbol
\begin{equation}\label{eq:correlation matrix simple case}
	\Gamma(k)=[\vartheta(k)-\vartheta(-k)]\mathrm I+[\vartheta(k)+\vartheta(-k)-1]\sigma^y
	\, .
\end{equation}
The generalisation to other correlation matrices is easy and will be discussed at the end of the section. 
First, we note that $\vartheta(k)$ can be identified with the filling function.
Second, from \eqref{eq:correlation matrix simple case} it follows that the Jordan-Wigner fermions $\bs c_\ell=(\bs a_{2\ell-1}+i\bs a_{2\ell})/2$ exhibit the correlations

\begin{equation}\label{eq:correlations filling function}
	\braket{\bs c_j \bs c_\ell^\dagger}=\int_{-\pi}^{\pi}\frac{dk}{2\pi}\vartheta(k)e^{ik(\ell-j)}, \quad \braket{\bs c_j \bs c_\ell}=0.
\end{equation}
If the filling function is smooth, correlations decay exponentially. A more interesting case arises when the filling function $\vartheta(k)$ is discontinuous. 
Let us assume that such discontinuities $k_r$, $r=1,2,\ldots, \Upsilon$ are rational multiples of $\pi$; hence, there is a positive integer $N$ and integers $m_r$, for $r=1,2,\ldots, \Upsilon$, such that $k_r=2\pi m_r/N$.  Note that, even if the condition of rationality is not satisfied, any Fermi momentum $k_r$ could be  approximated to arbitrary precision by taking sufficiently large $N$. From now on, in analogy with the ground-state physics, we will call Fermi momenta the momenta at the discontinuities. The next step is to divide the chain in blocks of length $N$ and define $N$ species of fermions $\bs d_\ell^{\nu}$ as the discrete Fourier transform of the original $\bs c_\ell$ in the blocks
\begin{equation}
	\bs d_\ell^{\nu}=\frac{1}{\sqrt{N}}\sum_{m=0}^{N-1}\bs c_{N\ell+m} e^{i\frac{2\pi \nu m}{N}}\, , \quad \nu=0,1,\ldots, N-1.
\end{equation}
We then define $N$ fields $\bs \psi_\nu(x)$ as the continuum limit of $\bs d_\ell^{\nu}$ at position $x=N a \ell$, by letting the lattice spacing $a$ to zero and promoting $x$ to a continuum variable,
\begin{equation}\label{eq:field decoupled definition}
	\bs \psi_\nu(x)\sim \frac{1}{\sqrt{Ma}} \bs d^{\nu}_{\frac{x}{Ma}}, \quad a\to 0.
\end{equation}
The reader will forgive us for the abuse of notations, in that from now on $x$ will indicate a position rather than the cross ratio as in the introduction. 
Note that $\bs \psi_\nu(x)$ satisfy the standard fermionic anticommutation relations
\begin{equation}
	\{\bs \psi_\mu(x),\bs \psi_\nu^\dagger(y)\}=\delta_{\mu,\nu}\delta(x-y), \qquad  \{\bs \psi_\mu(x),\bs \psi_\nu(y)\}=0.
\end{equation}
The 2-point correlation functions of the fields can be readily obtained by taking the continuum limit of the fermionic correlations. 
Specifically we have
\begin{multline}
	\braket{\bs \psi_\mu(x)\bs \psi_\nu^\dagger(y)}\sim \frac{1}{N a}\braket{\bs d^\mu_{\frac{x}{Na}}\bs d^{\nu\dag}_{\frac{y}{Na}}}=
	\frac{1}{N^2 a}\sum_{m,n=0}^{N-1}e^{i\frac{2\pi \mu m}{N}}e^{-i\frac{2\pi \nu n}{N}}\sum_{r=1}^\Upsilon\int_{k_r}^{k_{r+1}}\frac{dk}{2\pi}\vartheta(k)e^{ik(\frac{y-x}{a}+n-m)} \\
	=\delta_{\mu,\nu}\frac{\vartheta(\frac{2\pi\mu}{N}^+)-\vartheta(\frac{2\pi\mu}{N}^-)}{2}\frac{\chi_{K}(\frac{2\pi\mu}{N})}{i\pi(x-y)}+O(a)
\end{multline}
where $\chi_K$ is the indicator function of the set $K\equiv\{k_1,k_2,\ldots k_R\}$ of Fermi points, i.e., the correlation is nonzero only if $2\pi\mu/N$ is equal (up to $2\pi$) to some of the Fermi momenta $k_r$.
The short-distance behaviour  of the correlation (corresponding to distances approaching $0$ as $a\rightarrow 0$) can be obtained with a symmetric integration over one  position about the other (so that the contribution above vanishes)  
\begin{multline}
	\mathrm{p.v.}\int\mathrm d y\braket{\bs \psi_\mu(x)\bs \psi_\nu^\dagger(y)}\sim \lim_{\epsilon\rightarrow 0^+}\sum_{d=-\infty}^\infty\braket{\bs d^\mu_{\frac{x}{Na}}\bs d^{\nu\dag}_{\frac{x}{Na}+ d}}e^{-\epsilon |d|}=\\
	\frac{1}{N^2}\sum_{m,n=0}^{N-1}e^{i\frac{2\pi \mu m}{N}}e^{-i\frac{2\pi \nu n}{N}}\sum_{r=1}^\Upsilon\int_{k_r}^{k_{r+1}}dk\vartheta(k)\sum_{q=-\infty}^\infty \delta(k-\tfrac{2\pi q}{N})e^{\frac{2\pi i q(n-m)}{N}}=
	\delta_{\mu,\nu}\tfrac{\vartheta(\frac{2\pi\mu}{N}^+)+\vartheta(\frac{2\pi\mu}{N}^-)}{2}
\end{multline}
where $\vartheta(\frac{2\pi\mu}{N}^+)\neq 
\vartheta(\frac{2\pi\mu}{N}^-)$ only if $\frac{2\pi\mu}{N}\in K$. 
Putting all together we get
\begin{equation}\label{eq:fields decoupled correlations}
	\braket{\bs \psi_\mu(x)\bs \psi_\nu^\dagger(y)}=\delta_{\mu,\nu}\Bigl[\frac{\vartheta(\frac{2\pi\mu}{N}^+)+\vartheta(\frac{2\pi\mu}{N}^-)}{2}\delta(x-y)+\frac{\vartheta(\frac{2\pi\mu}{N}^+)-\vartheta(\frac{2\pi\mu}{N}^-)}{2}\frac{\chi_{K}(\frac{2\pi\mu}{N})}{i\pi(x-y)}\Bigr]\, .
\end{equation}
From this expression we see that each discontinuity is associated with a field with non-trivial correlations but still decoupled from the rest; all the remaining fields with $\frac{2\pi \mu}{N}\notin K$ have trivial correlations, proportional to the identity. 
Such a factorisation allows us to focus on the contribution from a particular Fermi point  with $k_F\in K$, taking care only eventually of the presence of more fields. For the sake of compactness, we introduce the parametrization $\tpm=\lim_{k\to k_F^\pm}[2\vartheta(k)-1]$, in terms of which the 2-point function of the field $\bs \psi$ corresponding to a particular discontinuity $k_F$ (omitting the index of the field) reads
\begin{empheq}[box=\fbox]{equation}\label{eq:fields correlations each fermi point}
	\braket{\bs \psi(x) \bs \psi^\dagger(y)} =\frac{2+\tp+\tm}{4}\delta(x-y)+\frac{\tp-\tm}{2}\frac{1}{2\pi i(x-y)} \, .
\end{empheq}

The final step is to explain why these expressions hold true even when the symbol of the correlation matrix does not take the simple form~\eqref{eq:correlation matrix simple case}. In general $\Gamma(k)$ can not be diagonalised by a momentum-independent change of basis; it takes instead the form 
\begin{equation}\label{eq:correlation matrix general case}
	\Gamma(k)=e^{i W(k)}\Bigl\{[\vartheta(k)-\vartheta(-k)]\mathrm I+[\vartheta(k)+\vartheta(-k)-1]\sigma^y \Bigr\}e^{-i W(k)}
	\, ,
\end{equation}
where $W(k)=W^\dag(k)$ and $W^t(-k)=-W(k)$. This is a general way to parametrise the famous Bogoliubov transformation required to diagonalise systems such as the transverse-field Ising model. In fact, $W(k)$ is nothing but the symbol of the quadratic operator that generates the transformation mapping the post-quench Hamiltonian $\bs H$ to an operator commuting with the Hamiltonian of the XX model. Such a transformation exists because all quadratic systems have an eigenbasis of Slater determinants. In addition, if $\bs H$ has a conservation law with a non-critical ground state, the generator $\bs W$ can be chosen in such a way that $e^{i  W(k)}$ is smooth.
Since such a quasilocal transformation does not affect the notion of space in the continuum limit, it 
can be simply incorporated in the definition of the fields; specifically, we can redefine the $\bs c$-fermions as follows
\be\label{eq:ca}
\bs c_\ell=\sum_{n=-\infty}^\infty\frac{1}{2}\begin{pmatrix}
	1&i
\end{pmatrix}\int_{-\pi}^\pi\frac{\mathrm d k}{2\pi}e^{i k(\ell-n)}e^{iW(k)}\begin{pmatrix}
	\bs a_{2n-1}\\
	\bs a_{2n}
\end{pmatrix}
\ee
and leave the relation between the $\bs d$-fermions and the $\bs c$ ones unchanged. Mapping~\eqref{eq:ca} manifests the quasilocality of the transformation: while $\bs c_\ell$ is a superposition of fermions at different sites, we are still labelling it using a single index $\ell$, generating in turn a spatial indetermination 
\be 
\Delta x\sim \frac{a}{-\lim_{r\rightarrow\infty}\frac{1}{r}\log \left|\int_{-\pi}^\pi\frac{\mathrm d k}{2\pi}e^{i k r}e^{i W(k)}\right|}
\ee
which however approaches zero in the continuum limit. 
As a final remark we point out that even with a non-smooth $W(k)$ one would generally obtain an expression similar to \eqref{eq:fields decoupled correlations}, though with a different parametrisation. Since $W(k)$ is smooth in all the quenches that we consider, we do not develop further for that special situation; we just mention that a non-smooth $W(k)$ can be obtained by quenching to the critical Ising (or XY) model.

\subsection{Tripartite information}
Plugging the asymptotic behaviour of the late-time correlation functions into the formulas reviewed in Section~\ref{ss:Renyi} would be sufficient to obtain the asymptotic behaviour of the entanglement entropies of the subsystems entering in the tripartite information. The nonlinear way in which the correlation matrix appears in the determinants describing the various terms of the expansion of the entropy of disjoint blocks is however not ideal. By deriving again the formulas within a quantum field theory, we have found a more convenient representation, cf. \eqref{eq:trace of a product of gaussians lattice P on the left}; we sketch in this section the method employed.

\subsubsection{Entanglement entropies in  a QFT at equilibrium: a brief overview}
In a quantum field theory at equilibrium there is a standard formalism to express the R\'enyi entropies in terms of partition functions \cite{Cardy2008Form,Calabrese2009Entanglement,Casini2009Entanglement}, which starts with interpreting the density matrix of a thermal state at inverse temperature $\beta$ as a path integral on the imaginary time interval $(0,\beta)$. Here we briefly review the formalism, focusing on fermionic fields (for bosons see e.g. \cite{Calabrese2009Entanglement,Casini2009Entanglement}) in the ground state $\ket{0}$ ($\beta=\infty$), which is the relevant case for this work.  

For the sake of simplicity we assume that the system is described by a single fermionic field $\bs \psi(x)$, satisfying the standard anticommutation relations $\{\bs \psi(x),\bs \psi^\dagger(y)\}=\delta(x-y)$.
We employ the fermionic coherent state path integral formalism based on Grassmann variables~\cite{Kleinert2009}. In this formalism the brakets $\braket{\psi|0}$  and $\braket{0|\psi'}$ can be represented as path integrals in the lower and upper half plane respectively. If $L=\cup_i^n(u_i,v_i)$ is a subsystem (which we have assumed to be a union of several intervals $(u_i,v_i)$) the reduced density matrix of $\bs\rho=\ket{0}\bra{0}$ is obtained by tracing out the degrees of freedom in the complement of $L$, that we denote by $\overline{L}$,
\begin{equation}
	\rho_L(\psi_L,\psi'_L)=\int \mathcal{D}\psi^\dagger_{\overline{L}}\mathcal{D}\psi_{\overline{L}}\bra{-\psi_{\overline{L}},\psi_L}\bs\rho\ket{\psi_{\overline{L}},\psi_L'},
\end{equation}
where a minus sign appears, that is a well-known peculiarity of the representation of trace with Grassmann variables. After an integral transformation $\psi\to-\psi$ in the lower half plane, the reduced density matrix reads
\be\label{eq:rhoPsiPsi}
\rho_L(\psi_L;\psi_L')= \frac{1}{Z_1(L)}\int_{\psi(x,0^-)=\psi_L(x), \; x\in L}^{\psi(x,0^+)= -\psi_L'(x), \; x\in L}\mathcal D \psi^\dagger \mathcal D \psi \ e^{-S_E[\psi^\dagger,\psi]}\, ,
\ee
where $S_E[\psi^\dagger,\psi]$ is the Euclidean action and $\psi(x)$, $\psi^\dag(x)$ are Grassmann fields. It corresponds to gluing the upper and lower half plane on $\overline{L}$, leaving a cut on $L$. The partition function $Z_1(L)$, which ensures the normalization $\tr[ \rho_L]=1$, reads
\be\label{eq:ZGrass}
Z_1(L)= \int\mathcal D \psi_L^\dagger \mathcal{D} \psi_L \  \rho_L(\psi_L;-\psi_L).
\ee

The R\'enyi entropies with integral index can be obtained by cyclically gluing together a given number of copies of the cylinder, so that one obtains
\be
\tr[\rho_L^\alpha]=\frac{\mathrm Z_\alpha(L)}{Z_1^\alpha(L)}
\ee
with $\mathrm Z_\alpha(L)$ the partition function on an $\alpha$-sheeted Riemann surface with branch points at $u_j$ and $v_j$. A convenient alternative representation introduces $\alpha$ copies of the field, $\psi^{(i)\dagger}(x)$, $\psi^{(i)}(x)$, for $i=1,2,\ldots,\alpha$, described by decoupled equivalent actions (so that the total action is the sum of the actions of the copies), moving the original complication  of having to deal with a nontrivial Riemann surface to the boundary conditions satisfied by the copies. Specifically, in the simple case that we are describing, subsequent copies are identified at the branch cuts, that is to say,
\be\label{eq:boundarystandard}
\ba
&\psi^{(i)}(x,0^-)=\psi^{(i+1)}(x,0^+), \quad i=1,2,\ldots,\alpha-1\\
&\psi^{(\alpha)}(x,0^-)=(-1)^{\alpha+1} \psi^{(1)}(x,0^+)
\ea
\ee
for $x\in L$, and $\psi^{(i)}(x,0^-)=\psi^{(i)}(x,0^+)$ for $x\notin L$, for any $i$. Here, we have changed the integration variables (redefining them with appropriate minus signs) so as to glue the copies with no additional signs except possibly for the last one. The factor $(-1)^{\alpha+1}$  comes from collecting one minus sign for each product in $\rho_L^\alpha$ and one from the final trace.
For future convenience, we define the (anti)circulant matrix
\begin{equation}\label{eq:matrix T definition}
	T_{\ell,n}=\delta_{\ell+1,n}+(-1)^{\alpha+1}\delta_{\ell,\alpha}\delta_{n,1}\, , \quad \ell,n=1,2,\ldots ,\alpha\, ,
\end{equation}
in terms of which the boundary conditions read
\begin{equation}\label{eq:boundarystandardcompact}
	\psi^{(i)}(x,0^-)=\sum_{j=1}^\alpha T_{ij} \psi^{(j)}(x,0^+).
\end{equation}

The advantage of this formulation stands in the fact that $\tr[\rho_L^\alpha]$ can be interpreted as a $2n$-points correlation function of special fields, called ``twist fields'', where $n$ is the number of disjoint intervals. Importantly, boundary conditions~\eqref{eq:boundarystandard} ensure the locality of the twist fields~\cite{Cardy2008Form}. In the presence of conformal symmetry such a locality condition allows one to use the power of conformal field theory to predict the asymptotic behaviour of $\tr[\rho_L^\alpha]$.

As presented so far, this approach is effective to compute the R\'enyi entropies of fermionic subsystems. 
In order to compute the entropies of spin blocks in spin chains, however,  this picture should be slightly generalised to take into account the nonlocality of the Jordan-Wigner transformation. In particular, also the trace of products of different density matrices appears in the R\'enyi entropies of disjoint blocks --- cf. \eqref{eq:trace of a product of gaussians}. A case that is relevant to our specific problem is when the reduced density matrices to multiply are equivalent to one another through a local unitary that, in each block of the subsystem, acts as the transformation associated with some global symmetry. One can then attach the corresponding unitary transformation to the boundary conditions  and still describe the system as consisting of $\alpha$ copies of the original one. For example, instead of \eqref{eq:boundarystandardcompact} one could have 
\be\label{eq:boundarynonstandard}
\psi^{(i)}(x,0^-)=U_i(x)\sum_{j=1}^\alpha T_{ij}\psi^{(j)}(x,0^+), \quad x\in L,
\ee
where the factors $U_i(x)$ depend on the sequence of reduced density matrices in the product.  

The reader is referred to Ref.~\cite{Coser2016Spin} for the application of this approach to the calculation of the R\'enyi-$\alpha$ tripartite information in the conformal critical ground state of a non-interacting spin chain.  For the analogous calculation after a global quench we face the additional problem that the formulation in terms of twist fields is not local (the energy density within $L$ does not commute with the twist fields), so we are not going to use the interpretation of the R\'enyi entropies as correlation functions.

\subsubsection{Popping down to a field theory on a ladder}

In view of the results reviewed in the previous section, we start by reinterpreting the late-time correlations calculated in Section~\ref{sec:continuum limit} as the correlations in the ground state of a field theory. This could sound impossible at first sight: the stationary state has extensive entropies, so it should not be possible to reinterpret it as a ground state of a local theory. This obstacle can be however overcome by immersing the system in a bigger one with unphysical degrees of freedom. Specifically, we propose to envelop the chain in a 3-legs ladder connected with one another through local interactions.
For this task we introduce three decoupled free fields, $\bs \psi_j(x)$, with $j=1,2,3$, characterised by the correlation functions
\begin{align}
	&\braket{\bs\psi_{j}(x)\bs  \psi_j^\dagger(y)}=\frac{1}{2}\delta(x-y)-\frac{(-1)^j}{2\pi i (x-y)}, \quad j=1,2\\
	& \braket{\bs\psi_{3}(x)\bs \psi^\dagger_3(y)}=\frac{1+\sigma}{2}\delta(x-y),
\end{align}
where $\sigma=\mathrm{sgn}[\tp+\tm]$. These correlations  are found in the ground state of the Hamiltonian
\begin{equation}\label{eq:Hamiltonian decoupled fields}
	\bs H=\int dx\left[ -i\bs\psi_1^\dagger(x)\partial_x\bs\psi_{1}(x) +i\bs\psi_2^\dagger(x) \partial_x\bs\psi_{2}(x) +\sigma\bs\psi_3^\dagger(x)\bs \psi_3(x)\right].
\end{equation}
Let us then introduce a $3\times 3$ unitary matrix $V$ with the following elements of the first column
\begin{equation} (V_{11},V_{21},V_{31})=\begin{cases}
		\left(\sqrt{\frac{1+\tp}{2}},\sqrt{\frac{1+\tm}{2}},\sqrt{-\frac{\tp+\tm}{2}}\right) , & \sigma< 0\\
		\left(\sqrt{\frac{1-\tm}{2}},\sqrt{\frac{1-\tp}{2}},\sqrt{\frac{\tp+\tm}{2}}\right) , & \sigma\geq0
	\end{cases}
\end{equation}
and define the fields
\begin{equation}\label{eq:ladder physical and unphysical fields}
	\begin{pmatrix}
		\bs\psi(x) \\ \bs\psi'(x) \\ \bs\psi''(x)
	\end{pmatrix}
	=V^{\dagger} \begin{pmatrix}
		\bs\psi_1(x)\\
		\bs\psi_2(x)\\
		\bs\psi_3(x)
	\end{pmatrix}\, .
\end{equation}
The first field $\bs\psi(x)$, given by
\begin{equation}
	\bs\psi(x)=V_{11}\bs\psi_1(x)+V_{21}\bs\psi_2(x)+V_{31}\bs \psi_3(x)\, ,
\end{equation}
has the desired correlation functions \eqref{eq:fields correlations each fermi point}. The remaining two components, $\bs\psi'(x)$ and $\bs\psi''(x)$, define two unphysical fields. Inverting the relation \eqref{eq:ladder physical and unphysical fields} enables us to interpret \eqref{eq:Hamiltonian decoupled fields} as a Hamiltonian on a three-leg ladder in which the physical field lives on the first leg and interacts locally with two unphysical fields living on the remaining two legs. In this interpretation, we are studying the entanglement of a subsystem immersed in one, infinite, leg of the ladder with the remaining part of the ladder. Note that this is a different problem from studying the entanglement between one whole leg of the ladder with the remaining legs, i.e. of one field with the other fields, that has been addressed in Refs.~\cite{Chen2013,Mollabashi2014,Furukawa2011,Xu2011}.

For future convenience, let us also introduce the following compact notation for the decoupled fields
\begin{equation}
	\vec{\bs \psi}\equiv \begin{pmatrix}
		\bs \psi_1\\
		\bs \psi_2\\
		\bs \psi_3
	\end{pmatrix} \; .
\end{equation}
Thus we have
\begin{equation}\label{eq:field vector correlator}
	\braket{\bs \psi(x)\bs  \psi^\dagger(y)}=\vec{u}^\dagger \braket{\vec{ \bs \psi}(x)\vec{\bs \psi}^\dag(y)}\vec{u}, \qquad \Vec{u}\equiv V \begin{pmatrix}
		1 \\ 0 \\ 0
	\end{pmatrix}=\begin{pmatrix}
		V_{11} \\ V_{21} \\ V_{31}
	\end{pmatrix}.
\end{equation}

In order to study the tripartite information at late time after a global quench we need to compute the entropy of a block of spins and also the one of two disjoint blocks --- cf. \eqref{eq:tripartite information definition}. The latter case requires the evaluation of \eqref{eq:trace of a product of gaussians} with $j_1,\ldots,j_\alpha \in\{1,2\}$. To that aim, we build on the procedure used in \cite{Casini2009reduced_density,Casini2009Entanglement,Fries2019,Blanco2022} for computing the fermionic entropies. As reviewed in the previous section, the reduced density matrix can be expressed as a path integral, with boundary conditions corresponding to gluing the space on the complement of the subsystem of interest. The trace of a power of the reduced density matrix is computed by introducing the appropriate number of copies of the fields, related by a suitable boundary conditions on the subsystem.
Our case is peculiar in that the boundary conditions are imposed only on the physical field, i.e., on one leg of the ladder, and should also account for the operator $\eqref{eq:operator P definition}$, which appears for every index $j_\ell=2$ in \eqref{eq:trace of a product of gaussians}.

Since the fields associated with different Fermi momenta are decoupled, their contribution in \eqref{eq:trace of a product of gaussians} is factorized.
In addition, the remaining fields that are not associated with Fermi momenta give contributions independent of $j_\ell$, which therefore cancel out in the ratios appearing in \eqref{eq:tripartite information universal ratios}. Without loss of generality we can then focus on a single field $\vec{\bs \psi}$, corresponding to a particular Fermi point, with correlation functions \eqref{eq:fields correlations each fermi point}. 
We introduce $\alpha$ copies of the field $\vec{\bs \psi}$, which we denote by $\vec{\bs\psi}^{(1)}, \ldots ,\vec{\bs \psi}^{(\alpha)}$, and the corresponding Grassmann fields $\vec{\psi}^{(1)}, \ldots ,\vec{\psi}^{(\alpha)}$. According to the representation reviewed in the previous section, the trace of a product of reduced density matrices is given by the path integral over those fields
\begin{equation}
	\{\Gamma_{j_1},\Gamma_{j_2},\ldots ,\Gamma_{j_\alpha}\}=\tr \left[\rho(\Gamma_{j_1})\rho(\Gamma_{j_2})\ldots\rho(\Gamma_{j_\alpha})\right]\sim \int \mathcal{D}\psi^\dagger \mathcal{D}\psi \exp{\left(-S_E^{(\alpha)}(k_F;\Sigma)\right)},
\end{equation}
where the Euclidean action follows from \eqref{eq:Hamiltonian decoupled fields} and reads
\be\label{eq:action copies}
S_E^{(\alpha)}(k_F;\Sigma)=\int\mathrm d \tau\int\mathrm d x\  \sum_{n=1}^\alpha\Bigl[\vec \psi^{(n)\dag}(x,\tau;\Sigma)
\begin{pmatrix}
	\partial_\tau -i\partial_x & 0 & 0\\
	0 & \partial_\tau +i\partial_x & 0 \\
	0 & 0 & \partial_\tau+\sigma 
\end{pmatrix}
\vec \psi^{(n)}(x,\tau;\Sigma)\, .
\ee
The boundary conditions read 
\begin{gather}
	\psi^{(\ell)}(x,0^-)=\begin{cases}
		\sum_{n=1}^\alpha T_{\ell n}\psi^{(n)}(x,0^+), &  x\in A\cup C\\
		\psi^{(\ell)}(x,0^+), & x\notin A\cup C
	\end{cases}
\end{gather}
for the copies of the physical field, with $T$ defined in \eqref{eq:matrix T definition}, and  
\be
\ba
\psi^{\prime(\ell)}(x,0^-)=&\psi^{\prime(\ell)}(x,0^+)\\ \psi^{\prime\prime(\ell)}(x,0^-)=&\psi^{\prime\prime(\ell)}(x,0^+)\, ,
\ea
\ee 
for the copies of the unphysical ones. We remind the reader that these boundary conditions and the factor $(-1)^{\alpha+1}$ come from the representation of the trace in the Grassmann formalism.
Note that $T$ is an (anti-)circulant matrix that can also be represented as follows
\be\label{eq:Tz}
T_{\ell n}=\frac{1}{\alpha}\sum_{z|z^\alpha=-1}(-z)^{\ell-n+1}\, .
\ee

We now come to the main complication of working with spins, i.e., $j_n$ not being generally equal to $1$. 
For each index $j_1,\ldots ,j_\alpha$ equal to $2$ 
there is an additional minus sign in the boundary conditions of the copies of the physical field on $x\in A$; this comes from matrix $\mathrm P$ --- \eqref{eq:operator P definition} --- which in the continuum limit reads
\begin{equation}\label{eq:operator P definition continuum}
	\bs P (x,y)=[-\chi_A(x)+\chi_C(x)]\delta(x-y)\, .
\end{equation}
We can then write the boundary conditions for the copies of the physical field  as follows
\begin{equation}
	\psi^{(\ell)}(x,0^-;\Sigma)=\begin{cases}
		\sum_{n=1}^\alpha (\Sigma T\Sigma)_{\ell n}\psi^{(n)}(x,0^+;\Sigma), &  x\in A\\
		\sum_{n=1}^\alpha  T_{\ell n}\psi^{(n)}(x,0^+;\Sigma), &  x\in C\\
		\psi^{(\ell)}(x,0^+;\Sigma), & x\notin A\cup C
	\end{cases},
\end{equation}
where 
\begin{empheq}[box=\fbox]{equation}\label{eq:Sigma definition}
	\Sigma_{\ell n}\equiv \delta_{\ell n}\sigma_n=-\delta_{\ell n}e^{i\pi j_\ell}\, . 
\end{empheq}
Using \eqref{eq:ladder physical and unphysical fields} the boundary conditions for the decoupled fields follow immediately
\begin{equation}\label{eq:boundary conditions decoupled fields}
	\begin{split}
		&\vec \psi^{(\ell)}(x,0^+;\Sigma)=\\
		&V\sum_n \begin{pmatrix}
			\chi_A(x)[\Sigma T \Sigma]_{\ell n} +\chi_C(x)T_{\ell n}+\chi_{\overline{A\cup C}}(x)\delta_{\ell n}&0 &0\\
			0& \delta_{\ell n} & 0\\
			0 & 0 & \delta_{\ell n}
		\end{pmatrix}V^\dagger\vec\psi^{(n)}(x,0^-;\Sigma) \, .
	\end{split}
\end{equation} 
The partition function for this system can be worked out using the method of Ref.~\cite{Casini2009Entanglement} and is  detailed in Appendix \ref{appendix: linearization of the determinant representation}. 
The result of the calculation can be expressed as follows
\begin{equation}\label{eq:trace of a product of gaussians factorized}
	\{\Gamma_{j_1},\Gamma_{j_2},\ldots, \Gamma_{j_\alpha}\}\sim\prod_{k_F}D^{(k_{F})}_{j_1,j_2,\ldots,j_\alpha},
\end{equation}
where we have written $\sim$ instead of the equality sign because the members of the equation are equal up to a multiplicative constant, which is independent of $j_1,j_2,\ldots,j_\alpha$ and hence does not contribute to the tripartite information.
The factor $D^{(k_F)}_{j_1,\ldots,j_{\alpha}}$ reads
\begin{equation}\label{eq:Dasym}
	D^{(k_F)}_{j_1,\ldots,j_{\alpha}}=\det\bs L_{\underline\sigma}^{(k_F)},
\end{equation}
where the operator $\bs L_{\underline\sigma}^{(k_F)}$ is given by
\begin{empheq}[box=\fbox]{equation}\label{eq:repalphaeven}
	\bs L_{\underline \sigma}^{(k_F)}=  2^{\frac{1-\alpha}{\alpha}}\mathrm I\otimes\bs 1-i 2^{\frac{1-\alpha}{\alpha}}
	\left[\B\otimes\frac{\bs 1-\bs P}{2}
	+\Sigma \B\Sigma\otimes\frac{\bs 1+\bs P}{2}
	\right]\mathrm I\otimes \Bigl( 2\bs C^{(k_F)}-\bs 1\Bigr )\, .
\end{empheq}
Here we have denoted by $\bs C^{(k_F)}$ the operator with the elements equal to the correlator \eqref{eq:fields correlations each fermi point} on $A\cup C$, i.e.
\be
\bs C^{(k_F)}(x,y)=\braket{\bs \psi(x) \bs \psi^\dagger(y)}, \qquad x,y\in A\cup C \, .
\ee
Matrix $\B$ is given by
\begin{empheq}[box=\fbox]{equation}\label{eq:Bdef}
	{}[\B]_{\ell n}=-i[e^{i\pi\frac{(n-\ell\mod \alpha)}{\alpha}}-\delta_{\ell n}]= \frac{1}{\alpha}\sum_{j=1}^\alpha\cot(\tfrac{\pi (j+\frac{1}{2})}{\alpha})e^{2\pi i j\frac{\ell-n}{\alpha}} 
	\qquad \ell,n\in\{1,\hdots,\alpha\}
\end{empheq}

As discussed in Appendix \ref{appendix generalisation and back to the chain}, this formula can be expressed in terms of the correlation matrix $\Gamma$ so as to become exact also on the lattice. This provides an indirect confirmation of the correctness of our procedure. 

Equations~\eqref{eq:trace of a product of gaussians factorized}, \eqref{eq:Dasym}, and \eqref{eq:repalphaeven} are the main results of this section and the very reason why we worked out a quantum field theory description. On the one hand, a numerical implementation of Eq.~\eqref{eq:Dasym} gives an efficient way to compute the tripartite information. On the other hand, the correlator appears linearly in Eq.~\eqref{eq:repalphaeven}, and this will allow us to map the calculation into a Riemann-Hilbert problem, which is the subject of the next section.

\section{Mapping to a Riemann-Hilbert problem}\label{sec:Riemann-Hilbert problem}
In this section we connect \eqref{eq:Dasym} to a Riemann-Hilbert problem with piece-wise constant matrix for a doubly connected domain. The starting point is to represent the determinant in terms of the resolvent $(\lambda \bs 1-\bs L_{\underline \sigma}^{(k_F)})^{-1}$ as follows:
\be\label{eq:fromDettoTr}
\det\bs L_{\underline \sigma}^{(k_F)}=\exp\tr[\log \bs L_{\underline \sigma}^{(k_F)}]=\exp\tr\Bigl[\int_{-\infty}^0\mathrm d \lambda \Bigl(\frac{\bs 1}{1-\lambda}+\frac{\bs 1}{\lambda \bs 1-\bs L_{\underline \sigma}^{(k_F)}}\Bigr)\Bigr]\, ,
\ee
where we have implicitly assumed that the eigenvalues of $\bs L_{\underline \sigma}$ do not fall on the negative real axis; we will comment later on the validity of this assumption. 
This representation traces the problem back to inverting a singular integral equation of the form $F(x)=\int\mathrm d y  K(x,y)\phi(y)$, with (cf. \eqref{eq:repalphaeven} and \eqref{eq:fields correlations each fermi point}) 
\begin{equation}
	\bs K(x,y)=\mathrm A(x)\delta(x-y)+\mathrm B(x)\frac{1}{\pi i (y-x)},
\end{equation}
where the coordinates $x,y$ belong to $A\cup C$ and, in our specific case, $\mathrm A(x)$ and $\mathrm B(x)$ are the $\alpha\times \alpha$ matrices
\be\label{eq:AB}
\ba
A(x)=&(\lambda-2^{\frac{1-\alpha}{\alpha}})\mathrm I+i2^{\frac{1-\alpha}{\alpha}}\frac{\gamma_++\gamma_-}{2}
\left[\B\chi_A(x)
+\Sigma \B\Sigma\chi_C(x)
\right]\\
B(x)=&i 2^{\frac{1-\alpha}{\alpha}}
\frac{\tm-\tp}{2}\left[\B\chi_A(x)
+\Sigma \B\Sigma\chi_C(x)
\right] \, ,
\ea
\ee
where $\chi_A(x)$ and $\chi_C(x)$ are the indicator functions of $A$ and $C$, respectively. 
Note that $A(x)$ and $B(x)$ are constant in the blocks, i.e. they are independent of $x$ in each block, and $\mathrm A(x)\pm \mathrm B(x)$ are invertible for $\lambda\leq 0$. In what follows the region associated with $A\cup C$ will be denoted by $L$.

The inverse $\bs K^{-1}$ can be recognised by solving the singular equation
\begin{equation}\label{eq:singeq}
	\mathrm A(x)\phi(x)+\frac{\mathrm B(x)}{\pi i}\int_{L}\frac{\phi(y)}{y-x}dy=\mathrm F(x), \quad x\in L,
\end{equation}
for an arbitrary matrix function $\mathrm F$. We indeed have
\begin{equation}\label{eq:phiKm1F}
	\phi(x)=\int_L \bs K^{-1}(x,y)\mathrm F(y)dy\, .
\end{equation}
The integrals here and in the following are understood as Cauchy principal values. In the remainder of the section we show, following \cite{Muskhelishvili1953}, how the singular equation is mapped to a Riemann-Hilbert problem and the inverse $\bs K^{-1}$ expressed in terms of its solution.

We define the sectionally holomorphic (holomorphic outside $L$, in which the limit to $L$ from up/down exists) matrix function
\begin{equation}
	\Phi(z)=\frac{1}{2\pi i}\int_L \frac{\phi(y)}{y-z}dy\qquad z\in\mathbb C
\end{equation}
and its limits to $L$ from up and down $\Phi^\pm(x)=\lim_{\varepsilon\to 0^+}\Phi(x\pm i\varepsilon)$. From now on, if not explicitly written otherwise, we will use $z$ for complex variables and $x$ for real ones. By virtue of the Plemelj formulae
\begin{equation}\label{eq:Plemelj}
	\Phi^\pm(x)=\pm \frac{1}{2}\phi(x)+\frac{1}{2\pi i}\int_L \frac{\phi(y)}{y-x}dy,
\end{equation}
we can recast the singular equation \eqref{eq:singeq} into the Riemann-Hilbert problem of finding the sectionally holomorphic matrix function $\Phi(z)$ that approaches the identity at infinity and satisfies on $L$ the equation
\begin{equation}\label{non-homogenous RH problem}
	\Phi^+(x)=\mathrm G(x)\Phi^{-}(x)+[\mathrm A(x)+\mathrm B(x)]^{-1}\mathrm F(x)\qquad x\in L\, ,
\end{equation}
where
\be
\mathrm G(x)=[\mathrm A(x)+\mathrm B(x)]^{-1}[\mathrm A(x)-\mathrm B(x)]\, . 
\ee
Let  $\mathrm X(z)$ be the solution to the homogeneous Riemann-Hilbert problem, i.e., the sectionally holomorphic matrix function that on $L$ satisfies
\begin{empheq}[box=\fbox]{equation}\label{eq:homRH}
	\mathrm X^+(x)=\mathrm G(x)\mathrm X^-(x)\qquad x\in L
\end{empheq}
and approaches the identity in the limit $|z|\rightarrow+\infty$. The Plemelj formula allows us to express the solution to the non-homogeneous problem \eqref{non-homogenous RH problem} as follows
\be
\Phi(z)=\frac{\mathrm X(z)}{2\pi i}\int_L\mathrm d x\frac{[\mathrm X^+(x)]^{-1}[\mathrm A(x)+\mathrm B(x)]^{-1}}{x-z}\mathrm F(x)\, .
\ee
Plugging this solution into \eqref{eq:Plemelj} and applying again the Plemelj formulae gives
\begin{multline}
	\phi(x)=\left[\tfrac{[\mathrm A(x)+\mathrm B(x)]^{-1}}{2}+\tfrac{[\mathrm A(x)-\mathrm B(x)]^{-1}}{2}\right]\mathrm F(x)
	+\tfrac{\mathrm X^+(x)}{2\pi i}\smallint_L\mathrm d y\tfrac{[\mathrm X^+(y)]^{-1}[\mathrm A(y)+\mathrm B(y)]^{-1}}{y-x}\mathrm F(y)\\
	-\tfrac{[\mathrm A(x)-\mathrm B(x)]^{-1}[\mathrm A(x)+\mathrm B(x)]\mathrm X^+(x)}{2\pi i}\smallint_L\mathrm d y\tfrac{[\mathrm X^+(y)]^{-1}[\mathrm A(y)+\mathrm B(y)]^{-1}}{y-x}\mathrm F(y)
\end{multline}
By comparison with \eqref{eq:phiKm1F} we identify the inverse of $\bs K$
\begin{multline}\label{eq:inverse of K not expanded}
	\bs K^{-1}(x,y)=\left[\tfrac{[\mathrm A(x)+\mathrm B(x)]^{-1}}{2}+\tfrac{[\mathrm A(x)-\mathrm B(x)]^{-1}}{2}\right]\delta(x-y)\\
	+\left[\tfrac{[\mathrm A(x)+\mathrm B(x)]^{-1}}{2}-\tfrac{[\mathrm A(x)-\mathrm B(x)]^{-1}}{2}\right]\tfrac{[\mathrm A(x)+\mathrm B(x)]\mathrm X^+(x)[\mathrm X^+(y)]^{-1}[\mathrm A(y)+\mathrm B(y)]^{-1}}{\pi i(y-x)}\, .
\end{multline}
In our specific problem, $\bs K=\lambda\bs 1-\bs L_\sigma^{(k_F)}$ and we only need the trace of $\bs K^{-1}$ --- cf. \eqref{eq:fromDettoTr}.  Therefore, we can expand $\bs K(x,y)$ for $y$ close to $x$ (a similar trick was used in Ref.~\cite{Blanco2022}); we obtain
\begin{multline}
	\bs K^{-1}(x,y)\sim \left[\tfrac{[\mathrm A(x)+\mathrm B(x)]^{-1}}{2}+\tfrac{[\mathrm A(x)-\mathrm B(x)]^{-1}}{2}\right]\delta(x-y)
	+\left[\tfrac{[\mathrm A(x)+\mathrm B(x)]^{-1}}{2}-\tfrac{[\mathrm A(x)-\mathrm B(x)]^{-1}}{2}\right]\frac{1}{\pi i(y-x)}\\
	+\left[\tfrac{[\mathrm A(x)-\mathrm B(x)]^{-1}}{2}-\tfrac{[\mathrm A(x)+\mathrm B(x)]^{-1}}{2}\right]\tfrac{[\mathrm A(x)+\mathrm B(x)]\partial_x\mathrm X^+(x)[\mathrm X^+(x)]^{-1}[\mathrm A(x)+\mathrm B(x)]^{-1}}{\pi i}\, ,
\end{multline}
where we used that our specific operator is such that $\mathrm A(x)$ and $\mathrm B(x)$ are constant in $A$ and $C$ and therefore their derivative is zero within $L$.
Plugging this into \eqref{eq:fromDettoTr} gives
\begin{empheq}[box=\fbox]{multline}
	\det\bs L_{\underline \sigma}^{(k_F)}=\exp\Bigl(\int_L\frac{\mathrm d x}{a}\frac{1}{2}\mathrm{tr}[\log(\tilde{\mathrm A}^2(x)-\mathrm B^2(x))]\Bigr)\\
	\times \exp\left(\int_{-\infty}^0\frac{\mathrm d \lambda}{2\pi i} \int_L\mathrm d x\tr\Bigl[[\partial_\lambda \log \mathrm G(x)][\partial_x \mathrm X^+(x)] [\mathrm X^+(x)]^{-1}\Bigr]\right)\, ,
\end{empheq}
where we made it explicit the dependency on $\lambda$ by defining the $\lambda$-independent matrix function $\tilde {\mathrm A}(x)=\mathrm{A}(x)-\lambda$ --- cf. \eqref{eq:AB} --- and used that, in our specific case, $[\tilde {\mathrm A}(x),\mathrm B(x)]=0$. We also regularised $\mathrm{tr}[\bs 1]$ by introducing a cut-off $a^{-1}\sim \delta(0)$, which can be interpreted as the inverse lattice spacing. The first exponential represents the leading extensive contribution to the entropy, which is not universal and therefore is not supposed to exactly match the lattice result. We note however that, as it should be, it is independent of $\underline \sigma$. The second exponential captures the entire universal part of the entropy of disjoint blocks and it is expressed in terms of the solution to the homogeneous Riemann-Hilbert problem~\eqref{eq:homRH} with $\log \mathrm G(x)$ the piecewise constant circulant matrix explicitly given by
\be
\log \mathrm G(x)=
\chi_A(x) 2\pi \mathrm R+\chi_C(x) 2\pi \Sigma \mathrm R\Sigma
\ee
where  (cf. \eqref{eq:AB} and \eqref{eq:Bdef})
\begin{empheq}[box=\fbox]{equation}\label{eq:R}
	\ba
	\mathrm R_{\ell n}=&[\mathrm R_-]_{\ell n}\\
	[\mathrm R_s]_{\ell n}=& \frac{1}{2\pi \alpha}\sum_{j=1}^\alpha\log\Bigl(\tfrac{2^{1-\frac{1}{\alpha}}\lambda-1 -is \tonly_{-s} \cot\bigl(\frac{\pi (j+\frac{1}{2})}{\alpha}\bigr)}{2^{1-\frac{1}{\alpha}}\lambda-1- i s\tonly_{s}\cot\bigl(\frac{\pi (j+\frac{1}{2})}{\alpha}\bigr)}\Bigr)e^{2\pi i j\frac{\ell -n}{\alpha}}\, .
	\ea
\end{empheq}
We note $[\partial_\lambda \mathrm R,\mathrm R]=0$ and
$
\mathrm R_s^\dag=-\mathrm R_{-s}
$.  

The $\alpha$-R\'enyi entropies of the disjoint blocks $A\cup C$ are finally given by
\begin{multline}\label{eq:SAC}
	S_\alpha[A\cup C]=\tfrac{1}{1-\alpha}\Bigl(\smallint_L\tfrac{\mathrm d x}{a}\sum_{k_F}\frac{1}{2}\mathrm{tr}[\log(\tilde{\mathrm A}^2(x)-\mathrm B^2(x))]\Bigr)\\
	+\tfrac{1}{1-\alpha}\log\Bigl[\tfrac{1}{2^\alpha}\sum_{\Sigma} \exp\Bigl(\sum_{k_F}\smallint_{-\infty}^0\tfrac{\mathrm d \lambda}{2\pi i} \smallint_L\mathrm d x\tr\Bigl[[\partial_\lambda \log \mathrm G(x)][\partial_x \mathrm X^+(x)] [\mathrm X^+(x)]^{-1}\Bigr]\Bigr)\Bigr]\, .
\end{multline}
The tripartite information is obtained by evaluating the corresponding linear combination of entropies or using the shortcut \eqref{eq:tripartite information universal ratios} based on the fact that replacing $\Sigma$ by $\mathrm I$ corresponds to computing the fermionic entropies, resulting in turn in a vanishing  tripartite information.

The next sections will be devoted to solve the homogeneous problem~\eqref{eq:homRH} and finally compute the tripartite information.

\section{Solution to the Riemann-Hilbert problem}\label{sec:Solution to the Riemann-Hilbert problem}
\subsection{Simple cases and Abelian approximation}
We start with the simple case in which 
\be\label{eq:Abcond}
[\mathrm R,\Sigma\mathrm R\Sigma]=0\, .
\ee
This happens, in particular, when $\Sigma=\mathrm I$ or the diagonal elements of $\Sigma$ have alternating sign. The solution to this Abelian homogeneous Riemann-Hilbert problem readily follows from the Plemelj formulae \cite{Muskhelishvili1953,Its2003TheRP}
\be
\label{eq:XpAb}
X^+(x)=\exp\Bigl(i\smallint_{0}^{1}\mathrm d t \tfrac{\mathrm R}{t-\frac{v_1-x-i\epsilon}{v_1-u_1}}-\tfrac{\Sigma  \mathrm R\Sigma}{t-\frac{x+i\epsilon-u_2}{v_2-u_2}} \Bigr)=
\exp\left[i R\log\Bigl(\tfrac{x-u_1+i\epsilon}{x-v_1+i\epsilon}\Bigr)-i \Sigma R\Sigma \log\Bigl(\tfrac{v_2-x-i\epsilon}{u_2-x-i\epsilon}\Bigr)\right]\, .
\ee
The universal contributions to the entropy can then be expressed as follows
\begin{multline}\label{eq:Abterm}
	\exp\Bigl(\smallint_{-\infty}^0\tfrac{\mathrm d \lambda}{2\pi i}\smallint_L\mathrm d x \tr\Bigl[[\partial_\lambda \log \mathrm G(x)][\partial_x \mathrm X^+(x)] [\mathrm X^+(x)]^{-1}\Bigr]\Bigr)=\\
	\exp\left(\tr[R_0^2]\log\Bigl(\tfrac{(v_1-u_1)(v_2-u_2)}{\epsilon^2}\Bigr)+\tr[(R_0\Sigma)^2]\log\Bigl(\tfrac{(v_2-u_1)(u_2-v_1)}{(v_2-v_1)(u_2-u_1)}\Bigr) \right)
\end{multline}
where $\mathrm R_0$ is $\mathrm R$ at $\lambda=0$, that is to say, 
\begin{empheq}[box=\fbox]{equation}\label{eq:R0}
	{}[R_0]_{\ell n}=\tfrac{1}{2\pi \alpha}\sum_{j=1}^\alpha\log\Bigl(\tfrac{1-i \tp\cot\bigl(\frac{\pi (j+\frac{1}{2})}{\alpha}\bigr)}{1-i\tm\cot\bigl(\frac{\pi (j+\frac{1}{2})}{\alpha}\bigr)}\Bigr)e^{2\pi i j\frac{\ell -n}{\alpha}}\, .
\end{empheq}
Eq.\eqref{eq:Abterm} describes the contribution to the entropy from terms satisfying \eqref{eq:Abcond}. Since such a condition is always satisfied for $\alpha=2$, we are in a position to compute the R\'enyi-$2$ tripartite information
\begin{empheq}[box=\fbox]{equation}
	I_3^{(2)}(A,B,C)=G_2(x_4)=-\log 2+\log\left[1+(1-x_4)^{\sum_{k_F}\frac{[\arctan(\tp)-\arctan(\tm)]^2}{\pi^2}}\right]\, ,
\end{empheq}
where, to avoid confusion with the position variable $x$, we have renamed the cross ratio $x_4$, i.e.,
\be
x_4=\frac{(v_1-u_1)(v_2-u_2)}{(u_2-u_1)(v_2-v_1)}\, .
\ee
This was first conjectured in Ref.~\cite{Maric2022Universality} and we have therefore proved its validity. 

If we enforce \eqref{eq:Abterm} also for the terms that do not satisfy \eqref{eq:Abcond}, we get what we call ``Abelian approximation'', which implicitly neglects any potential contribution coming from the commutator $[\mathrm R,\Sigma\mathrm R\Sigma]$. 
The Abelian approximation of the tripartite information is then given by
\be
G^{Abelian}_\alpha(x_4)=-\log 2+\tfrac{1}{\alpha-1}\log\Bigl[\tfrac{1}{2}\sum_{\Sigma} (1-x_4)^{\sum_{k_F}\tr[(R_0\Sigma)^2-R_0^2]}\Bigr]\, .
\ee
At first glance this might appear as a very crude simplification. Remarkably, instead, it turns out to be an excellent approximation whatever $\alpha$ and $\tpm$. We will come back to this point in Section~\ref{s:nonabelian}, where we will also work out the non-Abelian contributions. In the following we focus instead on the limit $\tp=-\tm\rightarrow\pm1$, in which we can compare the Abelian approximation with exact analytic results.
\paragraph{Pseudo-conformal limit.}

\begin{figure}[!t]
	\begin{center}
		\includegraphics[width=0.45\textwidth]{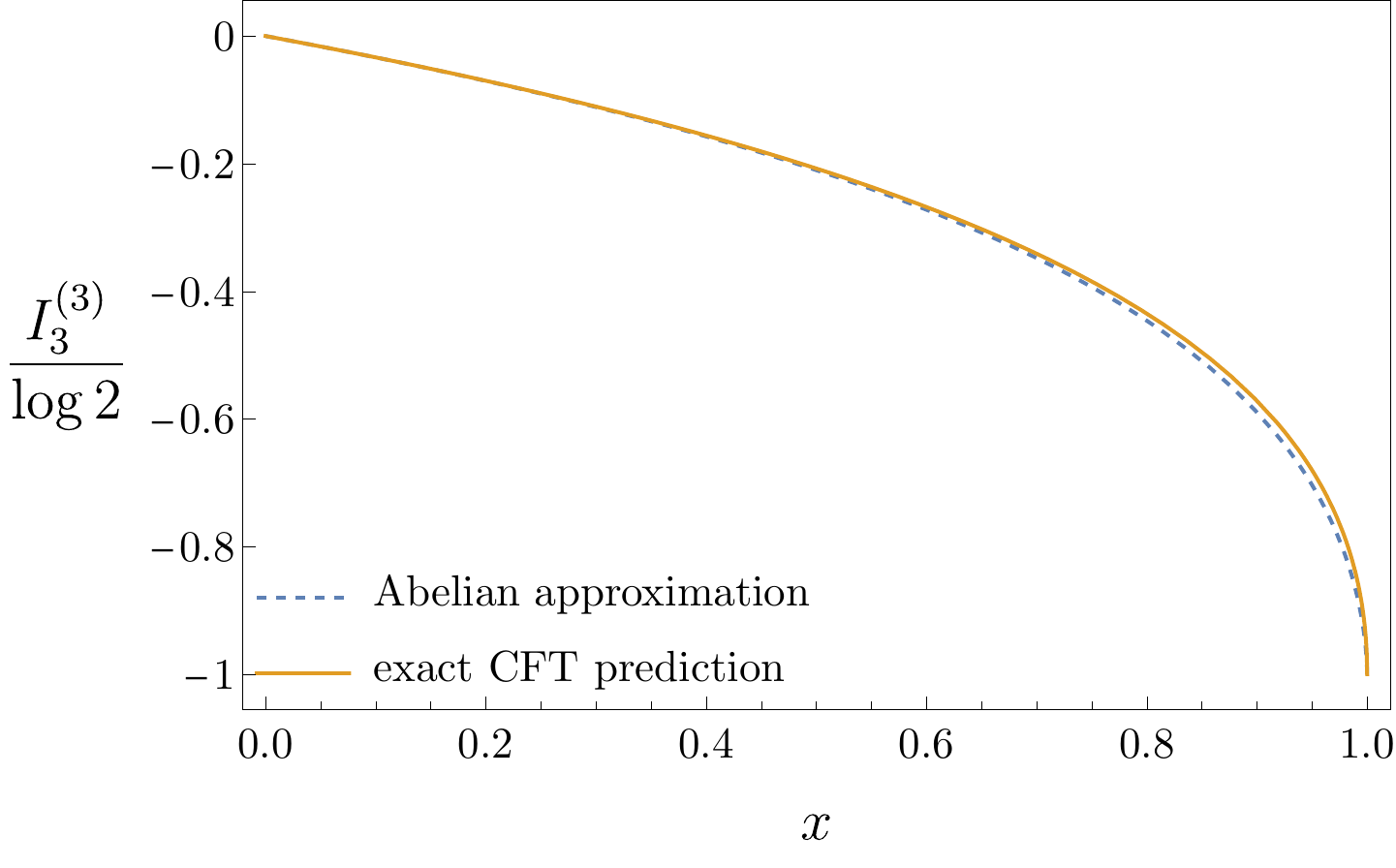}\qquad \includegraphics[width=0.45\textwidth]{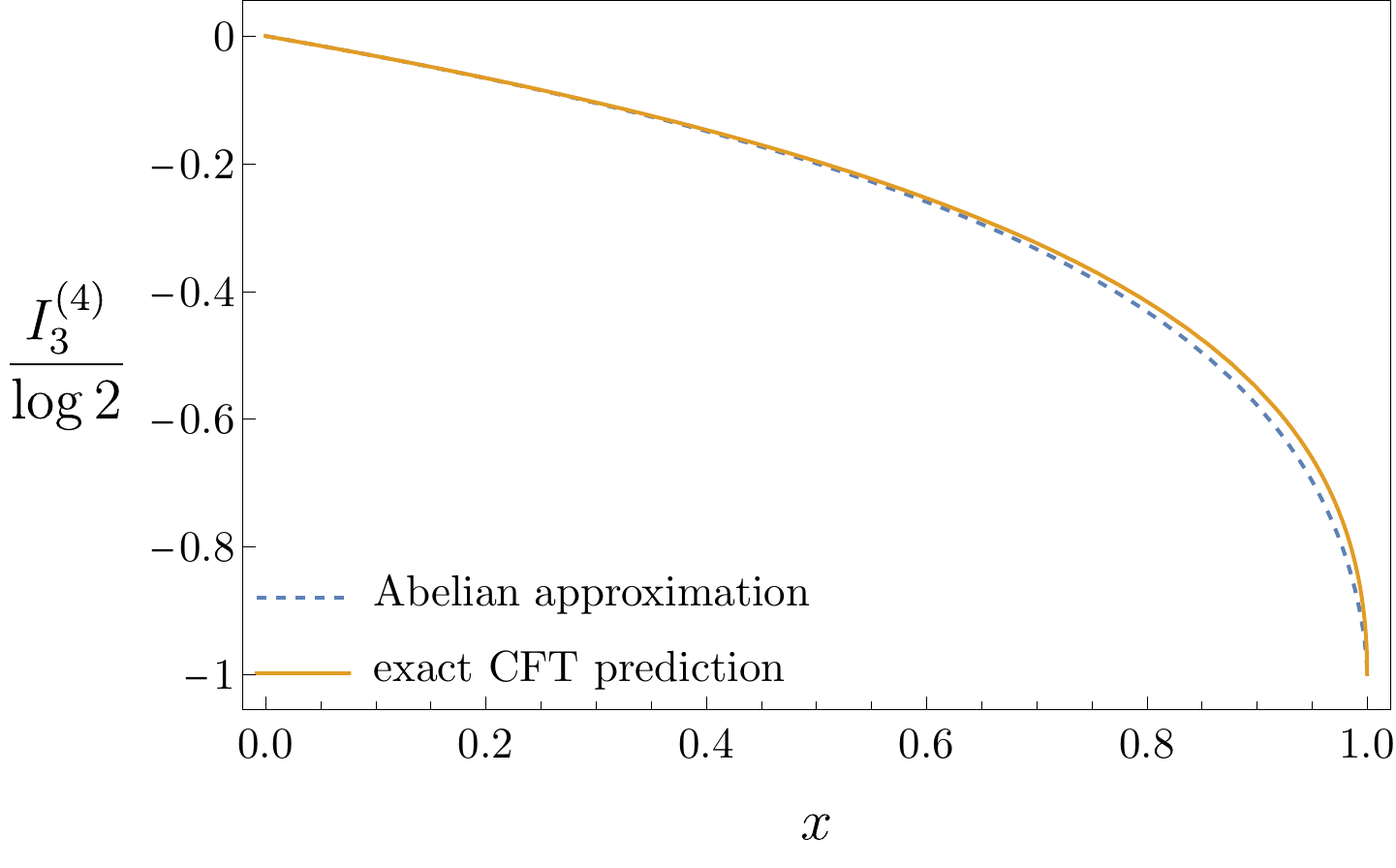}\\
		\includegraphics[width=0.45\textwidth]{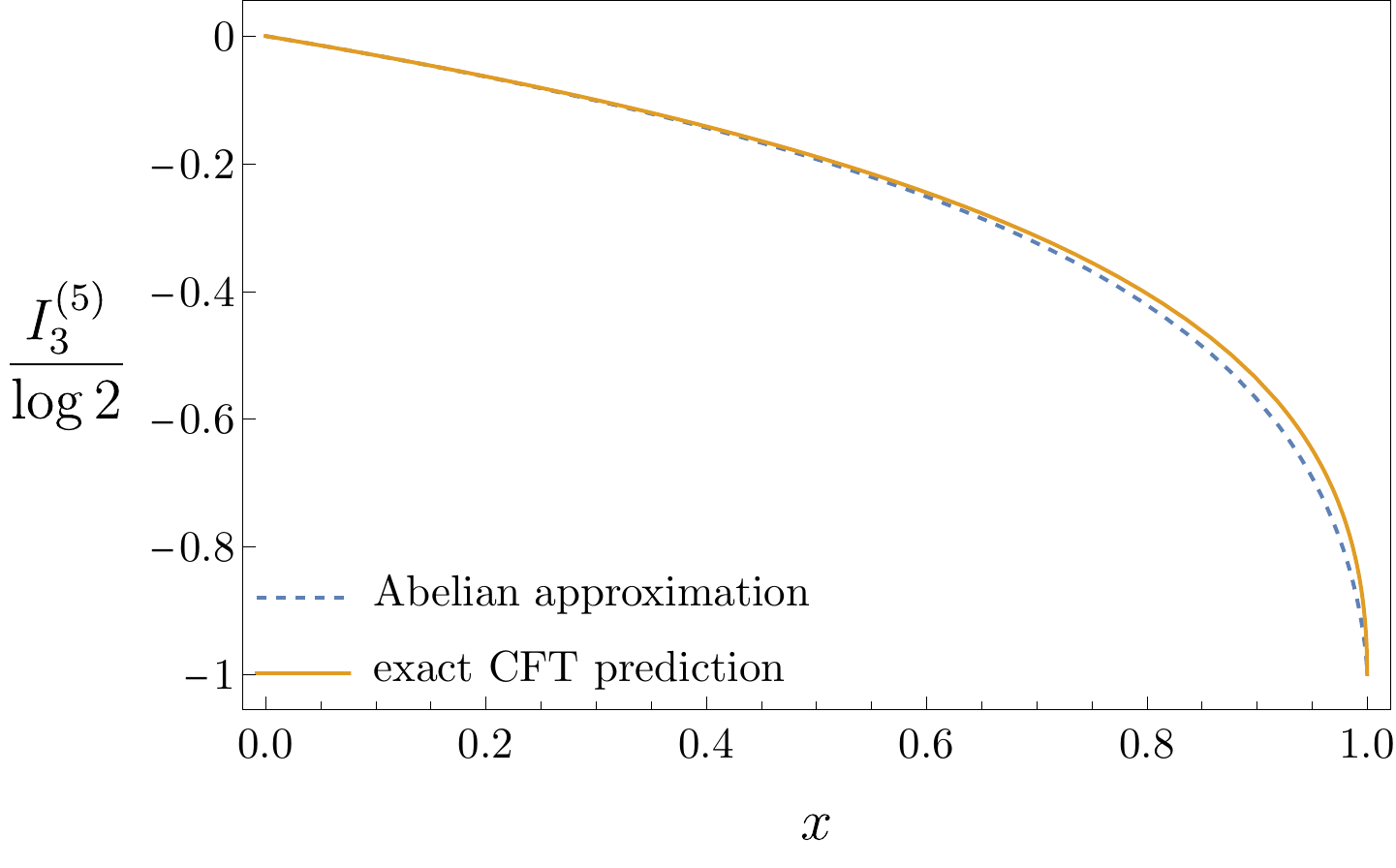}\qquad \includegraphics[width=0.45\textwidth]{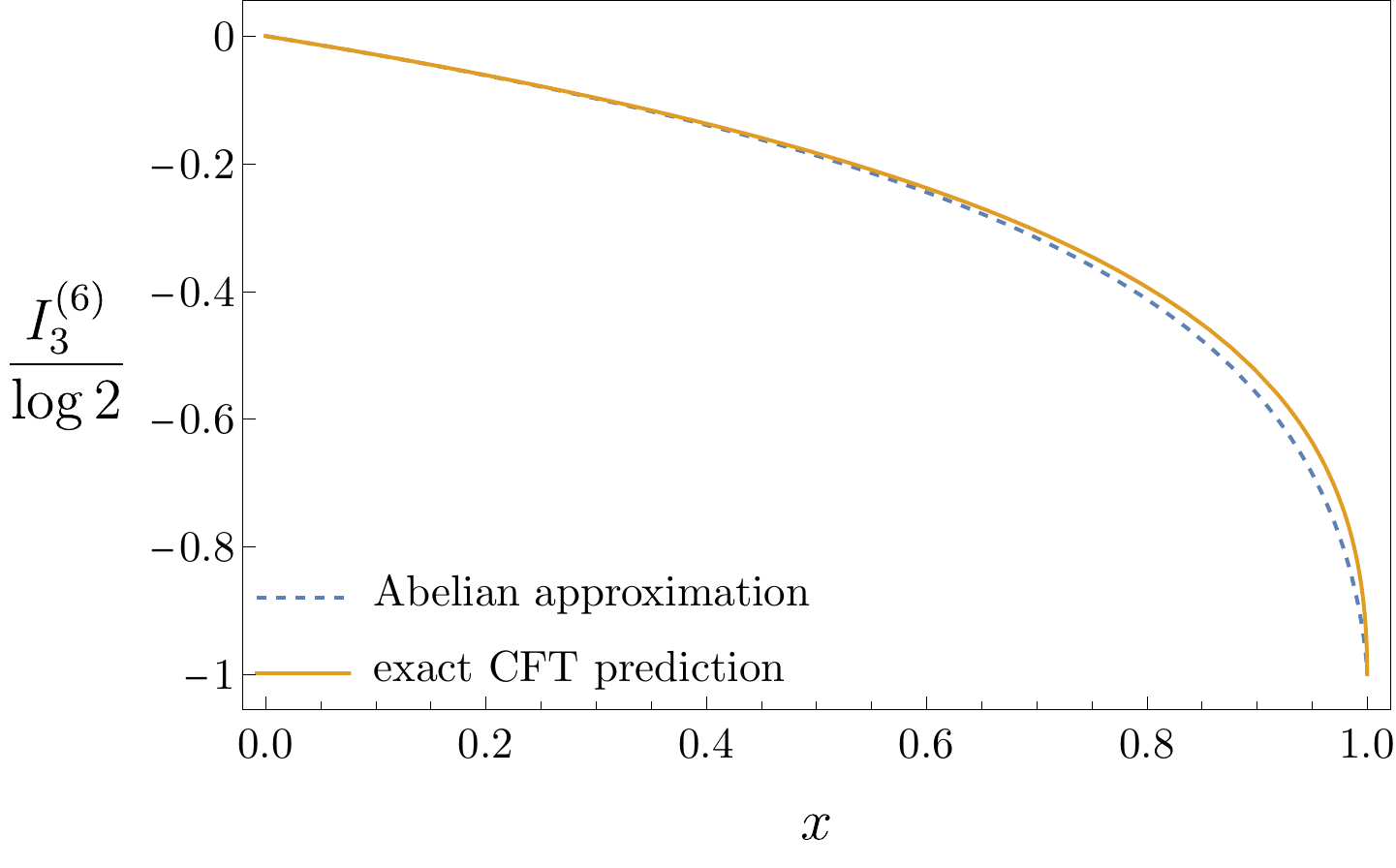}
		\caption{Abelian approximation of the R\'enyi-$\alpha$ tripartite information against the exact CFT prediction in a state with $2$ Fermi points (a standard Fermi sea) as a function of the cross ratio. The agreement is far beyond expectations. }\label{f:I3CFT}
	\end{center}
\end{figure}

With ``pseudo-conformal limit'' we mean the limit in which the Gaussian contributions~\eqref{eq:trace of a product of gaussians} to the tripartite information after a global quench match a subset of the terms characterising the tripartite information in the ground state of a conformal field theory. Strictly speaking, indeed, the state at late times after the quench does not exhibit  conformal symmetry, but when $\tp=-\tm\rightarrow\pm1$ the effect of such a symmetry breaking is completely captured by our preliminary selection of the Gaussian terms which are supposed to contribute at infinite time after the quench --- compare \eqref{eq:entropy disjoint blocks lattice after global quenches} with \eqref{eq:entropy disjoint blocks lattice exact}. 

Specifically, from the exact analytic results of Refs~\cite{Coser2016Spin} and as pointed out in Ref.~\cite{Maric2022Universality}, assuming $\tp=-\tm\rightarrow\pm1$ we have (for each Fermi point)
\be
\frac{\det \bs L_{\underline \sigma}^{(k_F)}}{\det \bs L_{\underline 1}^{(k_F)}}=\frac{\Theta(\vec \delta|\hat\tau^{(\alpha)}_{x_4})}{\Theta(\vec 0|\hat\tau^{(\alpha)}_{x_4})}\, ,\qquad
[\vec\delta]_j=\frac{1-\sigma_{j}\sigma_{j+1}}{4}\, ,
\ee
where $
\Theta(\vec z,M)=\sum_{\vec m\in \mathbb Z^{\alpha-1}}e^{i\pi \vec m^t M \vec m+2\pi i \vec m\cdot \vec\delta}
$ is the Siegel theta function and $\hat \tau_x$ is the $(\alpha-1)\times (\alpha-1)$ period matrix of the Riemann surface $\mathcal R_\alpha$ with elements
\be
[\hat \tau^{(\alpha)}_x]_{\ell n}=\frac{2i}{\alpha}\sum_{k=1}^{\alpha-1}\sin(\tfrac{\pi k}{\alpha})\cos(\tfrac{2\pi k(\ell-n)}{\alpha})\tfrac{P_{(k/\alpha)-1}(2x-1)}{P_{(k/\alpha)-1}(1-2x)}\, ,
\ee
and $P_\mu(z)$ are the Legendre polynomials.
Fig.~\ref{f:I3CFT} shows a comparison between the exact CFT predictions and the corresponding Abelian approximations, which are explicitly worked out for a few values of $\alpha$ in Appendix~\ref{a:CFT}. The agreement is excellent for small values of $\alpha$ and remains very good also increasing $\alpha$. It is nevertheless clear that the Abelian approximation is not exact.

\subsection{Beyond the Abelian approximation}\label{s:nonabelian}
The goodness of the Abelian approximation is striking, so one could wonder why is that the case. In addition, the Abelian approximation of the tripartite information is a function of  solely the cross ratio, which is not a property of the tripartite information that we can accept without skepticism: in our state there are also finite correlation lengths so, strictly speaking, there is no global conformal symmetry.  In order to shed light on these questions, we work out the solution to the Riemann-Hilbert problem even when $[\mathrm R,\Sigma\mathrm R\Sigma]\neq 0$ (we refer the reader to Ref. \cite{Its2003TheRP} for a general review on similar Riemann-Hilbert problems).

In such a non-Abelian case, \eqref{eq:XpAb} does not solve \eqref{eq:homRH} because the matrices do not appear in the correct order. Order of the operators apart, however, the structure of the solution should remain the same. Let us then start with the following Ansatz
\be
X(z)=\hat T^{-1}\exp\Bigl[i\smallint_{0}^{1}\mathrm d t \tfrac{\mathrm R_-(t)}{t-\frac{v_1-z}{v_1-u_1}}+\tfrac{\Sigma  \mathrm R^\dag_+ (t)\Sigma}{t-\frac{z-u_2}{v_2-u_2}} \Bigr]\qquad z\in \mathbb C\setminus L\, ,
\ee
where $\hat T$ ($\hat T^{-1}$) denotes the (anti) time ordering operator.
The matrix function $X^+(x)$ is defined as the limit $\epsilon\rightarrow 0^+$ for $z=x+i\epsilon$ and $x\in A\cup C$, that is to say, 
\be\label{eq:Xpx}
X^+(x)=\lim_{\epsilon\rightarrow 0^+}\hat T^{-1}\exp\Bigl[i\smallint_{0}^{1}\mathrm d t \tfrac{\mathrm R_-(t)}{t-\frac{v_1-x-i\epsilon}{v_1-u_1}}+\tfrac{\Sigma  \mathrm R_+^\dag (t)\Sigma}{t-\frac{x+i\epsilon-u_2}{v_2-u_2}} \Bigr]\qquad x\in  L\, .
\ee 
Plugging this into \eqref{eq:homRH} results in the consistency conditions
\be
\mathrm R_s(\tau)=\lim_{\epsilon\rightarrow 0^+}[\mathscr{L}_s(\tau-i\tfrac{\epsilon}{\ell_s},\tau)]^{-1}
\mathrm R_s\mathscr{L}_s(\tau-i\tfrac{\epsilon}{\ell_s},\tau)
\ee
where we introduced the shorthand
\be
\ell_-=v_1-u_1\qquad d=u_2-v_1\qquad \ell_+=v_2-u_2
\ee
and
\be
\mathscr{L}_s(z,\tau')=\hat T^{-1}\exp\Bigl[i\smallint_{0}^{\tau'}\mathrm d t \tfrac{\mathrm R_s(t)}{t-z}+\tfrac{\Sigma  \mathrm R_{-s}^\dag (t)\Sigma}{t+\frac{\ell_s}{\ell_{-s}}z+\frac{d}{\ell_{-s}}} \Bigr]
\ee
This implicitly gives $X^+(x)$ as a function of $\mathrm R_s$ (note, in particular, that $\mathrm R_s(\tau)$ could be constructed starting from $\mathrm R_s(0)=\mathrm R_s$).
En passant, we can already argue that our preliminary assumption about the eigenvalues of $\bs L_{\underline\sigma}^{(k_F)}$ not falling on the negative real axis is satisfied: the inverse \eqref{eq:inverse of K not expanded} of $\lambda\bs 1-\bs L_{\underline\sigma}^{(k_F)}$ depends on $\lambda$ only through $[A(x)\pm B(x)]^{-1}$ and $R_\pm$, which are smooth matrix functions of $\lambda$ on the real line, justifying in turn representation~\eqref{eq:fromDettoTr}.

We report two identities that are particularly useful for the upcoming calculation: 
\be
\Sigma [\mathscr{L}_s(-\tfrac{d}{\ell_s}-\tfrac{\ell_{-s}}{\ell_s}z,\tau')]^\dag\Sigma =[\mathscr{L}_{-s}(z^\ast,\tau')]^{-1}
\ee
and
\be
{}[\partial_z \mathscr{L}_s(z,\tau')][\mathscr{L}_s(z,\tau')]^{-1}=i \smallint_0^{\tau'}\mathrm d\tau  
\Bigl[\tfrac{\mathrm R_s(z,\tau)}{(\tau-z)^2}-\tfrac{\ell_s}{\ell_{-s}}\tfrac{\Sigma [\mathrm R_{-s}(-\frac{d}{\ell_{-s}}-\frac{\ell_s}{\ell_{-s}}z^\ast,\tau)]^{\dag}\Sigma}{(\tau+\frac{\ell_s}{\ell_{-s}}z+\frac{d}{\ell_{-s}})^2} \Bigr]\, .
\ee
Here we introduced what will become the main ingredient of our analytical analysis, i.e., the auxiliary matrix function
\be
R_s(z,\tau')=\mathscr{L}_s(z,\tau')\mathrm R_s(\tau')[\mathscr{L}_s(z,\tau')]^{-1}\, ,
\ee
which satisfies the following integro-differential equation
\be\label{eq:intdifR}
\ba
{}&\partial_{\tau_1} \mathrm R_s(\tau_1-i\tfrac{\epsilon}{\ell_s},\tau_2)=\\
&\qquad i \left[\smallint_0^{\tau_2}\mathrm d\tau_2'  
\Bigl[\tfrac{\mathrm R_s(\tau_1-i\tfrac{\epsilon}{\ell_s},\tau_2')}{(\tau_2'-\tau_1+i\tfrac{\epsilon}{\ell_s})^2}-\tfrac{\ell_s}{\ell_{-s}}\tfrac{\Sigma [\mathrm R_{-s}(-\frac{d}{\ell_{-s}}-\frac{\ell_s}{\ell_{-s}}\tau_1-i\frac{\epsilon}{\ell_{-s}},\tau_2')]^{\dag}\Sigma}{(\tau_2'+\frac{\ell_s}{\ell_{-s}}\tau_1-i\tfrac{\epsilon}{\ell_{-s}}+\frac{d}{\ell_{-s}})^2} \Bigr],\mathrm R_s(\tau_1-i\tfrac{\epsilon}{\ell_s},\tau_2)\right]\\
&\lim_{\epsilon\rightarrow 0^+}\mathrm R_s(\tau-i\tfrac{\epsilon}{\ell_s},\tau)=\lim_{\epsilon\rightarrow 0^+}\mathrm R_s(\tau -i\tfrac{\epsilon}{\ell_s},0)=\mathrm R_s\, .
\ea
\ee
We stress that, even if not manifest in our notations, $\mathrm R_s(\tau_1,\tau_2)$ depends on $\ell_{\pm}$ and $d$ independently of its arguments.
Since $X^+(x)=\mathscr{L}_-(\frac{v_1-x-i\epsilon}{\ell_-},1)$, we  have
\be
{}[\partial_x X^+(x)][X^+(x)]^{-1}=
i\smallint_0^1\mathrm d \tau' \Bigl[\tfrac{1}{\ell_+}\tfrac{\Sigma [\mathrm R_+ (\frac{x-u_2-i\epsilon}{\ell_+},\tau')]^\dag\Sigma}{(\tau'-\frac{x+i\epsilon-u_2}{\ell_+})^2}-
\tfrac{1}{\ell_-}\tfrac{\mathrm R_-(\frac{v_1-x-i\epsilon}{\ell_-},\tau')}{(\tau'-\frac{v_1-x-i\epsilon}{\ell_-})^2}\Bigr]\, ,
\ee
which finally gives
\begin{multline}\label{eq:trlog}
	\int_L\tfrac{\mathrm d x}{2\pi i} \mathrm{tr}[\partial_\lambda \log G(x) \partial_x X^+(x)][X^+(x)]^{-1} ]=\\
	\iint_{0}^{1}\mathrm d^2\tau\mathrm{tr}\left[(\partial_\lambda \mathrm R_-)\Bigl[\tfrac{\ell_-}{\ell_+}\tfrac{\Sigma [\mathrm R_+ (-\frac{\ell_-}{\ell_+}\tau_1-\frac{d}{\ell_+}-i\frac{\epsilon}{\ell_+},\tau_2)]^\dag\Sigma}{(\tau_2+\frac{\ell_-}{\ell_+}\tau_1+\frac{d}{\ell_+}-i\frac{\epsilon}{\ell_+})^2}-
	\tfrac{\mathrm R_-(\tau_1-i\frac{\epsilon}{\ell_-},\tau_2)}{(\tau_2-\tau_1+i\frac{\epsilon}{\ell_-})^2}\Bigr]\right]+\\
	\iint_{0}^{1}\mathrm d^2 \tau \left(\mathrm{tr}\left[(\partial_\lambda \mathrm R_+)  \Bigl[\tfrac{\ell_+}{\ell_-}\tfrac{\Sigma[\mathrm R_-(-\frac{\ell_+}{\ell_-}\tau_1-\frac{d}{\ell_-}-i\frac{\epsilon}{\ell_-},\tau_2)]^\dag\Sigma}{(\tau_2+\frac{\ell_+}{\ell_-}\tau_1+\frac{d}{\ell_-}-i\frac{\epsilon}{\ell_-})^2}-\tfrac{ \mathrm R_+ (\tau_1-i\frac{\epsilon}{\ell_+},\tau_2)}{(\tau_2-\tau_1+i \frac{\epsilon}{\ell_+})^2}\Bigr]\right]\right)^\ast
\end{multline}
Plugging this into \eqref{eq:SAC} gives the entropy of the disjoint blocks, from which we can finally obtain the tripartite information of adjacent blocks $A,B,C$
\begin{empheq}[box=\fbox]{multline}\label{eq:I3sol}
	e^{(\alpha-1)I_3^{(\alpha)}(A,B,C)}= \\
	\tfrac{1}{2^\alpha}\sum_\Sigma \exp \Bigl[
	\iint_0^1\!\!\!\mathrm d^2 \tau \int_{C_{[0,\mathrm R_0]}} \!\!\!\!\!\!\!\!\!\!\!\!\tr\Bigl[\mathrm d \mathrm R_-\ \Bigl(\tfrac{\ell_-}{\ell_+}\tfrac{\Sigma (\mathrm R_{+}(-\frac{\ell_-}{\ell_+}\tau_1-\frac{d}{\ell_+}-i\frac{\epsilon}{\ell_+},\tau_2))^\dag\Sigma+\mathrm R_-}{(\tau_2+\frac{\ell_-}{\ell_+}\tau_1+\frac{d}{\ell_+}-i\frac{\epsilon}{\ell_+})^2}-\tfrac{\mathrm R_-(\tau_1-i\frac{\epsilon}{\ell_-},\tau_2)-\mathrm R_-}{(\tau_1-\tau_2-i\frac{\epsilon}{\ell_-})^2}\Bigr)\Bigr]\Bigr]\\
	\exp \Bigl[
	\iint_0^1\!\!\!\mathrm d^2 \tau \Bigl(\int_{C_{[0,-\mathrm R_0^\dag]}} \!\!\!\!\!\!\!\!\!\!\!\!\tr\Bigl[\mathrm d \mathrm R_+\ \Bigl(\tfrac{\ell_+}{\ell_-}\tfrac{\Sigma (\mathrm R_{-} (-\frac{\ell_+}{\ell_-}\tau_1-\frac{d}{\ell_-}-i\frac{\epsilon}{\ell_-},\tau_2))^\dag\Sigma+\mathrm R_+}{(\tau_2+\frac{\ell_+}{\ell_-}\tau_1+\frac{d}{\ell_-}-i\frac{\epsilon}{\ell_-})^2}-\tfrac{\mathrm R_+(\tau_1-i\frac{\epsilon}{\ell_+},\tau_2)-\mathrm R_+}{(\tau_1-\tau_2-i\frac{\epsilon}{\ell_+})^2}\Bigr)\Bigr]\Bigr)^\ast\Bigr]\, .
\end{empheq}
\paragraph{Logarithmic (divergent) contributions.}
Since $R_\pm(z,\tau_2)$ is bounded (it is equivalent to $R_\pm$), for generic values of $\ell_\pm$ and $d$, the divergent contribution in \eqref{eq:trlog} comes from the neighbourhoods of $\tau_1=\tau_2$. By expanding about that line we identify the following asymptotic logarithmic divergency
\begin{multline}
	\int_{-\infty}^0\mathrm d \lambda\int_L\frac{\mathrm d x}{2\pi i} \mathrm{tr}[\partial_\lambda \log G(x) \partial_x X^+(x)][X^+(x)]^{-1} ]\sim\\
	-\int_{-\infty}^0\mathrm d \lambda\left[\iint_{0}^{1}\mathrm d^2\tau
	\tfrac{\mathrm{tr}[(\partial_\lambda \mathrm R_-)\mathrm R_-]}{(\tau_2-\tau_1+i\frac{\epsilon}{\ell_-})^2}+\iint_{0}^{1}\mathrm d^2 \tau \left( \tfrac{\mathrm{tr}(\partial_\lambda \mathrm R_+)  \mathrm R_+]}{(\tau_2-\tau_1+i \frac{\epsilon}{\ell_+})^2}\right)^\ast\right]=\mathrm{tr}[\mathrm R_0^2] \log\tfrac{\ell_+\ell_-}{\epsilon^2}\, . 
\end{multline}
This is completely captured by the Abelian approximation and, as expected, is independent of the particular value of $\Sigma$. In the chain it results in the asymptotic behaviour
\be
S_{A\cup C}\sim O(\ell_++\ell_-)+\frac{1}{1-\alpha}\mathrm{tr}[\mathrm R_0^2] \log(\ell_+\ell_-)+O(1)
\ee
where the leading non-universal extensive behaviour $O(\ell_++\ell_-)$ is partially captured by the first line of \eqref{eq:SAC}.

\paragraph{Residual tripartite information. }
Ref.~\cite{Maric2022Universality} introduced the concept of ``residual tripartite information'', which quantifies the non-commutativity of limits when $(a\ll) d\ll \ell_\pm$. Since we have already taken the continuum limit $a\rightarrow 0$, we can simply set $d=0$ in \eqref{eq:I3sol}, obtaining 
\begin{multline}
	\frac{1}{2^\alpha}\sum_\Sigma \exp \Bigl[
	\iint_0^1\mathrm d^2 \tau \int_{C_{[0,\mathrm R_0]}} \!\!\!\!\!\!\tr\Bigl[\mathrm d \mathrm R_-\ \Bigl(\tfrac{\ell_-}{\ell_+}\tfrac{\Sigma (\mathrm R_{+}(-\frac{\ell_-}{\ell_+}\tau_1-i\frac{\epsilon}{\ell_+},\tau_2))^\dag\Sigma+\mathrm R_-}{(\tau_2+\frac{\ell_-}{\ell_+}\tau_1-i\frac{\epsilon}{\ell_+})^2}-\tfrac{\mathrm R_-(\tau_1-i\frac{\epsilon}{\ell_-},\tau_2)-\mathrm R_-}{(\tau_1-\tau_2-i\frac{\epsilon}{\ell_-})^2}\Bigr)\Bigr]\Bigr]\\
	\exp \Bigl[
	\iint_0^1\mathrm d^2 \tau \Bigl(\int_{C_{[0,-\mathrm R_0^\dag]}} \!\!\!\!\!\!\tr\Bigl[\mathrm d \mathrm R_+\ \Bigl(\tfrac{\ell_+}{\ell_-}\tfrac{\Sigma (\mathrm R_{-} (-\frac{\ell_+}{\ell_-}\tau_1-i\frac{\epsilon}{\ell_-},\tau_2))^\dag\Sigma+\mathrm R_+}{(\tau_2+\frac{\ell_+}{\ell_-}\tau_1-i\frac{\epsilon}{\ell_-})^2}-\tfrac{\mathrm R_+(\tau_1-i\frac{\epsilon}{\ell_+},\tau_2)-\mathrm R_+}{(\tau_1-\tau_2-i\frac{\epsilon}{\ell_+})^2}\Bigr)\Bigr]\Bigr)^\ast\Bigr]
\end{multline}
with
\begin{multline}
	\mathrm R_s(\tau_1-i\tfrac{\epsilon}{\ell_s},\tau_2)=\mathrm R_s+\\
	\qquad i \int_{\tau_2}^{\tau_1}\mathrm d\tau_1'\left[\int_0^{\tau_2}\mathrm d\tau_2'  
	\Bigl[\tfrac{\mathrm R_s(\tau_1'-i\frac{\epsilon}{\ell_s},\tau_2')}{(\tau_2'-\tau_1'+i\frac{\epsilon}{\ell_s})^2}-\tfrac{\ell_s}{\ell_{-s}}\tfrac{\Sigma [\mathrm R_{-s}(-\frac{\ell_s}{\ell_{-s}}\tau_1'-i\frac{\epsilon}{\ell_{-s}},\tau_2')]^{\dag}\Sigma}{(\tau_2'+\frac{\ell_s}{\ell_{-s}}\tau_1'-i\frac{\epsilon}{\ell_{-s}})^2} \Bigr],\mathrm R_s(\tau_1'-i\tfrac{\epsilon}{\ell_s},\tau_2)\right]
\end{multline}
This produces a new singularity in the integrals at $\tau_2\sim\tau_1\sim 0$ for any $\Sigma\neq \mathrm I$, which appears already in the Abelian approximation. Contrary to the logarithmic divergency of the entropy of disjoint blocks, however, this singularity has also non-Abelian contributions. This can be seen by comparing the asymptotics (in the limit $d\rightarrow 0$) of the Abelian approximations in the pseudo-conformal limit, as worked out in the previous section, with the exact CFT predictions. 
Despite the Abelian approximation not capturing the singular behaviour exactly, we find that non-Abelian contributions are not strong enough to remove the divergency, with the remarkable consequence that the R\'enyi-$\alpha$ tripartite information approaches the finite value $-\log 2$ in the limit $d\rightarrow 0$, for any $\alpha$ and any $\tpm\neq 0$.

\subsection{Perturbation theory for the tripartite information
}\label{ss:perturbation}
Representation~\eqref{eq:I3sol} for the tripartite information is implicit, indeed the matrix functions $\mathrm R_s(z,\tau)$ are defined as the solutions to \eqref{eq:intdifR}. We develop in this section a perturbation theory to overcome this problem.   
Specifically, we carry out a formal expansion of $R_s(\tau_1,\tau_2)$ in $R_s$.

First, we rewrite \eqref{eq:intdifR} in integral form
\begin{multline}\label{eq:Rint}
	\mathrm R_s(\tau_1-i\tfrac{\epsilon}{\ell_s},\tau_2)=\mathrm R_s+\\
	i
	\smallint_{\tau_2}^{\tau_1}\mathrm d \tau_1'\Bigl[ \smallint_0^{\tau_2}\mathrm d\tau_2'\Bigl[\tfrac{\mathrm R_s(\tau_1'-i\frac{\epsilon}{\ell_s},\tau_2')}{(\tau_2'-\tau_1'+i\frac{\epsilon}{\ell_s})^2}-\tfrac{\ell_s}{\ell_{-s}}\tfrac{\Sigma \bigl(\mathrm R_{-s}(-\frac{d}{\ell_{-s}}-\frac{\ell_s}{\ell_{-s}}\tau_1'-i\frac{\epsilon}{\ell_{-s}},\tau_2')\bigr)^\dag\Sigma}{(\tau_2'+\frac{\ell_s}{\ell_{-s}}\tau_1'-i\frac{\epsilon}{\ell_{-s}}+\frac{d}{\ell_{-s}})^2}\Bigr],\mathrm R_s(\tau_1'-i\tfrac{\epsilon}{\ell_s},\tau_2)\Bigr]\, .
\end{multline}
It is clear that such a formal expansion is a triangularization of the integral equation, indeed the terms with a given number of $R_s$ matrices are quadratic functionals of the terms with a strictly lower number of $\mathrm R_s$ matrices.
In particular, the linear term of the expansion is trivially equal to $\mathrm R$. The quadratic terms can only come from the commutators of the linear term, and so on and so forth. 
From now on we assume $\tau_2\in \mathbb R_+$, $\tau_1\in\mathbb R$, 
and expand $\mathrm R_s(\tau_1-i\frac{\epsilon}{\ell_s},\tau_2)$ as follows:
\be\label{eq:Rexp}
\mathrm R_s(\tau_1-i\tfrac{\epsilon}{\ell_s},\tau_2)=\sum_{n=1}^\infty \mathrm R^{(n)}_s(\tau_1-i\tfrac{\epsilon}{\ell_s},\tau_2)\, ,
\ee
where $\mathrm R_s^{(n)}(\tau_1-i\tfrac{\epsilon}{\ell_s},\tau_2)$ is written in terms of products of matrices in which matrix $\mathrm R_s$ appears $n$ times.
Thus we have
\begin{empheq}[box=\fbox]{multline}
	\mathrm R_s^{(n)}(\tau_1-i\tfrac{\epsilon}{\ell_s},\tau_2)=\delta_{n1}\mathrm R_s-i\sum_{j=1}^{n-1}\int_{\tau_2}^{\tau_1}\mathrm d \tau_1'\Bigl[ \int_0^{\tau_2}\mathrm d\tau_2'\Bigl[\mathrm R_s^{(n-j)}(\tau_1'-i\tfrac{\epsilon}{\ell_s},\tau_2),\\
	\tfrac{\mathrm R_s^{(j)}(\tau_1'-i\tfrac{\epsilon}{\ell_s},\tau_2')}{(\tau_2'-\tau_1'+i\tfrac{\epsilon}{\ell_s})^2}-\tfrac{\ell_s}{\ell_{-s}}\tfrac{\Sigma \bigl(\mathrm R_{-s}^{(j)}(-\tfrac{d}{\ell_{-s}}-\tfrac{\ell_s}{\ell_{-s}}\tau_1'-i\tfrac{\epsilon}{\ell_{-s}},\tau_2')\bigr)^\dag\Sigma}{(\tau_2'+\frac{d}{\ell_{-s}}+\frac{\ell_s}{\ell_{-s}}\tau_1'-i\tfrac{\epsilon}{\ell_{-s}})^2}\Bigr]\, .
\end{empheq}
which also implies
\begin{multline}
	-\Sigma (\mathrm R_{-s}^{(n)}(-\tfrac{d}{\ell_{-s}}-\tfrac{\ell_{s}}{\ell_{-s}}\tau_1-i\tfrac{\epsilon}{\ell_{-s}},\tau_2))^\dag\Sigma =\delta_{n1}\Sigma\mathrm R_{s}\Sigma+\\
	i
	\sum_{j=1}^{n-1}\int_{\tau_1}^{-\tfrac{d}{\ell_{s}}-\tfrac{\ell_{-s}}{\ell_{s}}\tau_2}\mathrm d \tau_1'\Bigl[ \int_0^{\tau_2}\mathrm d\tau_2'\Bigl[\tfrac{ \mathrm R_{s}^{(j)}(\tau_1'-i\tfrac{\epsilon}{\ell_{s}},\tau_2')}{(\tau_2'-\tau_1'+i\tfrac{\epsilon}{\ell_{s}})^2}-\tfrac{\ell_{s}}{\ell_{-s}}\tfrac{\Sigma (\mathrm R_{-s}^{(j)}(-\tfrac{d}{\ell_{-s}}-\tfrac{\ell_{s}}{\ell_{-s}}\tau_1'-i\tfrac{\epsilon}{\ell_{-s}},\tau_2'))^\dag\Sigma}{(\tau_2'+\tfrac{d}{\ell_{-s}}+\tfrac{\ell_{s}}{\ell_{-s}}\tau_1'-i\tfrac{\epsilon}{\ell_{-s}})^2}\Bigr],\\
	\Sigma (\mathrm R_{-s}^{(n-j)}(-\tfrac{d}{\ell_{-s}}-\tfrac{\ell_{s}}{\ell_{-s}}\tau_1'-i\tfrac{\epsilon}{\ell_{-s}},\tau_2)^\dag \Sigma\Bigr]\, .
\end{multline}
This expansion can be used to set up a perturbation theory for the tripartite information
\begin{empheq}[box=\fbox]{equation}
	\begin{gathered}\label{eq:KnSigma}
		e^{(\alpha-1)I_3^{(\alpha)}}= \frac{1}{2^\alpha}\sum_\Sigma\prod_{n=1}^\infty K^{(n,\Sigma)}_\alpha\, , \qquad
		\text{with}\\
		K^{(n,\Sigma)}_\alpha=
		\exp \Bigl[
		\iint_0^1\!\!\!\mathrm d^2 \tau \!\!\!\int_{C_{[0,\mathrm R_0]}} \!\!\!\!\!\!\!\!\!\!\!\!\tr\Bigl[\mathrm d \mathrm R_- \Bigl(\tfrac{\Sigma (\mathrm R_{+}^{(n)} (-\frac{\ell_-}{\ell_+}\tau_1-\frac{d}{\ell_+}-i\frac{\epsilon}{\ell_+},\tau_2))^\dag\Sigma+\mathrm R_-}{\frac{\ell_+}{\ell_-}(\tau_2+\frac{\ell_-}{\ell_+}\tau_1+\frac{d}{\ell_+}-i\frac{\epsilon}{\ell_+})^2}-\tfrac{\mathrm R^{(n)}_-(\tau_1-i\frac{\epsilon}{\ell_-},\tau_2)-\mathrm R_-}{(\tau_1-\tau_2-i\frac{\epsilon}{\ell_-})^2}\Bigr)\Bigr]\Bigr]\\
		\exp \Bigl[
		\iint_0^1\mathrm d^2 \tau \Bigl(\int_{C_{[0,-\mathrm R_0^\dag]}} \!\!\!\!\!\!\!\!\!\!\!\!\!\!\!\tr\Bigl[\mathrm d \mathrm R_+ \Bigl(\tfrac{\Sigma (\mathrm R_{-}^{(n)} (-\frac{\ell_+}{\ell_-}\tau_1-\frac{d}{\ell_-}-i\frac{\epsilon}{\ell_-},\tau_2))^\dag\Sigma+\mathrm R_+}{\frac{\ell_-}{\ell_+}(\tau_2+\frac{\ell_+}{\ell_-}\tau_1+\frac{d}{\ell_-}-i\frac{\epsilon}{\ell_-})^2}-\tfrac{\mathrm R^{(n)}_+(\tau_1-i\frac{\epsilon}{\ell_+},\tau_2)-\mathrm R_+}{(\tau_1-\tau_2-i\frac{\epsilon}{\ell_+})^2}\Bigr)\Bigr]\Bigr)^\ast\Bigr]\, .
	\end{gathered}
\end{empheq}
Since the dependency  on $\lambda$ is just through $\mathrm R_\pm$, the variables $\tau_1$ and $\tau_2$ turn out to be completely decoupled from $\lambda$, and we can work out the respective integrals separately. The integral in $\lambda$ encodes all the system details, whereas the integrals in $\tau_1$ and $\tau_2$ depend only on the configuration of the subsystems, i.e., on $\ell_\pm$ and $d$. For this reason, we call the former the ``physical part'' and the latter the ``configurational part''. 
A priori one could expect this expansion to be effective only in the limit of small $\tonly$, indeed $ \mathrm R_s\rightarrow 0$ as $\tonly\rightarrow 0$ --- cf. \eqref{eq:R}. It turns out, instead, that such a perturbation theory is quickly convergent for any value of $\tonly$, including the limit $\tonly\rightarrow 1$. In the latter case, as discussed in the previous section, the tripartite information is a function of the cross ratio $x_4$, therefore we have
\be
e^{(\alpha-1)G^{\tonly\rightarrow 1}_\alpha(x_4)}= \frac{1}{2^\alpha}\sum_\Sigma\prod_{n=1}^\infty K^{(n,\Sigma)}_\alpha(\tfrac{\ell_-}{d},\tfrac{\ell_+}{d};\tonly\rightarrow 1)
\ee
and we can decompose $K^{(n,\Sigma)}_\alpha(\tfrac{\ell_-}{d},\tfrac{\ell_+}{d};\tonly\rightarrow 1)$ as the product of a finite numbers of terms, let us say $M_n$ (at the first orders of the expansion,  $M_n=1$), corresponding to independent configurational parts, as follows 
\be
K^{(n,\Sigma)}_\alpha(\tfrac{\ell_-}{d},\tfrac{\ell_+}{d};\tonly\rightarrow 1)= \prod_{j=1}^{M_n}[k_{n,j}(\tfrac{\ell_-}{d},\tfrac{\ell_+}{d})]^{\beta_{n,j}^{(\alpha)}(\Sigma)}\, .
\ee
Here $\beta_{n,j}^{(\alpha)}$ are exponents characterising the physical part and $k_{n,j}(\ell_-,d,\ell_+)$ the corresponding functions associated with the configurational part. Since the exponents can be changed by changing $\alpha$ and $\Sigma$, it is reasonable to expect that the dependency on $x$ is not peculiar to the limit $\tonly\rightarrow 1$, whereas we have
\begin{empheq}[box=\fbox]{equation}
	\textsc{Conjecture:}\quad k_{n,j}(\tfrac{\ell_-}{d},\tfrac{\ell_+}{d})\equiv k_{n,j}(x_4)\, .
\end{empheq}
The configurational part is however independent of the physical details, i.e., it does not change when $\tpm$ are finite. Consequently, the tripartite information is expected to be a function of the cross ratio alone in all the quench protocols investigated.
In the next section we will check the validity of this conjecture at the first orders of the perturbative expansion. 

\paragraph{First order (Abelian approximation).}
At the first order we can neglect the commutator on the right hand side of \eqref{eq:Rint}. Thus we have
\be
\ba
R^{(1)}_s(\tau_1-i\tfrac{\epsilon}{\ell_s},\tau_2)= & \mathrm R_s\\
-\Sigma (\mathrm R_{-s}^{(1)}(-\tfrac{d}{\ell_{-s}}-\tfrac{\ell_{s}}{\ell_{-s}}\tau_1-i\tfrac{\epsilon}{\ell_{-s}},\tau_2))^\dag\Sigma =&\Sigma\mathrm R_{s}\Sigma\, .
\ea
\ee
The physical part of $K^{(1,\Sigma)}_\alpha$ is then captured by the following (identical) integrals ($M_1=1$)
\be\label{eq:k1x}
\beta_{1,1}^{(\alpha)}(\Sigma)=\int^{\mathrm R_0}_0\mathrm{tr}[\mathrm d \mathrm R_- (\mathrm R_--\Sigma \mathrm R_-\Sigma)]=
\Bigl(\int^{-\mathrm R_0^\dag}_0\mathrm{tr}[\mathrm d \mathrm R_+ (\mathrm R_+-\Sigma \mathrm R_+\Sigma)]\Bigr)^\ast=\tfrac{1}{2}\tr[\mathrm R_0^2-(\Sigma \mathrm R_0)^2]\, ,
\ee
where we didn't specify the paths $C_{[0,\mathrm R_0]}$ and $C_{[0,-\mathrm R_0^\dag]}$ as in \eqref{eq:KnSigma} because the integrals are independent of the particular path.

Given that the two integrals of the physical part match each other, the configurational part requires just the evaluation of the following integral
\be\label{eq:k11}
\log k_{1,1}=\iint_0^1\mathrm d^2 \tau \tfrac{\ell_-}{\ell_+}\tfrac{1}{(\tau_2+\frac{\ell_-}{\ell_+}\tau_1+\frac{d}{\ell_+}-i\frac{\epsilon}{\ell_+})^2}+
\tfrac{\ell_+}{\ell_-}\tfrac{1}{(\tau_2+\frac{\ell_+}{\ell_-}\tau_1+\frac{d}{\ell_-}+i\frac{\epsilon}{\ell_-})^2}=-2\log(1-x_4)\, .
\ee
Putting all together we recover  the ``Abelian approximation'' that we  worked out in the previous section:
\begin{empheq}[box=\fbox]{equation}\label{eq:G4abelian}
	e^{(\alpha-1)G_\alpha(x)}\approx
	\frac{1}{2^\alpha}\sum_\Sigma (1-x_4)^{\tr[(\Sigma \mathrm R_0)^2-\mathrm R_0^2]}\, .
\end{empheq}
Identifying the Abelian approximation as the leading order of a perturbation theory helps understand why such a simple approximation turns out to be so good when compared with exact results, for example, in the CFT limit. 

\paragraph{Second and third order.} At the second order in $\mathrm R$ we have
\be
\ba
\mathrm R_s^{(2)}(\tau_1-i\tfrac{\epsilon}{\ell_s},\tau_2)= & \chi^{(s)}_{1}(\tau_1,\tau_2) i[\Sigma\mathrm R_s\Sigma,\mathrm R_s]\\
-\Sigma (\mathrm R_{-s}^{(2)}(-\tfrac{d}{\ell_{-s}}-\tfrac{\ell_{s}}{\ell_{-s}}\tau_1-i\tfrac{\epsilon}{\ell_{-s}},\tau_2))^\dag\Sigma =&\chi^{(-s)\ast}_{1}(-\tfrac{d}{\ell_{-s}}-\tfrac{\ell_{s}}{\ell_{-s}}\tau_1,\tau_2)  i[\Sigma\mathrm R_{s}\Sigma,\mathrm R_{s}]] 
\ea
\ee
where
\begin{multline}
	\chi_{1}^{(s)}(\tau_1,\tau_2)=\frac{\ell_s}{\ell_{-s}}\int_{\tau_2}^{\tau_1} \int_0^{\tau_2}\tfrac{\mathrm d \tau_1'\mathrm d \tau_2'}{(\tau_2'+\frac{d}{\ell_{-s}}+\frac{\ell_s}{\ell_{-s}}\tau_1'-i\tfrac{\epsilon}{\ell_{-s}})^2}=\\
	\log(d+\ell_s\tau_1-i\epsilon)+\log(d+(\ell_s+\ell_{-s})\tau_2-i\epsilon)-\log(d+\ell_s\tau_2-i\epsilon)-\log(d+\ell_s\tau_1+\ell_{-s}\tau_2-i\epsilon)
	\, .
\end{multline}
Since $\mathrm{tr}[\mathrm d \mathrm R [\mathrm R,\mathrm Y]]=0$ for any $\mathrm Y$, however, the physical part of $K^{(2,\Sigma)}_\alpha$ vanishes ($M_2=0$), therefore the leading correction to  \eqref{eq:G4abelian} is higher order. Let us then consider the third order. We have
\be
\mathrm R_s^{(3)}(\tau_1-i\tfrac{\epsilon}{\ell},\tau_2)=  -\chi_{21}^{(s)}(\tau_1,\tau_2)[\Sigma \mathrm R_s\Sigma,[\Sigma\mathrm R_s\Sigma,\mathrm R_s]]-\chi_{22}^{(s)}(\tau_1,\tau_2)[ \mathrm R_s,[\mathrm R_s,\Sigma\mathrm R_s\Sigma]]
\ee
\begin{multline}
	-\Sigma (\mathrm R_{-s}^{(3)}(-\tfrac{d}{\ell_{-s}}-\tfrac{\ell_{s}}{\ell_{-s}}\tau_1-i\tfrac{\epsilon}{\ell_{-s}},\tau_2))^\dag\Sigma =-\chi^{(-s)\ast}_{21}(-\tfrac{d}{\ell_{-s}}-\tfrac{\ell_{s}}{\ell_{-s}}\tau_1,\tau_2)[ \mathrm R_s,[\mathrm R_s,\Sigma \mathrm R_s\Sigma ]]\\
	-\chi^{(-s)\ast}_{22}(-\tfrac{d}{\ell_{-s}}-\tfrac{\ell_{s}}{\ell_{-s}}\tau_1,\tau_2)[ \Sigma \mathrm R_s\Sigma ,[\Sigma \mathrm R_s\Sigma ,\mathrm R_s]]\, ,
\end{multline}
where
\be
\ba
\chi_{21}^{(s)}(\tau_1,\tau_2)=&\tfrac{\ell_s}{\ell_{-s}}\int_{\tau_2}^{\tau_1}\mathrm d \tau_1' \int_0^{\tau_2}\mathrm d\tau_2'\tfrac{\chi^{(s)}_{1}(\tau_1',\tau_2)}{(\tau_2'+\frac{d}{\ell_{-s}}+\frac{\ell_s}{\ell_{-s}}\tau_1'-i\frac{\epsilon}{\ell_{-s}})^2}\\
\chi_{22}^{(s)}(\tau_1,\tau_2)=&\int_{\tau_2}^{\tau_1}\mathrm d \tau_1'\int_0^{\tau_2}\mathrm d\tau_2'\Bigl[\tfrac{\chi^{(s)}_{1}(\tau_1',\tau_2')-\chi^{(s)}_{1}(\tau_1',\tau_2)}{(\tau_2'-\tau_1'+i\frac{\epsilon}{\ell_s})^2}+\tfrac{\ell_s}{\ell_{-s}}\tfrac{\chi^{(-s)\ast}_{1}(-\frac{d}{\ell_{-s}}-\frac{\ell_{s}}{\ell_{-s}}\tau_1',\tau_2') }{(\tau_2'+\frac{d}{\ell_{-s}}+\frac{\ell_s}{\ell_{-s}}\tau_1'-i\frac{\epsilon}{\ell_{-s}})^2}\Bigr]
\ea
\ee
Concerning the physical part of $K^{(3,\Sigma)}_\alpha(x_4)$, this is captured again by two identical integrals ($M_3=1$):
\begin{multline}\label{eq:3rdexponent}
	\beta_{3,1}^{(\alpha)}(\Sigma)=-\int_0^{R_0} \tr[\mathrm d\mathrm R_- [\Sigma \mathrm R_-\Sigma,[\Sigma\mathrm R_-\Sigma,\mathrm R_-]]]=-\Bigl(\int_0^{-R_0^\dag} \tr[\mathrm d\mathrm R_+ [\Sigma \mathrm R_+\Sigma,[\Sigma\mathrm R_+\Sigma,\mathrm R_+]]]\Bigr)^\ast\\
	-\frac{1}{2}\tr[(\Sigma R_0^2)^2-(\Sigma R_0)^4]\equiv \frac{1}{4}\tr[([\Sigma\mathrm R_0\Sigma,\mathrm R_0])^2]\, ,
\end{multline}
where, as before, we omitted the paths because the integrals do not depend on it. As manifest in the rightmost expression, this terms capture a purely non-Abelian contribution.
As before, given that the two integrals of the physical part match each other, the configurational part  is  characterised by a single integral
\be\label{eq:k3x}
\log k_{3,1}=-\iint_0^1\mathrm d^2 \tau  \tfrac{\ell_-}{\ell_+}\tfrac{\chi^{(+)\ast}_{22}(-\frac{d}{\ell_{+}}-\frac{\ell_{-}}{\ell_{+}}\tau_1,\tau_2)}{(\tau_2+\frac{\ell_-}{\ell_+}\tau_1+\frac{d}{\ell_+}-i\frac{\epsilon}{\ell_+})^2}+\tfrac{\chi_{21}^{(-)}(\tau_1,\tau_2)}{(\tau_1-\tau_2-i\frac{\epsilon}{\ell_-})^2}
+\Bigl(\tfrac{\ell_+}{\ell_-}\tfrac{\chi^{(-)\ast}_{22}(-\frac{d}{\ell_{-}}-\frac{\ell_{+}}{\ell_{-}}\tau_1,\tau_2)}{(\tau_2+\frac{\ell_+}{\ell_-}\tau_1+\frac{d}{\ell_-}-i\frac{\epsilon}{\ell_-})^2}+\tfrac{\chi_{21}^{(+)}(\tau_1,\tau_2)}{(\tau_1-\tau_2-i\frac{\epsilon}{\ell_+})^2}\Bigr)^\ast\, .
\ee
Despite this could be calculated analytically, the result is cumbersome and we have not been able to simplify it so as to exhibit it here. We have not even found a satisfactory way to make it manifest that it depends only on the cross ratio. We have however made some simplifications that allow for an efficient numerical evaluation. Specifically, the integrals above can be reduced to the following:
\begin{multline}
	\iint_0^1\mathrm d^2 \tau  \Bigl(\tfrac{\chi_{21}^{(s)}(\tau_1,\tau_2)}{(\tau_1-\tau_2-i\frac{\epsilon}{\ell_s})^2}+\tfrac{\ell_s}{\ell_{-s}}\tfrac{\chi^{(-s)\ast}_{22}(-\frac{d}{\ell_{-s}}-\frac{\ell_{s}}{\ell_{-s}}\tau_1,\tau_2)}{(\tau_2+\frac{\ell_s}{\ell_{-s}}\tau_1+\frac{d}{\ell_{-s}}-i\frac{\epsilon}{\ell_{-s}})^2}\Bigr)\xrightarrow{\epsilon\rightarrow 0^+}\\
	\tfrac{\ell_s}{\ell_{-s}}\iint_0^1\mathrm d^2 \tau \int_{\tau_2}^{\tau_1}\mathrm d \tau_1' \int_0^{\tau_2}\mathrm d\tau_2'\tfrac{\log\left(\frac{(d+\ell_s\tau_1')(d+(\ell_s+\ell_{-s})\tau_2)}{(d+\ell_s\tau_2)(d+\ell_s\tau_1'+\ell_{-s}\tau_2)}\right)}{(\tau_1-\tau_2)^2(\tau_2'+\frac{d}{\ell_{-s}}+\frac{\ell_s}{\ell_{-s}}\tau_1')^2}\\
	- \int_0^{\ell_{-s}}\mathrm d\tau_2'\int_{-d-\ell_s}^{\ell_{-s}}\mathrm d \tau_1'\tfrac{\left[\log
		\Bigl(\frac{d+(\frac{\ell_s}{\ell_{-s}}+1)\tau_2'}{d+\tau_2'}\Bigr)-\log(d+\tau_1'+\frac{\ell_{s}}{\ell_{-s}}\tau_2'+i\epsilon)\right]\log \left(\frac{(\max (0,-d-\tau_1')+d+\ell_{-s}) (\max
			(\tau_1',\tau_2')+d+\ell_s
			)}{(d+\ell_{-s}+\ell_s) (\max (0,-d-\tau_1')+\max
			(\tau_1',\tau_2')+d)}\right)}{(\tau_2'-\tau_1'-i \epsilon)^2}\\
	-\int_0^{\ell_s}\mathrm d\tau_1\int_0^{\ell_{-s}}\mathrm d \tau_2 \tfrac{\log
		\Bigl(\frac{(d+\tau_2)}{(d+(\frac{\ell_s}{\ell_{-s}}+1)\tau_2)}\Bigr)\Bigl[\log(i\epsilon)-\log(\tau_2+i\epsilon)+\log \Bigl(\frac{ d+\tau_1}{ d+\tau_1+\tau_2 }\Bigr)\Bigr]}{(\tau_2+\tau_1+d)^2}\\
	-\int_0^{\ell_{-s}}\mathrm d \tau_2\int_{-d-\ell_s}^{\tau_2}\mathrm d \tau_1'\tfrac{\tau_2(\ell_s-\max (0,-d-\tau_1'))\log
		\left(d+\tau_1'+\frac{\ell_{s}}{\ell_{-s}}\tau_2+i\epsilon\right)}{(\tau_1'+i
		\epsilon ) (\tau_1'-\tau_2+i \epsilon ) (d+\ell_s+\tau_2) (\max (0,-d-\tau_1')+d+\tau_2)}\\
	-\tfrac{\ell_s }{\ell_{-s}}\int_0^{\ell_{-s}}\!\!\!\!\!\!\mathrm d\tau_2'\int_{-d-\ell_s}^{\ell_{-s}}\!\!\!\!\!\!\mathrm d \tau_1'
	\tfrac{\Bigl[\log(\tau_1'+i\epsilon)-\log(\tau_1'-\tau_2'+i\epsilon)+\log   \Bigl(\frac{d+(\frac{\ell_s }{\ell_{-s}}+1)\tau_2'}{d+\frac{\ell_s }{\ell_{-s}}\tau_2'}\Bigr)\Bigr]\log \Bigl(\frac{(\max (0,-d-\tau_1')+d+\ell_{-s}) (\max
			(\tau_1',\tau_2')+d+\ell_s
			)}{(d+\ell_{-s}+\ell_s) (\max (0,-d-\tau_1')+\max
			(\tau_1',\tau_2')+d)}\Bigr)}{(\frac{\ell_s }{\ell_{-s}}\tau_2'+d+\tau_1'+i \epsilon)^2}
\end{multline}
The numerical evaluation of the integrals above is consistent with $\log k_{3,1}$ being a function of just the cross ratio $x_4$, i.e., $k_{3,1}=k_{3,1}(x_4)$. Figure~\ref{f:k31} shows a plot of $k_{3,1}(x_4)$. 
Importantly and contrary to  $k_{1,1}(x_4)$, $k_{3,1}(x_4)$ is smaller than $1$ and approaches zero as $x_4\rightarrow 1$. Since also $\beta_{3,1}^{(\alpha)}(\Sigma)$ is negative, the third order correction tends to increase the tripartite information. Nevertheless, such a correction is not strong enough to affect the residual tripartite information, which remains $-\log 2$ also at this perturbative level.

\paragraph{Inclusion of the leading non-Abelian correction.}

The leading non-Abelian correction is characterised by the function $k_{3,1}(x_4)$ defined in \eqref{eq:k3x} and the corresponding exponent \eqref{eq:3rdexponent}. We remedy not having provided an explicit expression for $k_{3,1}(x_4)$ by replacing $k_{3,1}(x_4)$ with a function $\tilde k_{3,1}(x_4)$ that is equivalent to it up to higher order corrections. We can indeed use the freedom in the choice of $\tilde k_{3,1}(x_4)$ to exhibit an explicit expression.
To that aim, we exploit the knowledge of the exact result in the limit $|\tpm|\rightarrow 1$ with the observation that truncating the perturbative series at the third order provides an excellent approximation. 
The function that replaces  $k_{3,1}(x_4)$ so as to make the pseudo-conformal limit $|\tpm |\rightarrow 1$ exact is indeed a captivating candidate for $\tilde k_{3,1}(x_4)$. We stress that, in this, we give up the independence of $k_{3,1}(x_4)$ from $\Sigma$ and $\alpha$.
For $\alpha=3$,  this results in  the following replacement 
\be
k_{3,1}(x_4)\rightarrow\tilde k_{3,1}(x_4) =(1-x_4)^{54}\Bigl[\tfrac{\Theta(0,0|\hat\tau_{x_4}^{(3)})}{\Theta(0,\frac{1}{2}|\hat\tau_{x_4}^{(3)})}\Bigr]^{\frac{729}{4}}\, ,
\ee
The corresponding estimation of the R\'enyi-$3$  tripartite information reads
\be
\tilde G_3(x_4)=-\log 2+\tfrac{1}{2}\log\Bigl(1+3(1-x_4)^{-2 \sum_{k_F}\beta_{1}(\tpm^{(k_F)})+54\sum_{k_F}\beta_{3}(\tpm^{(k_F)})}\Bigl[\tfrac{\Theta(0,0|\hat\tau_{x_4}^{(3)})}{\Theta(0,\frac{1}{2}|\hat\tau_{x_4}^{(3)})}\Bigr]^{\frac{729}{4}\sum_{k_F}\beta_{3}(\tpm^{(k_F)})}\Bigr)\, ,
\ee
where the exponents are explicitly given by --- cf. Eqs~\eqref{eq:k1x}~and~ \eqref{eq:k3x}
\be
\ba
\beta_{1}(\tpm)=&\tfrac{\log ^2\left(\frac{\sqrt{3} \tp-i}{\sqrt{3} \tm-i}\right)+\log
	^2\left(\frac{\sqrt{3} \tp+i}{\sqrt{3}
		\tm+i}\right)-\log \left(\frac{\sqrt{3}
		\tp+i}{\sqrt{3} \tm+i}\right) \log \left(\frac{\sqrt{3} \tp-i}{\sqrt{3} \tm-i}\right)}{9 \pi ^2}\\
\beta_{3}(\tpm)=&-\tfrac{1}{4}\beta_{1}^2(\tpm)\, .
\ea
\ee
Note that, having used the result for $\tpm\rightarrow 1$, this approximation is rougher for small $\tpm$, where it is expected to overestimate the correction to the Abelian approximation.

We can now check that the residual tripartite information does not depend on $\tpm$. To that aim, we should expand the prediction above around $x_4=1$. Since we have~\cite{Calabrese2011Entanglement}
\be
\tfrac{\Theta(0,0|\hat\tau_{x_4}^{(3)})}{\Theta(0,\frac{1}{2}|\hat\tau_{x_4}^{(3)})}\sim \frac{3^{3/4}}{2}(1-x_4)^{-\frac{1}{4}}
\ee
we find
\be
\tilde G_3(x_4)+\log 2\sim \tfrac{3}{2}(\tfrac{27}{16})^{-\frac{729}{64}\sum_{k_F}\beta_1^2(\tpm^{(k_F)})}(1-x_4)^{-2\sum_{k_F}\beta_1(\tpm^{(k_F)})-\frac{135}{64}\sum_{k_F}\beta_1^2(\tpm^{(k_F)})}\, 
\ee
where we used that, for any $\tpm\neq 0$, the exponent of $(1-x_4)$ is strictly positive ($-\frac{128}{135}<\beta_1(\tpm)<0$). 
We point out that such an estimate is perturbative and, in principle, is subject to corrections coming from higher order contributions.

\section{Conclusion}\label{sec:Conclusions}
In this work we have investigated  the tripartite information in the stationary state emerging after global quenches from two types of initial states: ground states of critical systems and inhomogeneous states obtained by  joining two macroscopically different parts, e.g., a domain wall. We have identified the quantum field theories capturing the limit of large subsystem's lengths and derived analytic expressions for the R\'enyi-$\alpha$ tripartite information. Specifically, using a mapping to a Riemann-Hilbert problem with a piecewise constant matrix for a doubly connected domain, we have proved the conjecture of Ref.~\cite{Maric2022Universality} for the asymptotic behaviour of the R\'enyi-$2$ tripartite information. We have also worked out an implicit expression for the R\'enyi-$\alpha$ tripartite information, eq.~\eqref{eq:I3sol}, which we have then unravelled within a rapidly convergent perturbation theory.
By keeping just the first three orders (where, in fact, the second order vanishes) of the expansion, we have obtained predictions that, in all quench settings investigated, approximate the R\'enyi-$\alpha$ tripartite information with an accuracy that easily surpasses the limitations of our numerical simulations.

Our findings have confirmed the conjectures of Ref~\cite{Maric2022Universality}: first,  the R\'enyi-$\alpha$ tripartite information approaches a function of the cross ratio, and, second, the limit in which the length of the second block $B$ becomes negligible with respect to that of $A$ and $C$ (but still much larger than the lattice spacing) is $-\log 2$ for any R\'enyi index $\alpha$, confirming the emergence of a negative residual tripartite information. The  vanishing of the residual tripartite information that is expected in thermal states (cf. sec. \ref{sec:Tripartite information}) suggests that a nonzero value is always associated with non-thermal stationary properties. In translationally invariant global quenches, this is generally expected only in systems with infinitely many conservation laws, which are usually integrable. Indeed, in one dimension, only with infinitely many local integrals of motion the system can retain memory of algebraically decaying correlations. In bipartitioning protocols, we do not have a comparably strong argument to exclude the emergence of a nonzero residual entropy in generic systems. 

This work leaves several open questions that range from improving our results to generalising them. 
If we overlook the powerful perturbation theory that we extracted from it, the exact implicit expression that we derived by formally solving the Riemann-Hilbert problem (for $\alpha>2$) is unsatisfactory: it does not make it manifest neither the dependency on the cross ratio nor the limit of the residual tripartite information. An explicit solution could then be desirable, both for confirming that our conjectures are exact and not just excellent approximations and for establishing an exact connection with the known CFT results. 

Second, we have assessed the universality of the residual tripartite information only in noninteracting models, and it is not even clear to us whether or not the residual tripartite information could take only discrete values (in our systems, $0$ or $-\log 2$). In this respect, generalising our findings to interacting integrable spin chains is not straightforward, but perhaps not impossible. For example, bosonisation could help in mapping the problem to a noninteracting one, in which the entropies are expressed in terms of just a correlation operator; carefully taking into account the nonlocality of the Jordan-Wigner transformation might then allow one to carry out the analogous calculation. 

In addition, we have  investigated only models that do not allow for string order at infinite times after the quench.
There are however spin-chain systems in which symmetry protected topological order can emerge even after a global quench~\cite{Fagotti2022Global,FagottiMaricZadnik2022}. Considering that, in 2D, the tripartite information is sensitive to topological order, the question of whether the tripartite information could display some peculiar features in such systems is of central importance and still open.  

Another point that we overlooked is the actual relaxation dynamics: by assuming a description through a stationary state we have ignored that finite-time effects could be significant. With Ref.~\cite{Bocini2022Connected} in mind, we wonder whether some of those corrections could be included by taking into account how correlations change when the lengths of the subsystems are large enough not to allow us to ignore the inhomogeneity of the state.

Finally, two natural generalisations of our work are the analysis of the tripartite information of detached subsystems, as well as of entanglement measures such as the negativity, which could also help clarify whether the residual tripartite information is a  merely classical quantity or if instead it witnesses some exceptional entanglement structure.

\appendix

\section{Generalized XY model}
\label{appendix generalized XY models}
A generalisation of the quantum XY model~\cite{Lieb1961} was proposed in Ref.~\cite{Suzuki1971The} and is described by the Hamiltonian
\begin{equation}\label{eq:HgXY}
	\bs H=\sum_\ell \left(\sum_{\gamma_1,\gamma_2\in\{x,y\}}\sum_{n=1}^\infty J_{\gamma_1,\gamma_2}^{(n)} \bs \sigma^{\gamma_1}_\ell \left(\bs \prod_{j=1}^{n-1} \bs\sigma_{\ell+j}^z\right)\bs \sigma_{\ell+n}^{\gamma_2}+h\bs\sigma_\ell^z\right)\, .
\end{equation}
The common denominator of those models is that a Jordan-Wigner transformation maps the Hamiltonian into a quadratic form of fermions.
The XY model considered in \eqref{eq:XY} corresponds to the special case $J_{\gamma_1,\gamma_2}^{(k)}=\delta_{k,1}\delta_{\gamma_1,\gamma_2}\left(\delta_{\gamma_1,x}J_x+\delta_{\gamma_1,y}J_y\right)$. In this work we consider only local Hamiltonians, so we assume  $J^{(n)}_{\gamma_1,\gamma_2}=0$ for any $n$ larger than some given integer. The symbol of \eqref{eq:HgXY}, as defined in \eqref{eq:symbol of the Hamiltonian definition}, reads
\begin{align}
	& \mathcal{H}(k)=f_0(k)I+f_1(k)\sigma^x+f_2(k)\sigma^y+f_3(k)\sigma^z, \\
	& f_0(k)=2\sum_{n=1}^\infty \left(J^{(n)}_{x,y}-J^{(n)}_{y,x}\right)\sin(n k), \\
	& f_1(k)=2\sum_{n=1}^\infty \left(J^{(n)}_{x,x}-J^{(n)}_{y,y}\right)\sin(n k), \\
	& f_2(k)=2h-2\sum_{n=1}^\infty \left(J^{(n)}_{x,x}+J^{(n)}_{y,y}\right)\cos(n k), \\
	& f_3(k)=-2\sum_{n=1}^\infty \left(J^{(n)}_{x,y}+J^{(n)}_{y,x}\right)\sin(n k).
\end{align}
The excitation energies $\varepsilon(k)$ can be identified with the eigenvalues of $\mathcal H(k)$, i.e., $\varepsilon(k)=f_0(k)+\sqrt{\sum_{j=1}^3f_j^2(k)}$. We note that reflection symmetric Hamiltonians satisfy $f_0(k)=0$.

\section{Determinant representation}
\label{appendix: linearization of the determinant representation}

\subsection*{Integrating the fields out
}\label{appendix path integrals}
The partition function corresponding to the action \eqref{eq:action copies} with the boundary conditions \eqref{eq:boundary conditions decoupled fields} can be rewritten as a partition function in which the fields are continuous at the boundaries provided that the following term is added to the action
\begin{multline}
	\delta S_{E}^{(\alpha)}(k_F;\Sigma)=\int_{A}\mathrm d x \sum_{\ell,n}^\alpha(
	\delta_{\ell n}-[\Sigma T\Sigma]_{\ell n})\vec \psi^{(\ell)\dag}(x,0^+)\vec u\ \vec u^\dag \vec \psi^{(n)}(x,0^-)+\\
	\int_{C}\mathrm d x \sum_{\ell,n}^\alpha(
	\delta_{\ell n}-T_{\ell n})\vec \psi^{(\ell)\dag}(x,0^+)\vec u\ \vec u^\dag \vec \psi^{(n)}(x,0^-)\, .
\end{multline}
Here $\vec u$ is the normalised vector defined in \eqref{eq:field vector correlator}.
Indeed, this new action leads to the same classical solutions and boundary conditions as the original problem. Being now the dependency on $\Sigma$ explicit, we have eased the notations and replaced  $\vec \psi^{(n)}(x,\tau; \Sigma)$ by $\vec \psi^{(n)}(x,\tau)$.

We now linearise $\delta S_{E}^{(\alpha)}(k_F;\Sigma)$.  As shown below, this can be done by introducing an appropriate number of auxiliary Grassmann fields. Before doing it, however, we should take care of a subtlety. Since the eigenvalues $-z$ of $T$ satisfy $z^\alpha=-1$ (cf.~\eqref{eq:Tz}), the kernel of $\mathrm I- T$ is nonempty when $\alpha$ is odd.
For $\alpha$ even  we could linearise $\delta S_{E}^{(\alpha)}(k_F;\Sigma)$ by introducing $2\alpha$ auxiliary fields, half of them defined in $A$ and the rest in $C$ (see below);  for $\alpha$ odd, however, we would need only $2\alpha-2$ auxiliary fields. We choose here to overcome this discrepancy between even and odd $\alpha$ by introducing a small regulator $\epsilon$ and replacing  $I-T$ by $(1+\epsilon)I-T$. In this way the kernel is non-empty even when $\alpha$ is odd and we can treat the even and odd case on the same footing.
The regulator $\epsilon$ will be   sent to zero at the end of the calculation. For the reader's information, we mention that the  calculation could be effectively carried out also without introducing $\epsilon$, with the advantage of slightly reducing the size of the matrices appearing in the final result for $\alpha$ odd, at the cost however of having to make a distinction between even and odd $\alpha$.
That said, it is convenient to rewrite $\delta S_{E}^{(\alpha)}(k_F;\Sigma)$ in a basis diagonalising $T$:
\begin{multline}\label{eq:deltaSdiag}
	\int_{A}\mathrm d x \sum_{z|z^\alpha=-1}
	\Bigl[\frac{1}{\sqrt{\alpha}}\sum_{\ell=1}^\alpha \Sigma_{\ell\ell}(-z)^{\ell}
	\vec \psi^{(\ell)\dag}(x,0^+)\vec u\Bigr] (1+\epsilon+z) \Bigl[\frac{1}{\sqrt{\alpha}}\sum_{n=1}^\alpha \Sigma_{nn} (-1/z)^{n}\vec u^\dag \vec \psi^{(n)}(x,0^-)\Bigr]
	\\
	+\int_{C}\mathrm d x \sum_{z|z^\alpha=-1}
	\Bigl[\frac{1}{\sqrt{\alpha}}\sum_{\ell=1}^\alpha(-z)^{\ell}
	\vec \psi^{(\ell)\dag}(x,0^+)\vec u\Bigr] (1+\epsilon+z) \Bigl[\frac{1}{\sqrt{\alpha}}\sum_{n=1}^\alpha(-1/z)^{n}\vec u^\dag \vec \psi^{(n)}(x,0^-)\Bigr]
\end{multline}
where we used
\be
\bigl[\mathrm (1+\epsilon)I-T[\{0,\hdots,0\}]\bigr]_{\ell n}=
\frac{1}{\alpha}\sum_{z|z^\alpha=-1}(1+\epsilon+z)(-z)^{\ell-n}\, .
\ee
Let us then introduce the fields $\mathcal A(x;z)$ and $\mathcal C(x;z)$ with action
\be
\delta S_E^{(\alpha)}[\mathcal A,\mathcal C]=-f_\alpha\sum_{z|z^\alpha=-1}\left[\int_A\mathrm dx \mathcal A^\dag(x;z)\frac{1}{1+\epsilon+z}\mathcal A(x;z)+\int_C\mathrm dx \mathcal C^\dag(x;z)\frac{1}{1+\epsilon+z}\mathcal C(x;z)\right]
\ee
where the constant
\be
f_\alpha\equiv\Big[\prod_{z|z^\alpha=-1}(1+\epsilon+z)\Big]^{1/\alpha}
\ee
has been chosen in such a way that the total partition function is not affected by the addition of the new fields, indeed
\be
\idotsint\mathcal D\mathcal A^\dag\mathcal D\mathcal A\mathcal D\mathcal C^\dag\mathcal D\mathcal C e^{-\delta S_E^{(\alpha)}[\mathcal A,\mathcal C]}=1 \, .
\ee
The term $\delta S_{E}^{(\alpha)}(k_F;\Sigma)$ incorporating the boundary conditions can now be linearised by shifting the auxiliary fields $\mathcal A$ and $\mathcal C$. This results in the comprehensive action
\begin{multline}
	S_{E}^{(\alpha)}=\int\mathrm d \tau\int\mathrm d x\  \sum_{n=1}^\alpha\Bigl[\vec \psi^{(n)\dag}(x,\tau)
	\begin{pmatrix}
		\partial_\tau -i\partial_x & 0 & 0\\
		0 & \partial_\tau +i\partial_x & 0 \\
		0 & 0 & \partial_\tau+\sigma 
	\end{pmatrix}
	\vec \psi^{(n)}(x,\tau)\\
	-f_\alpha\sum_{z|z^\alpha=-1}\left[\int_A\mathrm dx \mathcal A^\dag(x;z)\frac{1}{1+\epsilon+z}\mathcal A(x;z)+\int_C\mathrm dx \mathcal C^\dag(x;z)\frac{1}{1+\epsilon+z}\mathcal C(x;z)\right]\\
	+\sqrt{\frac{f_\alpha}{\alpha}}\sum_{z|z^\alpha=-1}\int_A\mathrm dx \sum_{n=1}^\alpha\Bigl[\mathcal A^\dag(x;z) \Sigma_{nn} (-\tfrac{1}{z})^{n}\vec u^\dag \vec \psi^{(n)}(x,0^-)+\vec \psi^{(n)\dag}(x,0^+)\vec u \Sigma_{nn} (-z)^{n}\mathcal A(x;z)\Bigr]\\
	+\sqrt{\frac{f_\alpha}{\alpha}}\sum_{z|z^\alpha=-1}\int_C\mathrm dx \sum_{n=1}^\alpha\Bigl[\mathcal C^\dag(x;z)  (-\tfrac{1}{z})^{n}\vec u^\dag \vec \psi^{(n)}(x,0^-)+\vec \psi^{(n)\dag}(x,0^+)\vec u (-z)^{n}\mathcal C(x;z)\Bigr]
\end{multline}

The third step is to integrate over $\vec \psi^{(n)\dag}$ and $\vec \psi^{(n)}$. This produces the normalization factor associated with the path integral representation of $\mathrm{tr}[\rho_{A\cup C}]^\alpha$ multiplied by the partition function of the theory with Euclidean action
\begin{multline}
	S_E=-f_\alpha\sum_{z|z^\alpha=-1}\left[\int_A\mathrm dx \mathcal A^\dag(x;z)\frac{1}{1+\epsilon+z}\mathcal A(x;z)+\int_C\mathrm dx \mathcal C^\dag(x;z)\frac{1}{1+\epsilon+z}\mathcal C(x;z)\right]\\
	+\frac{f_\alpha}{\alpha}\int\mathrm d x\sum_{n=1}^\alpha  \sum_{z|z^\alpha=-1} \Bigl\{\int_A\mathrm d x'\mathcal A^\dag(x',z)\Sigma_{nn} (-\tfrac{1}{z})^n
	+\int_C\mathrm d x'\mathcal C^\dag(x',z) (-\tfrac{1}{z})^n  \Bigr\}\vec u^\dag \braket{\vec{\psi}^{(n)}(x')\vec{\psi}^{(n)\dagger}(x)} \vec u\\
	\sum_{w|w^\alpha=-1}\Bigl\{
	\chi_A(x) \Sigma_{nn}(-w)^n\mathcal A(x;w)
	+\chi_C(x) (-w)^n\mathcal C(x;w)
	\Bigr\}
\end{multline}
The correlation functions $\braket{\vec{\psi}^{(n)}(x)\vec{\psi}^{(n)\dagger}(y)}$ are independent of $n$ and from \eqref{eq:field vector correlator} it follows
\begin{equation}
	\vec{u}^\dagger\braket{\vec{\psi}^{(n)}(x)\vec{\psi}^{(n)\dagger}(y)}\vec{u}=\braket{\bs \psi(x)\bs \psi^\dagger(y)},
\end{equation}
where the right hand side is given by \eqref{eq:fields correlations each fermi point}.

Finally, the partition function \eqref{eq:trace of a product of gaussians} is the determinant of the operator $\bs L_{\underline{\sigma}}$ coupling the fields in $S_E$, i.e. 
\be
\{\Gamma_{j_1},\hdots,\Gamma_{j_\alpha}\}= \det
\bs L_{\underline \sigma}.
\ee
Defining the operator $\bs C$ by
\begin{equation}
	\bs C(x,y)=\braket{\bs \psi(x)\bs \psi^\dagger(y)}, \quad x,y\in A\cup C,
\end{equation}
where the right-hand side is given by \eqref{eq:fields correlations each fermi point}, gives
\be
\ba
\bs L_{\underline{\sigma}}(z,x;w,y)\sim &\begin{cases}
	f_\alpha\delta_{zw} \Bigl[\frac{\delta(x-y)}{1+\epsilon+z}-\bs C(x,y)\Bigr]&x,y\in A\bigvee x,y\in C\\
	-\frac{f_\alpha}{\alpha}\sum_{n=1}^\alpha \sigma_n(\tfrac{w}{z})^n\bs C(x,y)&x\in A, y\in C\bigvee x\in C, y\in A
\end{cases}\\
&z^\alpha=w^\alpha=-1. 
\ea
\ee
Here we used the symbol $\sim$ rather than $=$ because $\bs L_{\underline{\sigma}}(z,x;w,y)$ can be replaced by any other operator with the same determinant. 
More compactly, we have
\begin{empheq}[box=\fbox]{equation}
	\bs L_{\underline \sigma}\sim f_\alpha\left(\mathrm D\otimes \bs 1-\mathrm I\otimes \bs C+\frac{\mathrm I-\mathrm M_{\underline \sigma}}{2}\otimes (\bs C-\bs P \bs C\bs P)\right)
\end{empheq}
where
\be
\ba
[\mathrm D]_{zw}=&\frac{1}{1+\epsilon+z}\delta_{z w}\\
[\mathrm M_{\underline\sigma}]_{zw}=&\frac{1}{\alpha}\sum_{n=1}^\alpha\sigma_n(\tfrac{w}{z})^n
\ea
\ee
and $\bs P$ is defined in \eqref{eq:operator P definition continuum}.

\subsection*{Standard form}\label{appendix standard form}

To relate the calculation of the R\'enyi entropies to the solution of a Riemann-Hilbert problem with a piecewise constant matrix  it is convenient to represent $\bs L_{\underline\sigma}$ by an operator with a more standard form, specifically, an operator in which $\bs P$ acts only on the left of $\bs C$. As shown before long, bringing $\bs P$ to the left of $\bs C$ can be readily achieved by increasing the size of the representation of $\bs L_{\underline \sigma}$. Since, however, we would also like to eventually reduce the space back to (almost) its original size, it is firstly convenient to  diagonalise $\mathrm M_{\underline \sigma}$. This is straightforward, indeed, the matrix is circulant and the eigenvalues are equal to $\sigma_j$. We call $U_{\underline \sigma}$ a unitary matrix that diagonalises $\mathrm M_{\underline \sigma}$. Specifically, we choose the unitary transformation $\mathrm U_{\underline \sigma}$ that brings $\mathrm M_{\underline \sigma}$ into $\Sigma$:
\be
\mathrm M_{\underline \sigma}=\mathrm U_{\underline \sigma} \Sigma\mathrm U^\dag_{\underline \sigma}
\ee
where $j_{n}\in \{1,\hdots,\alpha\}$.
The change of basis results in
\be
\bs L_{\underline \sigma}\sim f_\alpha\left(\mathrm U^\dag_{\underline \sigma}\mathrm D\mathrm U_{\underline \sigma}\otimes \bs 1-\mathrm I\otimes \bs C+\frac{\mathrm I-\Sigma}{2}\otimes (\bs C-\bs P \bs C \bs P)\right)\, .
\ee
We now use the identity
\be
\det\begin{pmatrix}
	A_{11}&A_{12}\\
	A_{21}&A_{22}
\end{pmatrix}=\det A_{11}\det(A_{22}-A_{21}A_{11}^{-1}A_{12})
\ee
with 
\be
\ba
A_{11}=&\mathrm I\otimes \bs 1\\
A_{12}=&f_\alpha \mathrm I\otimes \bs P\\
A_{21}=&\frac{\mathrm I-\mathrm \Sigma}{2}\otimes \bs P \Bigl(\bs C-\tfrac{\bs 1}{2}\Bigr )\\
A_{22}=&f_\alpha\mathrm U^\dag_{\underline \sigma}(\mathrm D-\tfrac{\mathrm I}{2})\mathrm U_{\underline \sigma}\otimes \bs 1-f_\alpha\frac{\mathrm I+\Sigma}{2}\otimes\Bigl(\bs C-\tfrac{\bs 1}{2}\Bigr )
\ea
\ee
Thus we have
\begin{multline}
	\det \bs L_{\underline \sigma}=\det\left|\begin{pmatrix}
		\mathrm I\otimes \bs 1&f_\alpha \mathrm I\otimes \bs P\\
		0&f_\alpha\mathrm U^\dag_{\underline \sigma}(\mathrm D-\tfrac{\mathrm I}{2})\mathrm U_{\underline \sigma}\otimes \bs 1
	\end{pmatrix}+\right.\\
	\left.\begin{pmatrix}
		0&0\\
		\frac{\mathrm I-\mathrm \Sigma}{2}\otimes \bs P(\bs C-\tfrac{\bs 1}{2}\Bigr )&-f_\alpha\frac{\mathrm I+\Sigma}{2}\otimes\Bigl(\bs C-\tfrac{\bs 1}{2}\Bigr )
	\end{pmatrix}
	\right|\, ,
\end{multline}
in which, as desired, $\bs P$ does not appear to the right of $\bs C-\frac{\bs 1}{2}$.
We now simplify the diagonal part in this representation of $\bs L_{\underline \sigma}$. This can done by multiplying from the left by the operator
\be
\kappa_\alpha \begin{pmatrix}
	\mathrm I\otimes \bs 1&f_\alpha \mathrm I\otimes \bs P\\
	0&f_\alpha\mathrm U^\dag_{\underline \sigma}(\mathrm D-\frac{\mathrm I}{2})\mathrm U_{\underline \sigma}\otimes \bs 1
\end{pmatrix}^{-1}=\kappa_\alpha\begin{pmatrix}
	\mathrm I\otimes \bs 1&- \mathrm U^\dag_{\underline \sigma}(\mathrm D-\tfrac{\mathrm I}{2})^{-1}\mathrm U_{\underline \sigma}\otimes \bs P\\
	0&f_\alpha^{-1}\mathrm U^\dag_{\underline \sigma}(\mathrm D-\tfrac{\mathrm I}{2})^{-1}\mathrm U_{\underline \sigma}\otimes \bs 1
\end{pmatrix}
\ee
where $\kappa_\alpha$ ensures the transformation to have unitary determinant and is given by
\begin{equation}
	\kappa_\alpha=\frac{1}{\sqrt{2}} \bigg[\prod_{z|z^\alpha=-1}(1-\epsilon-z)\bigg]^{\frac{1}{2\alpha}}\xrightarrow[\epsilon\to 0]{} 2^{\frac{1-\alpha}{2\alpha}}
\end{equation}
Thus we have
\begin{multline}
	\bs L_{\underline \sigma}\sim \kappa_\alpha \begin{pmatrix}
		\mathrm I\otimes \bs 1&0\\
		0&\mathrm I\otimes \bs 1
	\end{pmatrix}+\\
	\kappa_\alpha \begin{pmatrix}
		- \mathrm U^\dag_{\underline \sigma}(\mathrm D-\tfrac{\mathrm I}{2})^{-1}\mathrm U_{\underline \sigma} \frac{\mathrm I-\mathrm \Sigma}{2}\otimes \bs 1&f_\alpha\mathrm U^\dag_{\underline \sigma}(\mathrm D-\tfrac{\mathrm I}{2})^{-1}\mathrm U_{\underline \sigma} \frac{\mathrm I+\mathrm \Sigma}{2}\otimes\bs P \\
		f_\alpha^{-1}\mathrm U^\dag_{\underline \sigma}(\mathrm D-\tfrac{\mathrm I}{2})^{-1}\mathrm U_{\underline \sigma}\frac{\mathrm I-\mathrm \Sigma}{2}\otimes \bs P&-\mathrm U^\dag_{\underline \sigma}(\mathrm D-\tfrac{\mathrm I}{2})^{-1}\mathrm U_{\underline \sigma}\frac{\mathrm I+\mathrm \Sigma}{2}\otimes \bs 1
	\end{pmatrix}\mathrm I\otimes \Bigl(\bs C-\tfrac{\bs 1}{2}\Bigr )\, .
\end{multline}
We can get rid of $f_\alpha$ by applying 
\be
\begin{pmatrix}
	f_\alpha^{-1}\mathrm I&0\\
	0&\mathrm I
\end{pmatrix}\otimes \bs 1\qquad \text{and}\qquad \begin{pmatrix}
	f_\alpha\mathrm I&0\\
	0&\mathrm I
\end{pmatrix}\otimes \bs 1
\ee
to the left and to the right, respectively. 
Making the dependency on the spatial coordinates explicit, we then find
\begin{multline}\label{eq:almost}
	\bs L_{\underline \sigma}(x;y)\sim \kappa_\alpha\delta(x-y) \begin{pmatrix}
		\mathrm I&0\\
		0&\mathrm I
	\end{pmatrix}-\kappa_\alpha\Bigl(\bs C(x,y)-\tfrac{1}{2}\delta(x-y)\Bigr )\\
	\times\left[ \chi_A(x) \begin{pmatrix}
		\mathrm U^\dag_{\underline \sigma}(\mathrm D-\tfrac{\mathrm I}{2})^{-1}\mathrm U_{\underline \sigma} \frac{\mathrm I-\mathrm \Sigma}{2}&\mathrm U^\dag_{\underline \sigma}(\mathrm D-\tfrac{\mathrm I}{2})^{-1}\mathrm U_{\underline \sigma} \frac{\mathrm I+\mathrm \Sigma}{2} \\
		\mathrm U^\dag_{\underline \sigma}(\mathrm D-\tfrac{\mathrm I}{2})^{-1}\mathrm U_{\underline \sigma}\frac{\mathrm I-\mathrm \Sigma}{2}&\mathrm U^\dag_{\underline \sigma}(\mathrm D-\tfrac{\mathrm I}{2})^{-1}\mathrm U_{\underline \sigma}\frac{\mathrm I+\mathrm \Sigma}{2}
	\end{pmatrix}\right. \\
	\left.+ \chi_C(x)\begin{pmatrix}
		\mathrm U^\dag_{\underline \sigma}(\mathrm D-\tfrac{\mathrm I}{2})^{-1}\mathrm U_{\underline \sigma} \frac{\mathrm I-\mathrm \Sigma}{2}&-\mathrm U^\dag_{\underline \sigma}(\mathrm D-\tfrac{\mathrm I}{2})^{-1}\mathrm U_{\underline \sigma} \frac{\mathrm I+\mathrm \Sigma}{2} \\
		-\mathrm U^\dag_{\underline \sigma}(\mathrm D-\tfrac{\mathrm I}{2})^{-1}\mathrm U_{\underline \sigma}\frac{\mathrm I-\mathrm \Sigma}{2}&\mathrm U^\dag_{\underline \sigma}(\mathrm D-\tfrac{\mathrm I}{2})^{-1}\mathrm U_{\underline \sigma}\frac{\mathrm I+\mathrm \Sigma}{2}
	\end{pmatrix} \right]
\end{multline}
The matrices appearing in the last two lines have $\alpha$ columns of zeroes, indeed, for any $\sigma_j$, either $\frac{1-\sigma_j}{2}=0$ or $\frac{1+\sigma_j}{2}=0$. 
Moving these columns to the left through a unitary transformation allows us to reduce the size of the matrix by means of the determinant identity
\be\label{eq:reduction}
\det\left|\begin{pmatrix}
	\lambda \mathrm I^{[d_0\times d_0]}\otimes \bs 1 &\lambda B^{[d_0\times d]}\otimes \bs B\\
	0^{[d\times d_0]}\otimes \bs 0&\lambda A^{[d\times d]}\otimes \bs D
\end{pmatrix}
\right|=\det[\lambda^{1+\frac{d_0}{d}} A^{[d\times d]}\otimes \bs D]\, ,
\ee
where we have specified the size of the matrices as superscripts.
The following idempotent transformation does the job
\be\label{eq:Vsigma}
\mathrm V_{\underline \sigma}= \begin{pmatrix}
	\frac{\mathrm I+\Sigma}{2}&\frac{\mathrm I-\Sigma}{2}\\
	\frac{\mathrm I-\Sigma}{2}&\frac{\mathrm I+\Sigma}{2}
\end{pmatrix}\otimes \bs I \, .
\ee
Finally, the operator is reduced to the following
\be
\bs L_{\underline \sigma}(x;y)\sim  2^{\frac{1-\alpha}{\alpha}}\delta(x-y)\mathrm I-i 2^{\frac{1-\alpha}{\alpha}}\Bigl(2\bs C(x,y)-\delta(x-y)\Bigr )\left[\chi_A(x)\B
+\chi_C(x)\Sigma \B\Sigma
\right]\, .
\ee
where $\B$ is the Hermitian circulant matrix
\be
{}[\B]_{\ell n}=-i[e^{i\pi\frac{(n-\ell\mod \alpha)}{\alpha}}-\delta_{\ell n}]= \frac{1}{\alpha}\sum_{j=1}^\alpha\cot(\tfrac{\pi (j+\frac{1}{2})}{\alpha})e^{2\pi i j\frac{\ell-n}{\alpha}} 
\qquad \ell,n\in\{1,\hdots,\alpha\}
\ee
Thus, we have shown that the operator $\bs L_{\underline \sigma}$ is equivalent to the following
\begin{empheq}[box=\fbox]{equation}
	\bs L_{\underline \sigma}\sim  2^{\frac{1-\alpha}{\alpha}}\mathrm I\otimes\bs 1-i 2^{\frac{1-\alpha}{\alpha}}
	\left[\B\otimes\frac{\bs 1-\bs P}{2}
	+\Sigma \B\Sigma\otimes\frac{\bs 1+\bs P}{2}
	\right]\mathrm I\otimes \Bigl( 2\bs C-\bs 1\Bigr )\, ,
\end{empheq}

\paragraph{Example: $\alpha=2$.}
Besides the ``fermionic'' configuration $\underline\sigma=(1,1)$, there is only one nontrivial configuration, $\underline\sigma=(-1,1)$, whose corresponding operator reads
\be
\bs L_{(-1,1)}\sim \frac{1}{\sqrt{2}}\begin{pmatrix}
	\bs 1&0\\
	0&\bs 1
\end{pmatrix}+\frac{i}{\sqrt{2}}\begin{pmatrix}
	0& \bs P\\
	\bs P&0
\end{pmatrix}(2\bs C-\bs 1)
\ee
This is a very special case where we can  reduce the problem even further (as it is always the case for the fermionic configuration). The matrices indeed commute and we can block-diagonalise the operator, which results in
\be\label{eq:Lalpha2}
\det \bs L_{(-1,1)}=\det\left| \frac{1}{\sqrt{2}}\bs 1+ \frac{i}{\sqrt{2}}
\bs P(2\bs C-1)\right|\det\left| \frac{1}{\sqrt{2}}\bs 1- \frac{i}{\sqrt{2}}
\bs P(2\bs C-1)\right|\, .
\ee

\paragraph{Example: $\alpha=3$.}
Besides the fermionic configuration, there is only one additional nonequivalent  configuration, $\underline\sigma=(-1,1,1)$,
whose corresponding operator can be represented as follows
\be\label{eq:Lalpha3}
\bs L_{(-1,1,1)}\sim 2^{-\frac{2}{3}}\left[\begin{pmatrix}
	\bs 1&0&0\\
	0&\bs 1&0\\
	0&0&\bs 1
\end{pmatrix}
+\begin{pmatrix}
	0& e^{i\frac{\pi}{3}} \bs P& e^{i\frac{2\pi}{3}} \bs P\\
	e^{i\frac{2\pi}{3}} \bs P &0&-e^{i\frac{\pi}{3}} \bs 1\\
	e^{i\frac{\pi}{3}} \bs P& -e^{i\frac{2\pi}{3}} \bs 1&0
\end{pmatrix}(2\bs C-\bs 1)\right]
\ee

\paragraph{Example: $\alpha=4$ and  $\underline\sigma=(-1,-1,1,1)$.}
We can exploit the simple structure of $\Sigma$ to halve the size of the auxiliary space. We find
\be
\det\bs L_{(-1,-1,1,1)}=\prod_{s=\pm 1}\det\left| 2^{-\frac{3}{4}} \begin{pmatrix}
	\bs 1&0\\
	0&\bs 1
\end{pmatrix}-i2^{-\frac{3}{4}}
\begin{pmatrix}
	s\bs P& e^{-i\frac{\pi}{4}}\bs 1+se^{i\frac{\pi}{4}}\bs P\\
	e^{i\frac{\pi}{4}}\bs 1+se^{-i\frac{\pi}{4}}\bs P&s\bs P
\end{pmatrix}\Bigl(2\bs C-1\Bigr )\right|
\ee

\subsection{Back to the chain}\label{appendix generalisation and back to the chain}
We have derived \eqref{eq:repalphaeven} starting from a specific class of correlation operators. We claim here that the final result is formally correct also relaxing those assumptions, and even reintroducing the chain. Following Ref.~\cite{Fagotti2010disjoint} (see also Sec. 3.1.2 of \cite{FagottiPhDThesis}), such a  generalisation can be done by the formal replacement of $2\bs C-\bs 1$ by the correlation matrix $\Gamma$, and of the determinant by its square root (in our specific case it is reasonable to expect that the sign is always positive, but in general the sign ambiguity can be lifted by computing the product of the eigenvalues with halved degeneracy). This prescription results in the formula \eqref{eq:trace of a product of gaussians lattice P on the left}. This representation is more useful than the one proposed in Ref.~\cite{Fagotti2010disjoint}: on the one hand, the matrix in the determinant is linear in $\Gamma_{A\cup C}$ and, on the other hand, it allows one to evaluate the R\'enyi entropy even for arbitrarily large values of $\alpha$ without the need of working out the recursive procedure of Ref.~\cite{Fagotti2010disjoint}.

We show the correctness of this representation for $\alpha=2$ and $\alpha=3$, which are the commonest cases investigated. 

Case $\alpha=2$ is actually trivial: by applying the prescription to \eqref{eq:Lalpha2} we immediately find
\be
\{\Gamma_1,\Gamma_2\}^2=\det\left| \frac{1}{\sqrt{2}}\mathrm I_{A\cup C}+ \frac{i}{\sqrt{2}}
P\Gamma_{A\cup C}\right|\det\left| \frac{1}{\sqrt{2}}\mathrm I_{A\cup C}- \frac{i}{\sqrt{2}}
P\Gamma_{A\cup C}\right|=\det\left|\frac{\mathrm I_{A\cup C}+P\Gamma_{A\cup C}P\Gamma_{A\cup C}}{2}\right|\, ,
\ee
which is the definition of $\{\Gamma_1,\Gamma_2\}$ in Ref.~\cite{Fagotti2010disjoint}. 

Case $\alpha=3$ is more interesting. We start applying the prescription to \eqref{eq:Lalpha3}
\be
\{\Gamma_2,\Gamma_1,\Gamma_1\}^2= \det\left|2^{-\frac{2}{3}}\begin{pmatrix}
	\mathrm I_{A\cup C}& e^{i\frac{\pi}{3}}P\Gamma_{A\cup C}& e^{i\frac{2\pi}{3}}P\Gamma_{A\cup C}\\
	e^{i\frac{2\pi}{3}}P\Gamma_{A\cup C} &\mathrm I_{A\cup C}&-e^{i\frac{\pi}{3}}\Gamma_{A\cup C}\\
	e^{i\frac{\pi}{3}}P\Gamma_{A\cup C}& -e^{i\frac{2\pi}{3}}\Gamma_{A\cup C} &\mathrm I_{A\cup C}
\end{pmatrix}\right|\, ,
\ee
express the determinant of the block matrix in terms of the blocks
\be
\{\Gamma_2,\Gamma_1,\Gamma_1\}^2=\det \left|2^{-\frac{2}{3}}\mathrm I_{A\cup C}\right| \det\left|2^{-\frac{2}{3}}\begin{pmatrix}
	\Gamma_{A\cup C}+ P\Gamma_{A\cup C}P\Gamma_{A\cup C}& -e^{i\frac{\pi}{3}}\left(\mathrm I_{A\cup C}+ P\Gamma_{A\cup C}P\Gamma_{A\cup C}\right)\\
	-e^{i\frac{2\pi}{3}}\left(\mathrm I_{A\cup C}- P\Gamma_{A\cup C}P\Gamma_{A\cup C}\right) & \Gamma_{A\cup C}+ P\Gamma_{A\cup C}P\Gamma_{A\cup C}
\end{pmatrix}\right|\, ,
\ee
recognize $\Gamma_2=P\Gamma_{A\cup C}P$, express the second determinant in terms of blocks, and compare the obtained expression with
\begin{equation}
	\{\Gamma_2,\Gamma_1,\Gamma_1\}^2=\{\Gamma_2\times\Gamma_1,\Gamma_1\}^2\{\Gamma_2,\Gamma_1\}^2,
\end{equation}
where the latter is evaluated using formulas \eqref{eq:product of two gaussians} and \eqref{eq:trace of a product of two gaussians determinant formula}.

\section{Abelian approximation in the pseudo-conformal limit}\label{a:CFT}
In this Appendix we work out the Abelian approximation when $\tp=-\tm\rightarrow\pm1$ at each Fermi point. 
\paragraph{R\'enyi-3 tripartite information.}
There is a single nontrivial contribution associated with
$
\underline \sigma=(1,1,-1)
$. By enforcing \eqref{eq:Abterm} we find
\be
\left.\tfrac{\det \bs L_{(1,1,-1)}^{(k_F)}}{\det \bs L_{(1,1,1)}^{(k_F)}}\right|_{Abelian}=(1-x_4)^{\frac{8}{27}}\, ,
\ee
which results in the following Abelian approximation
\be
G_3^{Abelian}(x_4)=-\log 2+\frac{1}{2}\log\Bigl(1+3(1-x_4)^{\frac{8\nu}{27}}\Bigr)\, .
\ee
where $\nu$ is the number of Fermi points. 
\paragraph{R\'enyi-4 tripartite information.}
There are three nontrivial contributions. By enforcing \eqref{eq:Abterm} we find
\be\label{eq:I3a4CFT}
\begin{gathered}
	\left.\tfrac{\det \bs L_{(1,1,1,-1)}^{(k_F)}}{\det \bs L_{(1,1,1,1)}^{(k_F)}}\right|_{Abelian}=(1-x_4)^{\frac{5}{16}}\qquad
	\left.\tfrac{\det \bs L_{(1,1,-1,-1)}^{(k_F)}}{\det \bs L_{(1,1,1,1)}^{(k_F)}}\right|_{Abelian}=(1-x_4)^{\frac{3}{8}}\\
	\tfrac{\det \bs L_{(1,-1,1,-1)}^{(k_F)}}{\det \bs L_{(1,1,1,1)}^{(k_F)}}=\left.\tfrac{\det \bs L_{(1,-1,1,-1)}^{(k_F)}}{\det \bs L_{(1,1,1,1)}^{(k_F)}}\right|_{Abelian}=(1-x_4)^{\frac{1}{2}}
\end{gathered}
\ee
This results in the following Abelian approximation
\be
G_4^{Abelian}(x_4)=-\log 2+\frac{1}{3}\log\Bigl(1+4(1-x_4)^{\frac{5\nu}{16}}+2(1-x_4)^{\frac{3\nu}{8}}+(1-x_4)^{\frac{\nu}{2}}\Bigr)\, .
\ee
\paragraph{R\'enyi-5 tripartite information.}
There are three nontrivial contributions. By enforcing \eqref{eq:Abterm} we find 
\be
\begin{gathered}
	\left.\tfrac{\det \bs L_{(1,1,1,1,-1)}^{(k_F)}}{\det \bs L_{(1,1,1,1,1)}^{(k_F)}}\right|_{Abelian}=(1-x_4)^{\frac{8}{25}}\qquad
	\left.\tfrac{\det \bs L_{(1,1,1,-1,-1)}^{(k_F)}}{\det \bs L_{(1,1,1,1,1)}^{(k_F)}}\right|_{Abelian}=(1-x_4)^{\frac{4}{125}(15-\sqrt{5})}\\
	\left.\tfrac{\det \bs L_{(1,1,-1,1,-1)}^{(k_F)}}{\det \bs L_{(1,1,1,1,1)}^{(k_F)}}\right|_{Abelian}=(1-x_4)^{\frac{4}{125}(15+\sqrt{5})}\, .
\end{gathered}
\ee
This results in the following Abelian approximation
\be
G_5^{Abelian}(x_4)=-\log 2+\frac{1}{4}\log\Bigl(1+5(1-x_4)^{\frac{8\nu}{25}}+5(1-x_4)^{\frac{4\nu(15-\sqrt{5})}{125}}+5(1-x_4)^{\frac{4\nu(15+\sqrt{5})}{125}}\Bigr)\, .
\ee
\paragraph{R\'enyi-6 tripartite information.}
There are seven nontrivial contributions. By enforcing \eqref{eq:Abterm} we find
\be
\begin{gathered}
	\left.\tfrac{\det \bs L_{(1,1,1,1,1,-1)}^{(k_F)}}{\det \bs L_{(1,1,1,1,1,1)}^{(k_F)}}\right|_{Abelian}=(1-x_4)^{\frac{35}{108}}\qquad
	\left.\tfrac{\det \bs L_{(1,1,1,1,-1,-1)}^{(k_F)}}{\det \bs L_{(1,1,1,1,1,1)}^{(k_F)}}\right|_{Abelian}=(1-x_4)^{\frac{23}{54}}\\
	\left.\tfrac{\det \bs L_{(1,1,1,-1,1,-1)}^{(k_F)}}{\det \bs L_{(1,1,1,1,1,1)}^{(k_F)}}\right|_{Abelian}=(1-x_4)^{\frac{31}{54}}\qquad
	\left.\tfrac{\det \bs L_{(1,1,-1,1,1,-1)}^{(k_F)}}{\det \bs L_{(1,1,1,1,1,1)}^{(k_F)}}\right|_{Abelian}=(1-x_4)^{\frac{16}{27}}\\
	\left.\tfrac{\det \bs L_{(1,1,1,-1,-1,-1)}^{(k_F)}}{\det \bs L_{(1,1,1,1,1,1)}^{(k_F)}}\right|_{Abelian}=(1-x_4)^{\frac{49}{108}}\qquad
	\left.\tfrac{\det \bs L_{(1,1,-1,1,-1,-1)}^{(k_F)}}{\det \bs L_{(1,1,1,1,1,1)}^{(k_F)}}\right|_{Abelian}=(1-x_4)^{\frac{67}{108}}\\
	\tfrac{\det \bs L_{(1,-1,1,-1,1,-1)}^{(k_F)}}{\det \bs L_{(1,1,1,1,1,1)}^{(k_F)}}=\left.\tfrac{\det \bs L_{(1,-1,1,-1,1,-1)}^{(k_F)}}{\det \bs L_{(1,1,1,1,1,1)}^{(k_F)}}\right|_{Abelian}=(1-x_4)^{\frac{3}{4}}
\end{gathered}
\ee
This results in the following Abelian approximation
\begin{multline}
	G_6^{Abelian}(x_4)=-\log 2+\frac{1}{5}\log\Bigl(1+6(1-x_4)^{\frac{35\nu}{108}}+6(1-x_4)^{\frac{23\nu}{54}}+6(1-x_4)^{\frac{31\nu}{54}}+6(1-x_4)^{\frac{16\nu}{27}}\\
	+3 (1-x_4)^{\frac{49\nu}{108}}+3(1-x_4)^{\frac{67\nu}{108}}+(1-x_4)^{\frac{3\nu}{4}}\Bigr)\, .
\end{multline}

\acknowledgments

\paragraph{Funding information}
This work was supported by the European Research Council under the Starting Grant No. 805252 LoCoMacro.

 \bibliographystyle{SciPost_bibstyle.bst}
 \bibliography{references.bib}

\begin{thebibliography}{100}
\providecommand{\url}[1]{\texttt{#1}}
\providecommand{\urlprefix}{URL }
\expandafter\ifx\csname urlstyle\endcsname\relax
  \providecommand{\doi}[1]{doi:\discretionary{}{}{}#1}\else
  \providecommand{\doi}{doi:\discretionary{}{}{}\begingroup
  \urlstyle{rm}\Url}\fi
\providecommand{\eprint}[2][]{\url{#2}}

\bibitem{Polkovnikov2011Colloquium}
A.~Polkovnikov, K.~Sengupta, A.~Silva and M.~Vengalattore,
\newblock \emph{{Colloquium: Nonequilibrium dynamics of closed interacting
  quantum systems}},
\newblock Rev. Mod. Phys. \textbf{83}, 863 (2011),
\newblock \doi{10.1103/RevModPhys.83.863}.

\bibitem{Eisert2015Quantum}
J.~Eisert, M.~Friesdorf and C.~Gogolin,
\newblock \emph{{Quantum many-body systems out of equilibrium}},
\newblock Nature Phys \textbf{11}, 124 (2015),
\newblock \doi{10.1038/nphys3215}.

\bibitem{Gogolin2016Equilibration}
C.~Gogolin and J.~Eisert,
\newblock \emph{{Equilibration, thermalisation, and the emergence of
  statistical mechanics in closed quantum systems}},
\newblock Rep. Prog. Phys. \textbf{79}(5), 056001 (2016),
\newblock \doi{10.1088/0034-4885/79/5/056001}.

\bibitem{Rigol2008}
M.~Rigol, V.~Dunjko and M.~Olshanii,
\newblock \emph{{Thermalization and its mechanism for generic isolated quantum
  systems}},
\newblock Nature \textbf{452}(7189), 854 (2008),
\newblock \doi{10.1038/nature06838}.

\bibitem{deutsch91}
J.~M. Deutsch,
\newblock \emph{{Quantum statistical mechanics in a closed system}},
\newblock Phys. Rev. A \textbf{43}, 2046 (1991),
\newblock \doi{10.1103/PhysRevA.43.2046}.

\bibitem{Srednicki1994Chaos}
M.~Srednicki,
\newblock \emph{{Chaos and quantum thermalization}},
\newblock Phys. Rev. E \textbf{50}, 888 (1994),
\newblock \doi{10.1103/PhysRevE.50.888}.

\bibitem{Deutsch_2018}
J.~M. Deutsch,
\newblock \emph{{Eigenstate thermalization hypothesis}},
\newblock Reports on Progress in Physics \textbf{81}(8), 082001 (2018),
\newblock \doi{10.1088/1361-6633/aac9f1}.

\bibitem{Kinoshita2006}
T.~Kinoshita, T.~Wenger and D.~S. Weiss,
\newblock \emph{A quantum newton's cradle},
\newblock Nature \textbf{440}(7086), 900 (2006),
\newblock \doi{10.1038/nature04693}.

\bibitem{Rigol2007Relaxation}
M.~Rigol, V.~Dunjko, V.~Yurovsky and M.~Olshanii,
\newblock \emph{{Relaxation in a Completely Integrable Many-Body Quantum
  System: An Ab Initio Study of the Dynamics of the Highly Excited States of 1D
  Lattice Hard-Core Bosons}},
\newblock Phys. Rev. Lett. \textbf{98}, 050405 (2007),
\newblock \doi{10.1103/PhysRevLett.98.050405}.

\bibitem{Vidmar2016Generalized}
L.~Vidmar and M.~Rigol,
\newblock \emph{{Generalized Gibbs ensemble in integrable lattice models}},
\newblock J. Stat. Mech. \textbf{2016}(6), 064007 (2016),
\newblock \doi{10.1088/1742-5468/2016/06/064007}.

\bibitem{Eisert2010Area_laws}
J.~Eisert, M.~Cramer and M.~B. Plenio,
\newblock \emph{Colloquium: Area laws for the entanglement entropy},
\newblock Rev. Mod. Phys. \textbf{82}, 277 (2010),
\newblock \doi{10.1103/RevModPhys.82.277}.

\bibitem{Maric2022Universality}
V.~Mari\'c and M.~Fagotti,
\newblock \emph{{Universality in the tripartite information after global
  quenches}},
\newblock \doi{10.48550/ARXIV.2209.14253},
\newblock \urlprefix\url{https://arxiv.org/abs/2209.14253} (2022),
  \eprint{2209.14253}.

\bibitem{Amico2008}
L.~Amico, R.~Fazio, A.~Osterloh and V.~Vedral,
\newblock \emph{Entanglement in many-body systems},
\newblock Rev. Mod. Phys. \textbf{80}, 517 (2008),
\newblock \doi{10.1103/RevModPhys.80.517}.

\bibitem{Horodecki2009Quantum}
R.~Horodecki, P.~Horodecki, M.~Horodecki and K.~Horodecki,
\newblock \emph{{Quantum entanglement}},
\newblock Rev. Mod. Phys. \textbf{81}, 865 (2009),
\newblock \doi{10.1103/RevModPhys.81.865}.

\bibitem{Kaufman2016}
A.~M. Kaufman, M.~E. Tai, A.~Lukin, M.~Rispoli, R.~Schittko, P.~M. Preiss and
  M.~Greiner,
\newblock \emph{Quantum thermalization through entanglement in an isolated
  many-body system},
\newblock Science \textbf{353}(6301), 794 (2016),
\newblock \doi{10.1126/science.aaf6725},
\newblock \eprint{https://www.science.org/doi/pdf/10.1126/science.aaf6725}.

\bibitem{Calabrese2004Entanglement}
P.~Calabrese and J.~Cardy,
\newblock \emph{{Entanglement entropy and quantum field theory}},
\newblock Journal of Statistical Mechanics: Theory and Experiment
  \textbf{2004}(06), P06002 (2004),
\newblock \doi{10.1088/1742-5468/2004/06/p06002}.

\bibitem{Calabrese2009Entanglement}
P.~Calabrese and J.~Cardy,
\newblock \emph{{Entanglement entropy and conformal field theory}},
\newblock Journal of Physics A: Mathematical and Theoretical \textbf{42}(50),
  504005 (2009),
\newblock \doi{10.1088/1751-8113/42/50/504005}.

\bibitem{Holzhey1994Geometric}
C.~Holzhey, F.~Larsen and F.~Wilczek,
\newblock \emph{Geometric and renormalized entropy in conformal field theory},
\newblock Nuclear Physics B \textbf{424}(3), 443 (1994),
\newblock \doi{https://doi.org/10.1016/0550-3213(94)90402-2}.

\bibitem{Korepin2004PRL}
V.~E. Korepin,
\newblock \emph{Universality of entropy scaling in one dimensional gapless
  models},
\newblock Phys. Rev. Lett. \textbf{92}, 096402 (2004),
\newblock \doi{10.1103/PhysRevLett.92.096402}.

\bibitem{Kitaev2006Topological}
A.~Kitaev and J.~Preskill,
\newblock \emph{{Topological Entanglement Entropy}},
\newblock Phys. Rev. Lett. \textbf{96}, 110404 (2006),
\newblock \doi{10.1103/PhysRevLett.96.110404}.

\bibitem{Cerf1998Information}
N.~J. Cerf and C.~Adami,
\newblock \emph{{Information theory of quantum entanglement and measurement}},
\newblock Physica D: Nonlinear Phenomena \textbf{120}(1), 62 (1998),
\newblock \doi{https://doi.org/10.1016/S0167-2789(98)00045-1},
\newblock Proceedings of the Fourth Workshop on Physics and Consumption.

\bibitem{Calabrese2005Evolution}
P.~Calabrese and J.~Cardy,
\newblock \emph{{Evolution of entanglement entropy in one-dimensional
  systems}},
\newblock J. Stat. Mech. \textbf{2005}(04), P04010 (2005),
\newblock \doi{10.1088/1742-5468/2005/04/p04010}.

\bibitem{Sotiriadis2008}
S.~Sotiriadis and J.~Cardy,
\newblock \emph{Inhomogeneous quantum quenches},
\newblock Journal of Statistical Mechanics: Theory and Experiment
  \textbf{2008}(11), P11003 (2008),
\newblock \doi{10.1088/1742-5468/2008/11/P11003}.

\bibitem{Bastianello2018Spreading}
A.~Bastianello and P.~Calabrese,
\newblock \emph{{Spreading of entanglement and correlations after a quench with
  intertwined quasiparticles}},
\newblock SciPost Phys. \textbf{5}, 033 (2018),
\newblock \doi{10.21468/SciPostPhys.5.4.033}.

\bibitem{Bertini2022Growth}
B.~Bertini, K.~Klobas, V.~Alba, G.~Lagnese and P.~Calabrese,
\newblock \emph{Growth of r\'enyi entropies in interacting integrable models
  and the breakdown of the quasiparticle picture},
\newblock Phys. Rev. X \textbf{12}, 031016 (2022),
\newblock \doi{10.1103/PhysRevX.12.031016}.

\bibitem{Alba2017Entanglement}
V.~Alba and P.~Calabrese,
\newblock \emph{{Entanglement and thermodynamics after a quantum quench in
  integrable systems}},
\newblock Proceedings of the National Academy of Sciences \textbf{114}(30),
  7947 (2017),
\newblock \doi{10.1073/pnas.1703516114},
\newblock \eprint{https://www.pnas.org/doi/pdf/10.1073/pnas.1703516114}.

\bibitem{Alba2018Entanglement}
V.~Alba and P.~Calabrese,
\newblock \emph{{Entanglement dynamics after quantum quenches in generic
  integrable systems}},
\newblock SciPost Phys. \textbf{4}, 017 (2018),
\newblock \doi{10.21468/SciPostPhys.4.3.017}.

\bibitem{Casini2016spread}
H.~Casini, H.~Liu and M.~Mezei,
\newblock \emph{Spread of entanglement and causality},
\newblock Journal of High Energy Physics \textbf{2016}(7), 77 (2016),
\newblock \doi{10.1007/JHEP07(2016)077}.

\bibitem{Liu2014}
H.~Liu and S.~J. Suh,
\newblock \emph{Entanglement growth during thermalization in holographic
  systems},
\newblock Phys. Rev. D \textbf{89}, 066012 (2014),
\newblock \doi{10.1103/PhysRevD.89.066012}.

\bibitem{Skinner2019}
B.~Skinner, J.~Ruhman and A.~Nahum,
\newblock \emph{Measurement-induced phase transitions in the dynamics of
  entanglement},
\newblock Physical Review X \textbf{9}(3) (2019),
\newblock \doi{10.1103/physrevx.9.031009}.

\bibitem{Alba2019QuantumInformationScrambling}
V.~Alba and P.~Calabrese,
\newblock \emph{Quantum information scrambling after a quantum quench},
\newblock Phys. Rev. B \textbf{100}, 115150 (2019),
\newblock \doi{10.1103/PhysRevB.100.115150}.

\bibitem{Bertini2022EntanglementNegativity}
B.~Bertini, K.~Klobas and T.-C. Lu,
\newblock \emph{Entanglement negativity and mutual information after a quantum
  quench: Exact link from space-time duality},
\newblock Phys. Rev. Lett. \textbf{129}, 140503 (2022),
\newblock \doi{10.1103/PhysRevLett.129.140503}.

\bibitem{Eisler2014}
V.~Eisler and Z.~Zimbor\'as,
\newblock \emph{Area-law violation for the mutual information in a
  nonequilibrium steady state},
\newblock Phys. Rev. A \textbf{89}, 032321 (2014),
\newblock \doi{10.1103/PhysRevA.89.032321}.

\bibitem{Parez2022}
G.~Parez and R.~Bonsignori,
\newblock \emph{Analytical results for the entanglement dynamics of disjoint
  blocks in the xy spin chain},
\newblock Journal of Physics A: Mathematical and Theoretical \textbf{55}(50),
  505005 (2023),
\newblock \doi{10.1088/1751-8121/acb097}.

\bibitem{Calabrese2011Quantum}
P.~Calabrese, F.~H.~L. Essler and M.~Fagotti,
\newblock \emph{{Quantum Quench in the Transverse-Field Ising Chain}},
\newblock Phys. Rev. Lett. \textbf{106}, 227203 (2011),
\newblock \doi{10.1103/PhysRevLett.106.227203}.

\bibitem{Calabrese2012}
P.~Calabrese, F.~H.~L. Essler and M.~Fagotti,
\newblock \emph{{Quantum quenches in the transverse field Ising chain: II.
  Stationary state properties}},
\newblock Journal of Statistical Mechanics: Theory and Experiment
  \textbf{2012}(07), P07022 (2012),
\newblock \doi{10.1088/1742-5468/2012/07/p07022}.

\bibitem{Casini2009reduced_density}
H.~Casini and M.~Huerta,
\newblock \emph{Reduced density matrix and internal dynamics for multicomponent
  regions},
\newblock Classical and Quantum Gravity \textbf{26}(18), 185005 (2009),
\newblock \doi{10.1088/0264-9381/26/18/185005}.

\bibitem{Casini2009Entanglement}
H.~Casini and M.~Huerta,
\newblock \emph{{Entanglement entropy in free quantum field theory}},
\newblock Journal of Physics A: Mathematical and Theoretical \textbf{42}(50),
  504007 (2009),
\newblock \doi{10.1088/1751-8113/42/50/504007}.

\bibitem{Fries2019}
P.~Fries and I.~A. Reyes,
\newblock \emph{Entanglement and relative entropy of a chiral fermion on the
  torus},
\newblock Phys. Rev. D \textbf{100}, 105015 (2019),
\newblock \doi{10.1103/PhysRevD.100.105015}.

\bibitem{Blanco2022}
D.~Blanco, T.~Ferreira~Chase, J.~Laurnagaray and G.~P\'erez-Nadal,
\newblock \emph{R\'enyi entropies of the massless dirac field on the torus},
\newblock Phys. Rev. D \textbf{105}, 045014 (2022),
\newblock \doi{10.1103/PhysRevD.105.045014}.

\bibitem{Igloi2010}
F.~Iglói and I.~Peschel,
\newblock \emph{On reduced density matrices for disjoint subsystems},
\newblock Europhysics Letters \textbf{89}(4), 40001 (2010),
\newblock \doi{10.1209/0295-5075/89/40001}.

\bibitem{Fagotti2010disjoint}
M.~Fagotti and P.~Calabrese,
\newblock \emph{{Entanglement entropy of two disjoint blocks in XY chains}},
\newblock Journal of Statistical Mechanics: Theory and Experiment
  \textbf{2010}(04), P04016 (2010),
\newblock \doi{10.1088/1742-5468/2010/04/p04016}.

\bibitem{Wolf2008}
M.~M. Wolf, F.~Verstraete, M.~B. Hastings and J.~I. Cirac,
\newblock \emph{Area laws in quantum systems: Mutual information and
  correlations},
\newblock Phys. Rev. Lett. \textbf{100}, 070502 (2008),
\newblock \doi{10.1103/PhysRevLett.100.070502}.

\bibitem{Kuwahara2021}
T.~Kuwahara, A.~M. Alhambra and A.~Anshu,
\newblock \emph{Improved thermal area law and quasilinear time algorithm for
  quantum gibbs states},
\newblock Phys. Rev. X \textbf{11}, 011047 (2021),
\newblock \doi{10.1103/PhysRevX.11.011047}.

\bibitem{Bernigau2015}
H.~Bernigau, M.~J. Kastoryano and J.~Eisert,
\newblock \emph{Mutual information area laws for thermal free fermions},
\newblock Journal of Statistical Mechanics: Theory and Experiment
  \textbf{2015}(2), P02008 (2015),
\newblock \doi{10.1088/1742-5468/2015/02/P02008}.

\bibitem{Lemm2022}
M.~Lemm and O.~Siebert,
\newblock \emph{Thermal area law for lattice bosons},
\newblock \doi{10.48550/ARXIV.2207.07760},
\newblock \urlprefix\url{https://arxiv.org/abs/2207.07760} (2022).

\bibitem{Alhambra2022}
A.~M. Alhambra,
\newblock \emph{Quantum many-body systems in thermal equilibrium},
\newblock \doi{10.48550/ARXIV.2204.08349},
\newblock \urlprefix\url{https://arxiv.org/abs/2204.08349} (2022).

\bibitem{Casini2005}
H.~Casini, C.~D. Fosco and M.~Huerta,
\newblock \emph{Entanglement and alpha entropies for a massive dirac field in
  two dimensions},
\newblock Journal of Statistical Mechanics: Theory and Experiment
  \textbf{2005}(07), P07007 (2005),
\newblock \doi{10.1088/1742-5468/2005/07/P07007}.

\bibitem{Calabrese2009Entanglement1}
P.~Calabrese, J.~Cardy and E.~Tonni,
\newblock \emph{{Entanglement entropy of two disjoint intervals in conformal
  field theory}},
\newblock Journal of Statistical Mechanics: Theory and Experiment
  \textbf{2009}(04), P11001 (2009),
\newblock \doi{10.1088/1742-5468/2009/11/P11001}.

\bibitem{Casini2007mutual}
H.~Casini,
\newblock \emph{Mutual information challenges entropy bounds},
\newblock Classical and Quantum Gravity \textbf{24}(5), 1293 (2007),
\newblock \doi{10.1088/0264-9381/24/5/013}.

\bibitem{Furukawa2009Mutual}
S.~Furukawa, V.~Pasquier and J.~Shiraishi,
\newblock \emph{{Mutual Information and Boson Radius in a $c=1$ Critical System
  in One Dimension}},
\newblock Phys. Rev. Lett. \textbf{102}, 170602 (2009),
\newblock \doi{10.1103/PhysRevLett.102.170602}.

\bibitem{Headrick2010}
M.~Headrick,
\newblock \emph{Entanglement r\'enyi entropies in holographic theories},
\newblock Phys. Rev. D \textbf{82}, 126010 (2010),
\newblock \doi{10.1103/PhysRevD.82.126010}.

\bibitem{Chen2013JHEP}
B.~Chen and J.~ju~Zhang,
\newblock \emph{On short interval expansion of r{\'{e}}nyi entropy},
\newblock Journal of High Energy Physics \textbf{2013}(11) (2013),
\newblock \doi{10.1007/jhep11(2013)164}.

\bibitem{Chen2014Holographic}
B.~Chen, F.-y. Song and J.-j. Zhang,
\newblock \emph{Holographic r{\'e}nyi entropy in ads3/lcft2 correspondence},
\newblock Journal of High Energy Physics \textbf{2014}(3), 137 (2014),
\newblock \doi{10.1007/JHEP03(2014)137}.

\bibitem{Headrick2015}
M.~Headrick, A.~Maloney, E.~Perlmutter and I.~G. Zadeh,
\newblock \emph{R{\'e}nyi entropies, the analytic bootstrap, and 3d quantum
  gravity at higher genus},
\newblock Journal of High Energy Physics \textbf{2015}(7), 59 (2015),
\newblock \doi{10.1007/JHEP07(2015)059}.

\bibitem{Casini2015Mutual}
H.~Casini, M.~Huerta, R.~C. Myers and A.~Yale,
\newblock \emph{Mutual information and the f-theorem},
\newblock Journal of High Energy Physics \textbf{2015}(10), 3 (2015),
\newblock \doi{10.1007/JHEP10(2015)003}.

\bibitem{Blanco2011}
D.~D. Blanco and H.~Casini,
\newblock \emph{Entanglement entropy for non-coplanar regions in quantum field
  theory},
\newblock Classical and Quantum Gravity \textbf{28}(21), 215015 (2011),
\newblock \doi{10.1088/0264-9381/28/21/215015}.

\bibitem{Arias2018}
R.~E. Arias, H.~Casini, M.~Huerta and D.~Pontello,
\newblock \emph{Entropy and modular hamiltonian for a free chiral scalar in two
  intervals},
\newblock Phys. Rev. D \textbf{98}, 125008 (2018),
\newblock \doi{10.1103/PhysRevD.98.125008}.

\bibitem{Blanco2019}
D.~Blanco, A.~Garbarz and G.~P{\'e}rez-Nadal,
\newblock \emph{Entanglement of a chiral fermion on the torus},
\newblock Journal of High Energy Physics \textbf{2019}(9), 76 (2019),
\newblock \doi{10.1007/JHEP09(2019)076}.

\bibitem{Ares2022}
F.~Ares, P.~Calabrese, G.~Di~Giulio and S.~Murciano,
\newblock \emph{Multi-charged moments of two intervals in conformal field
  theory},
\newblock Journal of High Energy Physics \textbf{2022}(9), 51 (2022),
\newblock \doi{10.1007/JHEP09(2022)051}.

\bibitem{Asplund2014}
C.~T. Asplund and A.~Bernamonti,
\newblock \emph{Mutual information after a local quench in conformal field
  theory},
\newblock Phys. Rev. D \textbf{89}, 066015 (2014),
\newblock \doi{10.1103/PhysRevD.89.066015}.

\bibitem{Molina-Vilaplana2011}
J.~Molina-Vilaplana and P.~Sodano,
\newblock \emph{Holographic view on quantum correlations and mutual information
  between disjoint blocks of a quantum critical system},
\newblock Journal of High Energy Physics \textbf{2011}(10), 11 (2011),
\newblock \doi{10.1007/JHEP10(2011)011}.

\bibitem{Lepori2022}
L.~Lepori, S.~Paganelli, F.~Franchini and A.~Trombettoni,
\newblock \emph{Mutual information for fermionic systems},
\newblock Phys. Rev. Res. \textbf{4}, 033212 (2022),
\newblock \doi{10.1103/PhysRevResearch.4.033212}.

\bibitem{Casini2009Remarks}
H.~Casini and M.~Huerta,
\newblock \emph{{Remarks on the entanglement entropy for disconnected
  regions}},
\newblock Journal of High Energy Physics \textbf{2009}(03), 048 (2009),
\newblock \doi{10.1088/1126-6708/2009/03/048}.

\bibitem{Shiba2012}
N.~Shiba,
\newblock \emph{Entanglement entropy of two spheres},
\newblock Journal of High Energy Physics \textbf{2012}(7), 100 (2012),
\newblock \doi{10.1007/JHEP07(2012)100}.

\bibitem{Cardy2013}
J.~Cardy,
\newblock \emph{Some results on the mutual information of disjoint regions in
  higher dimensions},
\newblock Journal of Physics A: Mathematical and Theoretical \textbf{46}(28),
  285402 (2013),
\newblock \doi{10.1088/1751-8113/46/28/285402}.

\bibitem{Allais2012}
A.~Allais and E.~Tonni,
\newblock \emph{Holographic evolution of the mutual information},
\newblock Journal of High Energy Physics \textbf{2012}(1), 102 (2012),
\newblock \doi{10.1007/JHEP01(2012)102}.

\bibitem{Agon2016}
C.~A. Ag{\'o}n and T.~Faulkner,
\newblock \emph{Quantum corrections to holographic mutual information},
\newblock Journal of High Energy Physics \textbf{2016}(8), 118 (2016),
\newblock \doi{10.1007/JHEP08(2016)118}.

\bibitem{Agon2016Large}
C.~A. Ag{\'o}n, I.~Cohen-Abbo and H.~J. Schnitzer,
\newblock \emph{Large distance expansion of mutual information for disjoint
  disks in a free scalar theory},
\newblock Journal of High Energy Physics \textbf{2016}(11), 73 (2016),
\newblock \doi{10.1007/JHEP11(2016)073}.

\bibitem{Chen2017PRD}
B.~Chen and J.~Long,
\newblock \emph{R\'enyi mutual information for a free scalar field in even
  dimensions},
\newblock Phys. Rev. D \textbf{96}, 045006 (2017),
\newblock \doi{10.1103/PhysRevD.96.045006}.

\bibitem{Casini2015area}
H.~Casini, F.~D. Mazzitelli and E.~Test\'e,
\newblock \emph{Area terms in entanglement entropy},
\newblock Phys. Rev. D \textbf{91}, 104035 (2015),
\newblock \doi{10.1103/PhysRevD.91.104035}.

\bibitem{Tonni2011}
E.~Tonni,
\newblock \emph{Holographic entanglement entropy: near horizon geometry and
  disconnected regions},
\newblock Journal of High Energy Physics \textbf{2011}(5), 4 (2011),
\newblock \doi{10.1007/JHEP05(2011)004}.

\bibitem{Groisman2005}
B.~Groisman, S.~Popescu and A.~Winter,
\newblock \emph{Quantum, classical, and total amount of correlations in a
  quantum state},
\newblock Phys. Rev. A \textbf{72}, 032317 (2005),
\newblock \doi{10.1103/PhysRevA.72.032317}.

\bibitem{Caraglio2008Entanglement}
M.~Caraglio and F.~Gliozzi,
\newblock \emph{{Entanglement entropy and twist fields}},
\newblock Journal of High Energy Physics \textbf{2008}(11), 076 (2008),
\newblock \doi{10.1088/1126-6708/2008/11/076}.

\bibitem{Rajabpour2012}
M.~A. Rajabpour and F.~Gliozzi,
\newblock \emph{Entanglement entropy of two disjoint intervals from fusion
  algebra of twist fields},
\newblock Journal of Statistical Mechanics: Theory and Experiment
  \textbf{2012}(02), P02016 (2012),
\newblock \doi{10.1088/1742-5468/2012/02/P02016}.

\bibitem{Calabrese2011Entanglement}
P.~Calabrese, J.~Cardy and E.~Tonni,
\newblock \emph{{Entanglement entropy of two disjoint intervals in conformal
  field theory: {II}}},
\newblock Journal of Statistical Mechanics: Theory and Experiment
  \textbf{2011}(01), P01021 (2011),
\newblock \doi{10.1088/1742-5468/2011/01/p01021}.

\bibitem{Coser2014OnRenyi}
A.~Coser, L.~Tagliacozzo and E.~Tonni,
\newblock \emph{On rényi entropies of disjoint intervals in conformal field
  theory},
\newblock Journal of Statistical Mechanics: Theory and Experiment
  \textbf{2014}(1), P01008 (2014),
\newblock \doi{10.1088/1742-5468/2014/01/P01008}.

\bibitem{Ruggiero2018}
P.~Ruggiero, E.~Tonni and P.~Calabrese,
\newblock \emph{Entanglement entropy of two disjoint intervals and the
  recursion formula for conformal blocks},
\newblock Journal of Statistical Mechanics: Theory and Experiment
  \textbf{2018}(11), 113101 (2018),
\newblock \doi{10.1088/1742-5468/aae5a8}.

\bibitem{Alba2010Entanglement}
V.~Alba, L.~Tagliacozzo and P.~Calabrese,
\newblock \emph{Entanglement entropy of two disjoint blocks in critical ising
  models},
\newblock Phys. Rev. B \textbf{81}, 060411 (2010),
\newblock \doi{10.1103/PhysRevB.81.060411}.

\bibitem{Alba2011Entanglement}
V.~Alba, L.~Tagliacozzo and P.~Calabrese,
\newblock \emph{{Entanglement entropy of two disjoint intervals in c=1
  theories}},
\newblock Journal of Statistical Mechanics: Theory and Experiment
  \textbf{2011}(06), P06012 (2011),
\newblock \doi{10.1088/1742-5468/2011/06/p06012}.

\bibitem{Fagotti2012New}
M.~Fagotti,
\newblock \emph{New insights into the entanglement of disjoint blocks},
\newblock {EPL} (Europhysics Letters) \textbf{97}(1), 17007 (2012),
\newblock \doi{10.1209/0295-5075/97/17007}.

\bibitem{Balasubramanian2011}
V.~Balasubramanian, A.~Bernamonti, N.~Copland, B.~Craps and F.~Galli,
\newblock \emph{Thermalization of mutual and tripartite information in strongly
  coupled two dimensional conformal field theories},
\newblock Phys. Rev. D \textbf{84}, 105017 (2011),
\newblock \doi{10.1103/PhysRevD.84.105017}.

\bibitem{Grava2021}
T.~Grava, A.~P. Kels and E.~Tonni,
\newblock \emph{Entanglement of two disjoint intervals in conformal field
  theory and the 2d coulomb gas on a lattice},
\newblock Phys. Rev. Lett. \textbf{127}, 141605 (2021),
\newblock \doi{10.1103/PhysRevLett.127.141605}.

\bibitem{Ares2021Crossing}
F.~Ares, R.~Santachiara and J.~Viti,
\newblock \emph{{Crossing-symmetric twist field correlators and entanglement
  negativity in minimal CFTs}},
\newblock Journal of High Energy Physics \textbf{2021}(10), 175 (2021),
\newblock \doi{10.1007/JHEP10(2021)175}.

\bibitem{Agon2022}
C.~A. Agón, P.~Bueno and H.~Casini,
\newblock \emph{{Tripartite information at long distances}},
\newblock SciPost Phys. \textbf{12}, 153 (2022),
\newblock \doi{10.21468/SciPostPhys.12.5.153}.

\bibitem{AliAkbari2021}
M.~Ali-Akbari, M.~Rahimi and M.~Asadi,
\newblock \emph{Holographic mutual and tripartite information in a
  non-conformal background},
\newblock Nuclear Physics B \textbf{964}, 115329 (2021),
\newblock \doi{https://doi.org/10.1016/j.nuclphysb.2021.115329}.

\bibitem{Hayden2013Holographic}
P.~Hayden, M.~Headrick and A.~Maloney,
\newblock \emph{Holographic mutual information is monogamous},
\newblock Phys. Rev. D \textbf{87}, 046003 (2013),
\newblock \doi{10.1103/PhysRevD.87.046003}.

\bibitem{Carollo2022}
F.~Carollo and V.~Alba,
\newblock \emph{Entangled multiplets and spreading of quantum correlations in a
  continuously monitored tight-binding chain},
\newblock Phys. Rev. B \textbf{106}, L220304 (2022),
\newblock \doi{10.1103/PhysRevB.106.L220304}.

\bibitem{Parez2022Multipartite}
G.~Parez, P.-A. Bernard, N.~Crampé and L.~Vinet,
\newblock \emph{Multipartite information of free fermions on hamming graphs},
\newblock Nuclear Physics B \textbf{990}, 116157 (2023),
\newblock \doi{https://doi.org/10.1016/j.nuclphysb.2023.116157}.

\bibitem{Hosur2016Chaos}
P.~Hosur, X.-L. Qi, D.~A. Roberts and B.~Yoshida,
\newblock \emph{{Chaos in quantum channels}},
\newblock Journal of High Energy Physics \textbf{2016}(2), 4 (2016),
\newblock \doi{10.1007/JHEP02(2016)004}.

\bibitem{Schnaack2019Tripartite}
O.~Schnaack, N.~B\"olter, S.~Paeckel, S.~R. Manmana, S.~Kehrein and M.~Schmitt,
\newblock \emph{{Tripartite information, scrambling, and the role of Hilbert
  space partitioning in quantum lattice models}},
\newblock Phys. Rev. B \textbf{100}, 224302 (2019),
\newblock \doi{10.1103/PhysRevB.100.224302}.

\bibitem{Sunderhauf2019Quantum}
C.~S{\"u}nderhauf, L.~Piroli, X.-L. Qi, N.~Schuch and J.~I. Cirac,
\newblock \emph{{Quantum chaos in the Brownian SYK model with large finite N :
  OTOCs and tripartite information}},
\newblock Journal of High Energy Physics \textbf{2019}(11), 38 (2019),
\newblock \doi{10.1007/JHEP11(2019)038}.

\bibitem{Kuno2022}
Y.~Kuno, T.~Orito and I.~Ichinose,
\newblock \emph{Purification and scrambling in a chaotic hamiltonian dynamics
  with measurements},
\newblock Phys. Rev. B \textbf{106}, 214304 (2022),
\newblock \doi{10.1103/PhysRevB.106.214304}.

\bibitem{Hastings2007}
M.~B. Hastings,
\newblock \emph{An area law for one-dimensional quantum systems},
\newblock Journal of Statistical Mechanics: Theory and Experiment
  \textbf{2007}(08), P08024 (2007),
\newblock \doi{10.1088/1742-5468/2007/08/P08024}.

\bibitem{Jin2004}
B.-Q. Jin and V.~E. Korepin,
\newblock \emph{{Quantum Spin Chain, Toeplitz Determinants and the
  Fisher---Hartwig Conjecture}},
\newblock Journal of Statistical Physics \textbf{116}(1), 79 (2004),
\newblock \doi{10.1023/B:JOSS.0000037230.37166.42}.

\bibitem{Fraenkel2021}
S.~Fraenkel and M.~Goldstein,
\newblock \emph{{Entanglement measures in a nonequilibrium steady state: Exact
  results in one dimension}},
\newblock SciPost Phys. \textbf{11}, 085 (2021),
\newblock \doi{10.21468/SciPostPhys.11.4.085}.

\bibitem{FagottiMaricZadnik2022}
M.~Fagotti, V.~Mari\'c and L.~Zadnik,
\newblock \emph{Nonequilibrium symmetry-protected topological order: emergence
  of semilocal gibbs ensembles},
\newblock \doi{10.48550/ARXIV.2205.02221},
\newblock \urlprefix\url{https://arxiv.org/abs/2205.02221} (2022),
  \eprint{2205.02221}.

\bibitem{Alba2009Entanglement}
V.~Alba, M.~Fagotti and P.~Calabrese,
\newblock \emph{{Entanglement entropy of excited states}},
\newblock Journal of Statistical Mechanics: Theory and Experiment
  \textbf{2009}(10), P10020 (2009),
\newblock \doi{10.1088/1742-5468/2009/10/p10020}.

\bibitem{Ares2014}
F.~Ares, J.~G. Esteve, F.~Falceto and E.~Sánchez-Burillo,
\newblock \emph{Excited state entanglement in homogeneous fermionic chains},
\newblock Journal of Physics A: Mathematical and Theoretical \textbf{47}(24),
  245301 (2014),
\newblock \doi{10.1088/1751-8113/47/24/245301}.

\bibitem{Bluhm2022exponentialdecayof}
A.~Bluhm, {\'{A}}.~Capel and A.~P{\'{e}}rez-Hern{\'{a}}ndez,
\newblock \emph{Exponential decay of mutual information for {G}ibbs states of
  local {H}amiltonians},
\newblock {Quantum} \textbf{6}, 650 (2022),
\newblock \doi{10.22331/q-2022-02-10-650}.

\bibitem{Hastings2006}
M.~B. Hastings and T.~Koma,
\newblock \emph{Spectral gap and exponential decay of correlations},
\newblock Communications in Mathematical Physics \textbf{265}(3), 781 (2006),
\newblock \doi{10.1007/s00220-006-0030-4}.

\bibitem{Suzuki1971The}
M.~Suzuki,
\newblock \emph{{The dimer problem and the generalized X-model}},
\newblock Physics Letters A \textbf{34}(6), 338 (1971),
\newblock \doi{https://doi.org/10.1016/0375-9601(71)90901-7}.

\bibitem{Lieb1961}
E.~Lieb, T.~Schultz and D.~Mattis,
\newblock \emph{{Two soluble models of an antiferromagnetic chain}},
\newblock Annals of Physics \textbf{16}(3), 407 (1961),
\newblock \doi{https://doi.org/10.1016/0003-4916(61)90115-4}.

\bibitem{Fagotti2016Charges}
M.~Fagotti,
\newblock \emph{{Charges and currents in quantum spin chains: late-time
  dynamics and spontaneous currents}},
\newblock J. Phys. A: Math. Theor. \textbf{50}(3), 034005 (2016),
\newblock \doi{10.1088/1751-8121/50/3/034005}.

\bibitem{Calabrese2016Quantum}
P.~Calabrese and J.~Cardy,
\newblock \emph{{Quantum quenches in 1+1 dimensional conformal field
  theories}},
\newblock J. Stat. Mech. \textbf{2016}(6), 064003 (2016),
\newblock \doi{10.1088/1742-5468/2016/06/064003}.

\bibitem{Essler2016Quench}
F.~H.~L. Essler and M.~Fagotti,
\newblock \emph{{Quench dynamics and relaxation in isolated integrable quantum
  spin chains}},
\newblock J. Stat. Mech. \textbf{2016}(6), 064002 (2016),
\newblock \doi{10.1088/1742-5468/2016/06/064002}.

\bibitem{Caux2016}
J.-S. Caux,
\newblock \emph{{The Quench Action}},
\newblock Journal of Statistical Mechanics: Theory and Experiment
  \textbf{2016}(6), 064006 (2016),
\newblock \doi{10.1088/1742-5468/2016/06/064006}.

\bibitem{Sabetta2013}
T.~Sabetta and G.~Misguich,
\newblock \emph{Nonequilibrium steady states in the quantum xxz spin chain},
\newblock Phys. Rev. B \textbf{88}, 245114 (2013),
\newblock \doi{10.1103/PhysRevB.88.245114}.

\bibitem{Eisler2014Entanglement}
V.~Eisler and Z.~Zimborás,
\newblock \emph{Entanglement negativity in the harmonic chain out of
  equilibrium},
\newblock New Journal of Physics \textbf{16}(12), 123020 (2014),
\newblock \doi{10.1088/1367-2630/16/12/123020}.

\bibitem{Mazza2018}
L.~Mazza, J.~Viti, M.~Carrega, D.~Rossini and A.~De~Luca,
\newblock \emph{Energy transport in an integrable parafermionic chain via
  generalized hydrodynamics},
\newblock Phys. Rev. B \textbf{98}, 075421 (2018),
\newblock \doi{10.1103/PhysRevB.98.075421}.

\bibitem{Bertini2018Entanglement}
B.~Bertini, M.~Fagotti, L.~Piroli and P.~Calabrese,
\newblock \emph{Entanglement evolution and generalised hydrodynamics:
  noninteracting systems},
\newblock Journal of Physics A: Mathematical and Theoretical \textbf{51}(39),
  39LT01 (2018),
\newblock \doi{10.1088/1751-8121/aad82e}.

\bibitem{Alba2019Entanglement}
V.~Alba, B.~Bertini and M.~Fagotti,
\newblock \emph{{Entanglement evolution and generalised hydrodynamics:
  interacting integrable systems}},
\newblock SciPost Phys. \textbf{7}, 5 (2019),
\newblock \doi{10.21468/SciPostPhys.7.1.005}.

\bibitem{Gruber2019Magnetization}
M.~Gruber and V.~Eisler,
\newblock \emph{{Magnetization and entanglement after a geometric quench in the
  XXZ chain}},
\newblock Phys. Rev. B \textbf{99}, 174403 (2019),
\newblock \doi{10.1103/PhysRevB.99.174403}.

\bibitem{Collura2020Domain}
M.~Collura, A.~De~Luca, P.~Calabrese and J.~Dubail,
\newblock \emph{{Domain wall melting in the spin-${\frac{1}{2}}$ XXZ spin
  chain: Emergent Luttinger liquid with a fractal quasiparticle charge}},
\newblock Phys. Rev. B \textbf{102}, 180409 (2020),
\newblock \doi{10.1103/PhysRevB.102.180409}.

\bibitem{Bastianello2022Introduction}
A.~Bastianello, B.~Bertini, B.~Doyon and R.~Vasseur,
\newblock \emph{Introduction to the special issue on emergent hydrodynamics in
  integrable many-body systems},
\newblock Journal of Statistical Mechanics: Theory and Experiment
  \textbf{2022}(1), 014001 (2022),
\newblock \doi{10.1088/1742-5468/ac3e6a}.

\bibitem{Alba2021Generalized}
V.~Alba, B.~Bertini, M.~Fagotti, L.~Piroli and P.~Ruggiero,
\newblock \emph{{Generalized-hydrodynamic approach to inhomogeneous quenches:
  correlations, entanglement and quantum effects}},
\newblock Journal of Statistical Mechanics: Theory and Experiment
  \textbf{2021}(11), 114004 (2021),
\newblock \doi{10.1088/1742-5468/ac257d}.

\bibitem{Bertini2016Determination}
B.~Bertini and M.~Fagotti,
\newblock \emph{{Determination of the Nonequilibrium Steady State Emerging from
  a Defect}},
\newblock Phys. Rev. Lett. \textbf{117}, 130402 (2016),
\newblock \doi{10.1103/PhysRevLett.117.130402}.

\bibitem{Bertini2016Transport}
B.~Bertini, M.~Collura, J.~De~Nardis and M.~Fagotti,
\newblock \emph{{Transport in Out-of-Equilibrium ${XXZ}$ Chains: Exact Profiles
  of Charges and Currents}},
\newblock Phys. Rev. Lett. \textbf{117}, 207201 (2016),
\newblock \doi{10.1103/PhysRevLett.117.207201}.

\bibitem{Castro-Alvaredo2016Emergent}
O.~A. Castro-Alvaredo, B.~Doyon and T.~Yoshimura,
\newblock \emph{{Emergent Hydrodynamics in Integrable Quantum Systems Out of
  Equilibrium}},
\newblock Phys. Rev. X \textbf{6}, 041065 (2016),
\newblock \doi{10.1103/PhysRevX.6.041065}.

\bibitem{Caux2013Time}
J.-S. Caux and F.~H.~L. Essler,
\newblock \emph{{Time Evolution of Local Observables After Quenching to an
  Integrable Model}},
\newblock Phys. Rev. Lett. \textbf{110}, 257203 (2013),
\newblock \doi{10.1103/PhysRevLett.110.257203}.

\bibitem{Fagotti2020}
M.~Fagotti,
\newblock \emph{{Locally quasi-stationary states in noninteracting spin
  chains}},
\newblock SciPost Phys. \textbf{8}, 48 (2020),
\newblock \doi{10.21468/SciPostPhys.8.3.048}.

\bibitem{Vidal2003}
G.~Vidal, J.~I. Latorre, E.~Rico and A.~Kitaev,
\newblock \emph{{Entanglement in Quantum Critical Phenomena}},
\newblock Phys. Rev. Lett. \textbf{90}, 227902 (2003),
\newblock \doi{10.1103/PhysRevLett.90.227902}.

\bibitem{Klich2014}
I.~Klich,
\newblock \emph{{A note on the full counting statistics of paired fermions}},
\newblock Journal of Statistical Mechanics: Theory and Experiment
  \textbf{2014}(11), P11006 (2014),
\newblock \doi{10.1088/1742-5468/2014/11/p11006}.

\bibitem{Affleck1988LesHouches}
I.~Affleck,
\newblock \emph{{Field Theory and Quantum Critical Phenomena}},
\newblock In \emph{{Les Houches Summer School in Theoretical Physics: Fields,
  Strings, Critical Phenomena}} (1988).

\bibitem{Kadanoff1966}
L.~P. Kadanoff,
\newblock \emph{Scaling laws for ising models near ${T}_{c}$},
\newblock Physics Physique Fizika \textbf{2}, 263 (1966),
\newblock \doi{10.1103/PhysicsPhysiqueFizika.2.263}.

\bibitem{Wilson1974}
K.~G. Wilson,
\newblock \emph{Renormalization group and critical phenomena. i.
  renormalization group and the kadanoff scaling picture},
\newblock Phys. Rev. B \textbf{4}, 3174 (1971),
\newblock \doi{10.1103/PhysRevB.4.3174}.

\bibitem{Cardy2008Form}
J.~L. Cardy, O.~A. Castro-Alvaredo and B.~Doyon,
\newblock \emph{Form factors of branch-point twist fields in quantum integrable
  models and entanglement entropy},
\newblock Journal of Statistical Physics \textbf{130}(1), 129 (2008),
\newblock \doi{10.1007/s10955-007-9422-x}.

\bibitem{Kleinert2009}
H.~Kleinert,
\newblock \emph{{Path Integrals in Quantum Mechanics, Statistics, Polymer
  Physics, and Financial Markets}},
\newblock World Scientific, 5th edn. (2009).

\bibitem{Coser2016Spin}
A.~Coser, E.~Tonni and P.~Calabrese,
\newblock \emph{{Spin structures and entanglement of two disjoint intervals in
  conformal field theories}},
\newblock Journal of Statistical Mechanics: Theory and Experiment
  \textbf{2016}(5), 053109 (2016),
\newblock \doi{10.1088/1742-5468/2016/05/053109}.

\bibitem{Chen2013}
X.~Chen and E.~Fradkin,
\newblock \emph{Quantum entanglement and thermal reduced density matrices in
  fermion and spin systems on ladders},
\newblock Journal of Statistical Mechanics: Theory and Experiment
  \textbf{2013}(08), P08013 (2013),
\newblock \doi{10.1088/1742-5468/2013/08/P08013}.

\bibitem{Mollabashi2014}
A.~Mollabashi, N.~Shiba and T.~Takayanagi,
\newblock \emph{Entanglement between two interacting cfts and generalized
  holographic entanglement entropy},
\newblock Journal of High Energy Physics \textbf{2014}(4), 185 (2014),
\newblock \doi{10.1007/JHEP04(2014)185}.

\bibitem{Furukawa2011}
S.~Furukawa and Y.~B. Kim,
\newblock \emph{Entanglement entropy between two coupled tomonaga-luttinger
  liquids},
\newblock Phys. Rev. B \textbf{83}, 085112 (2011),
\newblock \doi{10.1103/PhysRevB.83.085112}.

\bibitem{Xu2011}
C.~Xu,
\newblock \emph{Entanglement entropy of coupled conformal field theories and
  fermi liquids},
\newblock Phys. Rev. B \textbf{84}, 125119 (2011),
\newblock \doi{10.1103/PhysRevB.84.125119}.

\bibitem{Muskhelishvili1953}
N.~I. Muskhelishvili,
\newblock \emph{Singular Integral Equations},
\newblock P. Noordhoff N.V. - Groningen-Holland (1953).

\bibitem{Its2003TheRP}
A.~Its,
\newblock \emph{{The Riemann-Hilbert Problem and Integrable Systems}},
\newblock Notices of the American Mathematical Society \textbf{50}(11) (2003).

\bibitem{Fagotti2022Global}
M.~Fagotti,
\newblock \emph{{Global Quenches after Localized Perturbations}},
\newblock Phys. Rev. Lett. \textbf{128}, 110602 (2022),
\newblock \doi{10.1103/PhysRevLett.128.110602}.

\bibitem{Bocini2022Connected}
S.~Bocini,
\newblock \emph{Connected correlations in partitioning protocols: a case study
  and beyond},
\newblock \doi{10.48550/ARXIV.2212.07151},
\newblock \urlprefix\url{https://arxiv.org/abs/2212.07151} (2022).

\bibitem{FagottiPhDThesis}
M.~Fagotti,
\newblock \emph{{Entanglement \& Correlations in exactly solvable models}},
\newblock Ph.D. thesis (2011).

\end{thebibliography}
\end{document}